\documentclass[useAMS,usenatbib]{mn2e}
\usepackage{multicol}
\usepackage{multirow}
\usepackage{natbib}

\usepackage{amsmath} 
\usepackage{graphicx}
\usepackage{txfonts}
\usepackage{psfig}
\usepackage{subfigure}
\usepackage{natbib}
\usepackage{aas_macros}

\title[Tree-based solvers for AMR code FLASH - I]
{Tree-based solvers for adaptive mesh refinement code FLASH - I:\\
gravity and optical depths.}
\author[W\"unsch et al.] {R.~W\"unsch$^{1}$ \thanks{E-mail: richard@wunsch.cz},
S. Walch$^{2,3}$, F. Dinnbier$^{1,3,5}$, A. Whitworth$^{4}$ \\
$^{1}$Astronomical Institute, Czech Academy of Sciences, Bo\v{c}n\'\i\ II 1401, 141 00 Prague, Czech Republic\\
$^{2}$Max-Planck-Institute for Astrophysics, Karl-Schwarzschild-Str. 1, 85741 Garching, Germany\\
$^{3}$1 Physikalisches Institut, Universit{\"a}t zu K{\"o}ln, Z{\"u}lpicher Str. 77, 50937 K{\"o}ln, Germany\\
$^{4}$School of Physics \& Astronomy, Cardiff University, The Parade, Cardiff CF24 3AAQ, Wales, UK \\
$^{5}$Faculty of Mathematics and Physics, Charles University in Prague, V Hole\v{s}ovi\v{c}k\'{a}ch 2, CZ-180 00 Prague, Czech Republic}

\newcommand{\erfc}{\mathrm{erfc}}
\newcommand{\erf}{\mathrm{erf}}
\newcommand{\req}[1]{Equation (\ref{#1})}
\newcommand{\du}{\mathrm{du}}
\newcommand{\erfcx}{\mathrm{erfcx}}
\newcommand{\sqrtij}{\sqrt{l_1^2+(l_2/b)^2}}
\newcommand{\NHipix}{N_{{\rm H},i_{\rm PIX}}}
\newcommand{\Aew}{\mathbf{A}}
\newcommand{\Aewa}[1]{A_{#1}}
\newcommand{\eit}{e_{l_1,l_2,l_3}}
\newcommand{\uij}{u_{i_1}}
\newcommand{\uid}{u_{i_1,i_2}}
\newcommand{\uit}{u_{i_1,i_2,i_3}}
\newcommand{\vlj}{v_{l_1}}
\newcommand{\vld}{v_{l_1,l_2}}
\newcommand{\vlt}{v_{l_1,l_2,l_3}}
\newcommand{\dd}{\mathrm{d}}
\newcommand{\dq}{\mathrm{dq}}

\begin{document}

\date{Accepted 2017 December 27. Received 2017 November 28; in original form 2017 August 03}

\pagerange{\pageref{firstpage}--\pageref{lastpage}} \pubyear{2017}

\maketitle

\label{firstpage}

\begin{abstract}

We describe an OctTree algorithm for the MPI-parallel, adaptive mesh-refinement
code {\sc FLASH}, which can be used to calculate the gas self-gravity, and also
the angle-averaged local optical depth, for treating ambient diffuse radiation.
The algorithm communicates to the different processors only those parts of the
tree that are needed to perform the tree walk locally. The advantage of this
approach is a relatively low memory requirement, important in particular for the
optical depth calculation, which needs to process information from many
different directions. This feature also enables a general tree-based radiation
transport algorithm that will be described in a subsequent paper, and delivers
excellent scaling up to at least 1500 cores. Boundary conditions for gravity can
be either isolated or periodic, and they can be specified in each direction
independently, using a newly developed generalisation of the Ewald method. The
gravity calculation can be accelerated with the {\em adaptive block update}
technique by partially re-using the solution from the previous time-step.
Comparison with the {\sc Flash} internal multi-grid gravity solver shows that
tree based methods provide a competitive alternative, particularly for problems
with isolated or mixed boundary conditions. We evaluate several multipole
acceptance criteria (MACs) and identify a relatively simple APE MAC which
provides high  accuracy at low computational cost. The optical depth estimates
are found to agree very well with those of the {\sc RADMC-3D} radiation
transport code, with the tree solver being much faster. Our algorithm is
available in the standard release of the {\sc FLASH} code in version 4.0 and
later.

\end{abstract}


\section{Introduction}%


Solving Poisson's equation for general mass distributions is a common problem in
numerical astrophysics. Grid-based hydrodynamic codes frequently use iterative
multi-grid or spectral methods for that purpose. On the other hand, particle
codes often use tree-based algorithms. The extensive experience with tree
gravity solvers in particle codes can be transferred to the domain of grid-based
codes. Here we describe an implementation of the tree-based gravity solver for
the Adaptive Mesh Refinement (AMR) code {\sc Flash} \citep{Fryxell2000} and show
that its efficiency is comparable to the {\sc Flash} intrinsic multi-grid solver
\citep{Ricker2008}. An advantage of this approach is that the tree code can be
used for more general calculations performed in parallel with the gravity; in
particular, calculation of the optical depth in every cell of the computational
domain with the algorithm developed by \citet{Clark2012a} and general radiation
transport with the TreeRay algorithm (described in Paper~II; W\"unsch et al., in
prep.).

Hierarchically structured, tree-based algorithms represent a well-established
technique for solving the gravitational N-body problem at reduced computational
cost \citep[][hereafter BH86]{Barnes1986}. Many Lagrangian codes implement trees
to compute the self-gravity of both collisionless (stars or dark matter) and
collisional (gas) particles, e.g. {\sc Gadget-2} \citep{Springel2005}, {\sc
Vine} \citep{Wetzstein2009, Nelson2009}; {\sc EvoL} \citep{Merlin2010}, {\sc
Seren}, \citep{Hubber2011}, {\sc Gandalf} \citep{Hubber2018}. The three most
important characteristics of the tree algorithm are the tree structure (also
called the grouping strategy), the multipole acceptance criterion (MAC) deciding
whether to open a child-node or not, and the order of approximation of the
integrated quantity within nodes (e.g. mass distribution).

{\sc Tree structure:} Each node on the tree represents a part of the
computational domain, hereafter a volume, and the child-nodes of a given
parent-node collectively represent the same volume as the parent-node. The most
common 'OctTree' structure is built by a recursive subdivision of the
computational domain, where every parent-node is split into eight
equal-volume child-nodes, until we reach the last generation. The nodes of
the last generation are called leaf-nodes and they cover the whole computational
domain. 

Tree structures other than the OctTree are also often used. \citet{Bentley1979}
constructs a balanced "k-d" binary tree by recursively dividing parent-nodes so
that each of the resulting child-nodes contains half ($\pm 1$) of the particles
in the parent-node; this tree structure is used in the codes {\sc pkdgrav}
\citep{Stadel2001} and {\sc Gasoline} \citep{Wadsley2004}. In contrast,
\citet{Press1986} constructs a binary tree, from the bottom up, by successively
amalgamating nearest neighbour particles or nodes into parent nodes. This
"Press-tree" has been further improved by \citet{JerniganPorter1989}, and is
used, for instance, by \citet{Benz1990} and \citet{Nelson2009}. More complex
structures have been suggested. For example, \citet{Ahn2008} describe the
"k-means" algorithm, in which a parent-node is adaptively divided into $k$ child-nodes 
according to the particle distribution in the parent-node.

There seems to be no unequivocally superior tree structure. \citet{Waltz2002}
compare OctTrees with binary trees, and find that OctTrees provide slightly
better performance with the same accuracy. On the other hand,
\citet{Anderson1999} argues, on the basis of an analytical study, that certain
types of binary trees should provide better performance than OctTrees.
\citet{Makino1990} points out that differences in performance are mainly in the
tree construction part, and that the tree-walk takes a comparable amount of time
in either type of tree-structure. Therefore, the choice of tree-structure should
be informed by more technical issues, like the architecture of the computer to
be used, other software to which the tree will be linked, and so on.

{\sc Multipole acceptance criterion:} Another essential part of a tree code is
the criterion, or criteria, used to decide whether a given node can be used to
calculate the gravitational field, or whether its child-nodes should be
considered instead. This is a key factor determining the accuracy and
performance of the code. Since this criterion often reduces to deciding whether
the multipole expansion representing the contribution from the node in question
provides a sufficiently accurate approximation for the calculation of the
gravitational potential, it is commonly referred to as the multipole acceptance
criterion (MAC). We retain this terminology even though nodes in the code
presented here may possess more general properties than just a multipole
expansion. 

The original BH86 geometric MAC uses a simple criterion, which is purely based
on the ratio of the angular size of a given node and its distance to the cell at
which the gravitational potential should be computed. More elaborate methods
also take into account the mass distribution within a particular node or even
constrain the allowed total acceleration error \citep[][ SW94; see
\S\ref{sec_MAC}]{SalmonWarren1994}.

{\sc Order of approximation:}  \citet{Springel2001} suggest that if the
gravitational acceleration is computed using multipole moments up to order $p$,
then the maximum error is of the order of the contribution from the $(\!p\!+\!1\!)^{\rm
th}$ multipole moment. There is no consensus on where to terminate the multipole
expansion of the mass distribution in a node. The original BH86 tree code uses
moments up to second order ($p\!=\!2$), i.e. quadrupoles, and many authors
follow this choice. \citet{Wadsley2004} find the highest efficiency using
$p\!=\!4$ in the {\sc Gasoline} code. On the other hand, SW94 find that their
code using the SumSquare MAC is most efficient with $p\!=\!1$, i.e. just
monopole moments. This suggests that the optimal choice of $p$ may depend
strongly on other properties of the code and its implementation, and possibly
also on the architecture of the computer. \citet{Springel2005} advocates using
just monopole moments on the basis of memory and cache usage efficiency. We
follow this approach and consider only monopole moments, i.e. $p=1$ for all
implemented MACs.

{\sc Further improvements:} Tree codes have often been extended with new
features or modified to improve their behaviour. \citet{Barnes1990} noted that
neighbouring particles interact with essentially the same nodes, and introduced
interaction lists that save time during a tree-walk. This idea was further
extended by \citet{Dehnen2000,Dehnen2002} who describes a tree with mutual
node-node interactions. This greatly reduces the number of interactions that
have to be calculated, leading -- in theory -- to an ${\cal O}({\cal N})$
CPU-time dependence on the number of particles, ${\cal N}$. Dehnen's
implementation also symmetrizes the gravitational interactions to ensure
accurate momentum conservation, which is in general not guaranteed with
tree-codes. Recently, \citet{2017ComAC...4....2P} develop this so called Fast
Multipole Method (FMM) further and implement it into massively parallel
cosmological N-body code {\sc PKDGRAV3}.

{\sc Hybrid codes.} Tree-codes are also sometimes combined with other algorithms
into 'hybrid' codes. For example, \citet{Xu1995} describes a {\sc TreePM} code
which uses a tree to calculate short-range interactions, and a particle-mesh
method \citep{HockneyEastwood1981} to calculate long-range interactions. The
{\sc TreePM} code has been developed further by \citet{Bode2000,Bagla2002,
BodeOstriker2003, Bagla2009,Khandai2009}. There are also general purpose
tree codes available, that can work with both N-body and grid-based codes, e.g.
the MPI parallel tree gravity solver {\sc FLY} \citet{2007CoPhC.176..211B}.

In this paper we describe a newly developed, cost-efficient, tree-based solver
for self-gravity and diffuse radiation that has been implemented into the {\sc
AMR} code {\sc FLASH}. This code has been developed since 2008, and since {\sc
FLASH} version 4.0 it is a part of the official release. The GPU accelerated
tree gravity solver, based on the early version of the presented code, has been
developed by \citet{2016NewA...45...14L}. The paper is organized as follows: In
\S\ref{sec:alg} we describe the implemented algorithm, which splits up into the
tree-solver (\S\ref{sec:treesolver}), the gravity module (\S\ref{sec:gravity})
and the optical depth module (\S\ref{sec:treecol}). Accuracy and performance for
several static and dynamic tests are discussed in \S\ref{sec:accperf}, and we
conclude in \S\ref{sec:summary}. In appendix \ref{ap:accEwald} we
provide formulae for acceleration in computational domains with periodic and
mixed boundary conditions, and in appendix \ref{ap:runpar} we give runtime
parameters of the code.


\section{The algorithm}
\label{sec:alg}


The {\sc flash} code \citep{Fryxell2000} is a complex framework consisting of
many inter-operable modules that can be combined to solve a specific problem.
The tree code described here can only be used with a subset of the possible {\sc
flash} configurations. The basic requirement is usage of the {\sc
paramesh}-based grid unit (see \citealt{MacNeice2000} for a description of the
{\sc paramesh} library); support for other grid units (uniform grid, Chombo) can
be added in future. Furthermore, the grid geometry must be 3D Cartesian.

The {\sc paramesh} library defines the computational domain as a collection of
blocks organised into a tree data structure which we refer to as the {\em
amr-tree}. Each node on the amr-tree represents a block. The block at the top of
the amr-tree, corresponding to the entire computational domain, is called the
{\em root-block} and represents refinement level $\ell = 1$. The root-block is
divided into eight equal-volume blocks having the same shape and orientation as
the root-block, and these blocks represent refinement level $\ell=2$. This
process of block division is then repeated recursively until the blocks created
satisfy an adaptive-mesh refinement criterion. The blocks at the bottom of the
tree, which are not divided, are called {\em leaf-blocks}, and the refinement
level of a leaf-block is labelled $\ell_{\rm lb}$. In regions where the AMR
criterion requires higher spatial resolution, the leaf-blocks are smaller and
their refinement level, $\ell_{\rm lb}$, is larger (i.e. they are further down
the tree).

The number of grid cells in a block (a logically cuboidal collection of cells; see
below) must be the same in each direction and equal to $2^{\ell_\mathrm{bt}}$
where $\ell_\mathrm{bt}$ is an arbitrary integer number. In practice, it should
be $\ell_\mathrm{bt} \ge 3$, because most hydrodynamic solvers do not allow
blocks containing fewer than $8^3$ cells, in order to avoid overlapping of ghost
cells. Note that the above requirements do not exclude non-cubic computational
domains, because such domains can be created either by setting up blocks with
different physical sizes in each direction or by using more than one root
block\footnote{If there is more than one root block, the single tree structure
becomes a forest. This decreases the efficiency of the gravity solver, and
therefore the number of root blocks should be kept as small as possible.} in
each direction \citep{Walch2015}. 

Within each leaf-block is a local {\em block-tree} which extends the amr-tree
down to the level of individual grid cells. All block-trees have the same number
of levels, $\ell_{\rm bt}\;(\geq\!3)$. The nodes on a block-tree represent
refinement levels $\ell_{\rm lb}+1\;\;(8$ nodes here),  $\ell_{\rm
lb}+2\;\;(8^2\!=\!64$ nodes here), $\ell_{\rm lb}+3\;\;(8^3\!=\!512$ nodes
here), and so on. The nodes at the bottom of the block-tree are {\em
leaf-nodes}, and represent the grid cells on which the equations of
hydrodynamics are solved.

Each node -- both the nodes on the amr-tree, and the nodes on the local block-trees -- stores 
collective information about the set of grid cells that it contains, e.g. their total
mass, the position of the centre of mass, etc.

Our algorithm consists of a general {\em tree-solver} implementing the tree
construction, communication and tree-walk, and modules which include the
calculations of specific physical equations, e.g. gravitational accelerations or
optical depths. The tree-solver communicates with the physical modules by means
of interface subroutines which allow physical modules, on the one hand to store
various quantities on the nodes, and on the other hand to walk the tree
accessing the quantities stored on the nodes. When walking the tree, physical
modules may use different MACs that reflect the nature of the quantity they are
seeking to evaluate. An advantage of this approach is that it makes code
maintenance more straightforward and efficient. Moreover, new functionality can
be added easily by writing new physical modules or extending existing ones,
without needing to change the relatively complex tree-solver algorithm. 

The boundary conditions can be either isolated or periodic, and they can be specified 
in each direction independently, i.e. mixed boundary conditions with one or two directions 
periodic and the remaining one(s) isolated are allowed (see \S\ref{sec:gravity}).

In the following \S\ref{sec:treesolver}, we describe the tree-solver, and in
\S\ref{sec:gravity} and \S\ref{sec:treecol}, respectively, we give
descriptions of the gravity module and the module (called {\sc OpticalDepth})
which calculates heating by the interstellar radiation field.


\subsection{Tree-solver}
\label{sec:treesolver}


The {\em tree-solver} creates and utilises the tree data structure described
above. Maintaining a copy of the whole tree on each processor would incur
prohibitively large memory requirements. Therefore, only the amr-tree (i.e. the
top part of the tree, between the root-block node and the leaf-block nodes) is
communicated to all processors. The block-tree within a leaf-block  is held on
the processor whose domain contains that leaf-block, and communicated wholly or
partially to another processor only if it will be needed by that processor
during a subsequent tree-walk. The tree-solver itself stores in each tree-node
-- with the exception of the leaf-nodes -- the total mass of the node and the
position of its centre of mass, i.e. four floating point numbers. For leaf-nodes
(the nodes corresponding to individual grid cells) only their masses are stored,
because the positions of their centres of mass are identical to their
geometrical centres and are already known. Additionally, each physical module
can store any other required quantity on the tree-nodes.

The tree-solver consists of three steps: tree-build, communication and
tree-walk. In the tree-build step, the tree is built from bottom up by
collecting information from the individual grid cells, summing it, and
propagating it to the parent tree-nodes. The initial stages of this step, those
that involve the block-trees within individual leaf-blocks, are performed
locally. However, as soon as the leaf-block nodes are reached, information has
to be exchanged between processors because parent-nodes are not necessarily
located on the same processor. At the end of this step, each processor possesses
a copy of the amr-tree {\em plus} all the block-trees corresponding to
leaf-blocks that are located on that processor.

\begin{figure}
\includegraphics[width=\columnwidth]{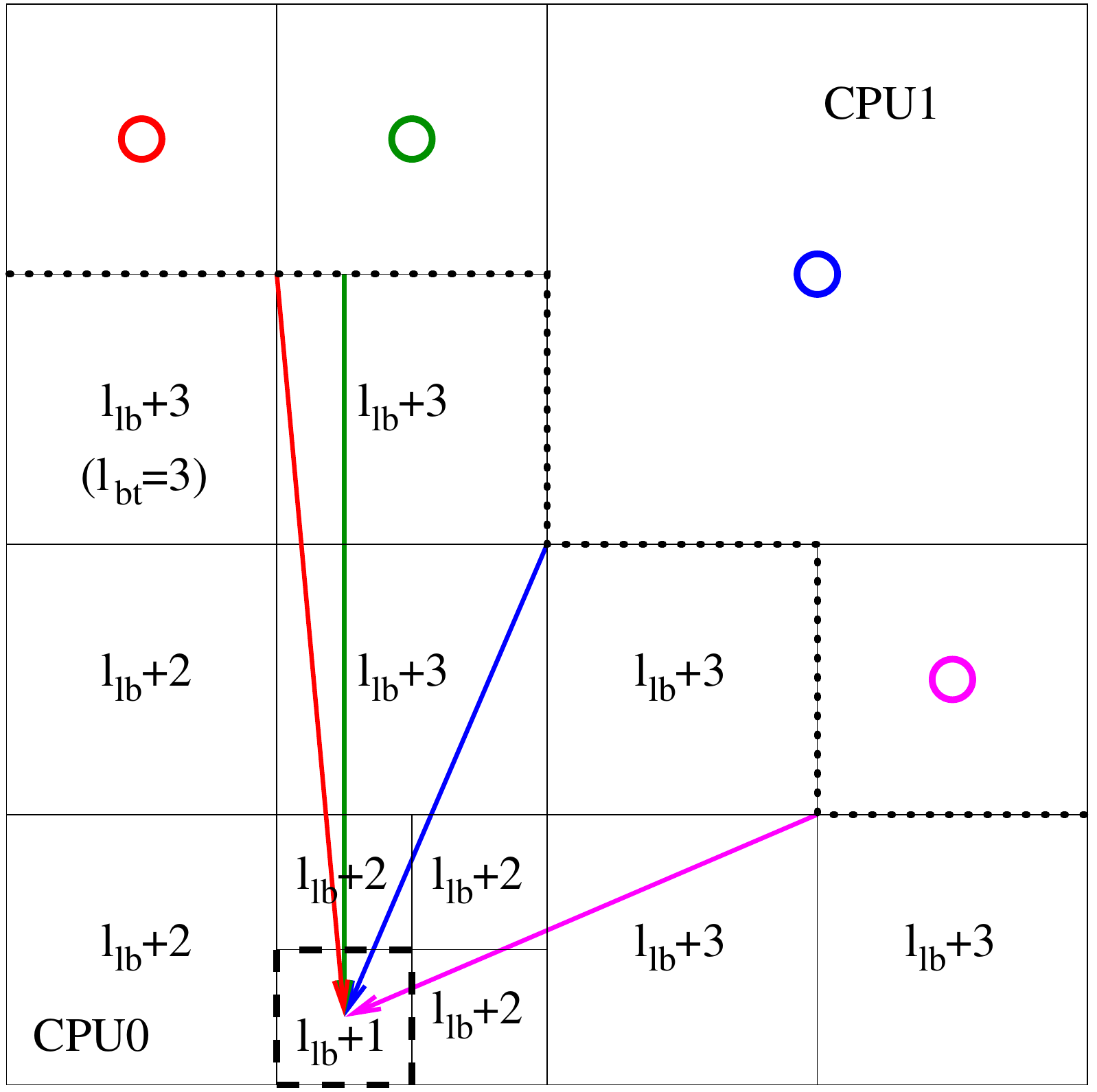}
\caption{Determining the block-tree levels that need to be 
exported from the leaf-blocks in the spatial domain of processor CPU0 
to processor CPU1. In this case the spatial domains of the two processors 
are adjacent, and are separated by the thick dotted line. For each 
leaf-block on CPU0 (for example, the one enclosed by a thick dashed line) 
its block-tree is traversed and the MAC is evaluated in relation to all 
the leaf-blocks on processor CPU1; for this purpose the code uses the 
distance from the {\em centre of mass} of a node of the block-tree on CPU0, 
to the {\em closest point} of a leaf-block on CPU1, as illustrated by the 
coloured arrows. The level of detail communicated to CPU1 is then set by 
the finest level reached during this procedure. In the case illustrated, 
the leaf-block on CPU0 that is furthest from the leaf-blocks on CPU1 (the 
one enclosed by a thick dashed line) exports only the first two levels of 
its block-tree, i.e. from level $\ell_{\rm lb}$ to $\ell_{\rm lb}\!+\!1$. 
In contrast, the leaf-blocks on CPU0 that are closest to the leaf-blocks 
on CPU1 export their full block-trees, i.e. from level $\ell_{\rm lb}$ to 
level $\ell_{\rm lb}\!+\!3$.}
\label{fig:treecomm}
\end{figure}

\begin{figure*}
\includegraphics[width=0.8\textwidth]{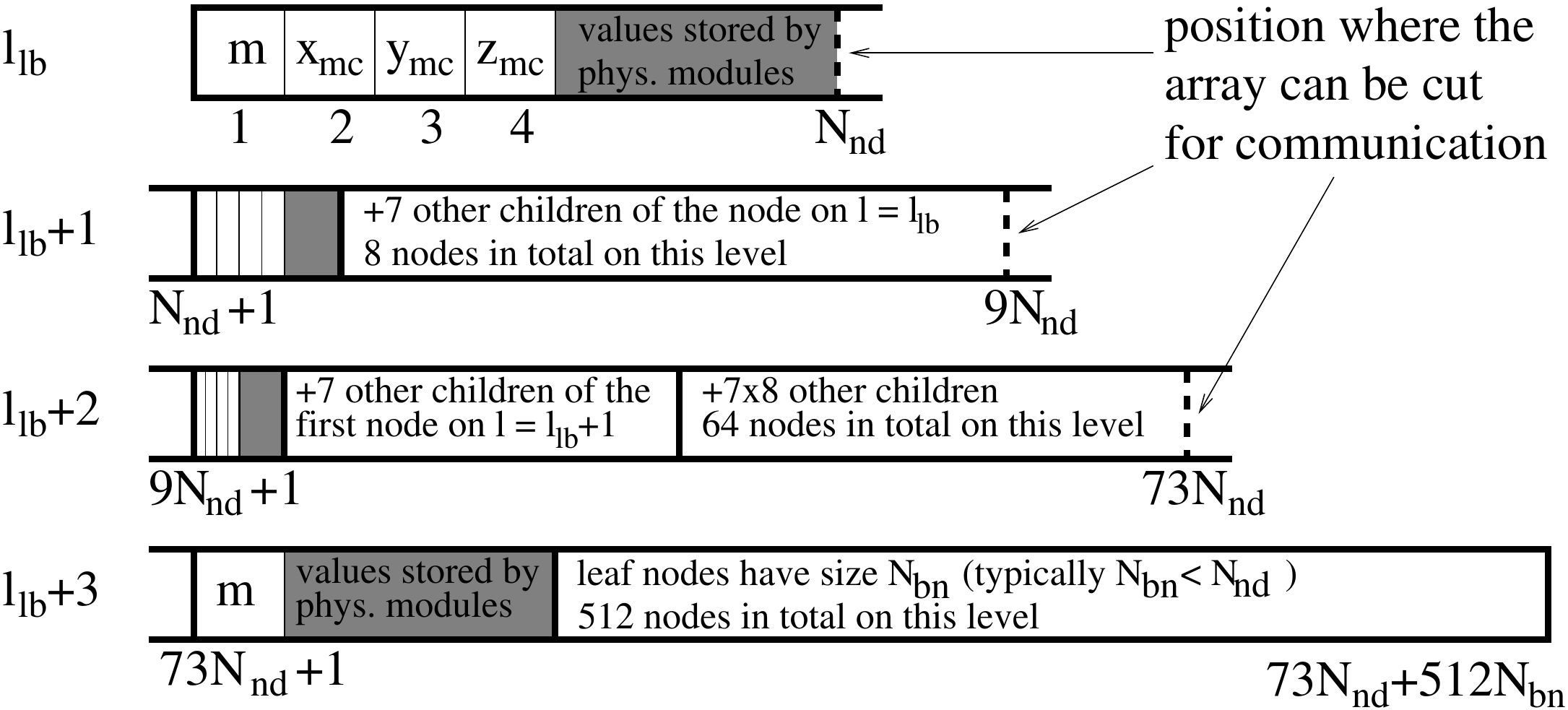}
\caption{Organization of a block-tree within a block in memory. It is a 1D array
sorted by levels, starting from $\ell\!=\!\ell_\mathrm{lb}$.}
\label{fig:treeinram}
\end{figure*}

The communication step ensures that each processor imports from all other
processors all the information that it will need for the tree-walks, which are
subsequently called by the physical modules. To this end, the code considers all
pairs of processors, and determines what tree information the one processor (say
CPU0; see Figure~\ref{fig:treecomm}) needs to export to the other processor (say
CPU1). To do this, the code walks the block-trees of all the leaf-blocks on
CPU0, and applies a suite of MACs (required by the tree-solver itself and the
used physical modules)  in relation to all the leaf-blocks on CPU1. This suite
of MACs determines for each leaf-block on CPU0, the level of its block-tree that
delivers sufficient detail to CPU1 to satisfy the resolution requirements of all
the physical modules that will be called before the tree is rebuilt. Thus, a
leaf-block on CPU0 that has very little physical influence on any of the
leaf-blocks on CPU1 (for example by virtue of being very distant or of low mass)
may only need to send CPU1 the information stored on its lowest (i.e. coarsest
resolution) level, $\ell_{\rm lb}$. Conversely, a leaf-block on CPU0 that has a
strong influence on at least one of the leaf-blocks on CPU1 (for example by
virtue of being very close or very massive)  may need to send the information
stored on its highest (finest resolution) level, $\ell_{\rm lb}\!+\!\ell_{\rm
bt}$. In order to simplify communication, the required nodes of each block-tree
on CPU0 are then stored in a one-dimensional array, ordered by level, starting
at $\ell = \ell_\mathrm{lb}$ and proceeding to higher levels (see Figure
\ref{fig:treeinram}). Finally, the arrays from all the block-trees on CPU0 are
collated into a single message and sent to CPU1. This minimizes the number of
messages sent, thereby ensuring efficient communication, even on networks with
high latency.

Note that this communication strategy in which tree-nodes are communicated
differs from a commonly used one in which particles (equivalents of grid cells)
are communicated instead \citep[e.g. {\sc Gadget}][]{Springel2005}. In this way,
the communication is completed before the tree-walk is executed and the
tree-walk runs locally, i.e. separately on each processor. The communication
strategy adopted in this work provides a significant benefit for the
OpticalDepth and the TreeRay modules as they work with a large amount of
additional information per grid cell (or particle), which does not have to be
stored and communicated (see \S\ref{sec:treecol}).

The final step is a tree-walk, in which the whole tree is traversed in a
depth-first manner for each grid cell or in general for an arbitrary {\em target
point} (e.g. the position of a sink particle). During the process, the suite of
MACs is evaluated recursively for each node and if it is acceptable for the
calculation, subroutines of physical modules that do the calculation are called,
otherwise its child-nodes are opened.

The tree-solver itself only implements a simple geometric MAC
\citep{Barnes1986}, which accepts a node if its angular size, as seen from the
target point, ${\bf r}$, is smaller than a user-set limit, $\theta_\mathrm{lim}$.
Specifically, if $h$ is the linear size of the node and ${\bf r}_{\rm a}$ is the
position of the centre of mass of the node, the node is accepted (and so its
child-nodes need not be considered) if
\begin{equation}
\label{eq:bhmac}
\frac{h}{|{\bf r}-{\bf r}_{\rm a}|} < \theta_\mathrm{lim}\,.
\end{equation}
It has been shown by \citet[][hereafter SW94]{SalmonWarren1994} that the BH86
MAC can lead to unexpectedly large errors when the target point is relatively
far from the centre of mass of the node but very close to its edge. Several
alternative geometric MACs were suggested to mitigate this problem
\citep{SalmonWarren1994,Dubinski1996}. Following \citet{Springel2005}, we extend
the geometric MAC by setting the parameter $\eta_\mathrm{SB}$ such that a node
is only accepted if the target point lies outside a cuboid $\eta_\mathrm{SB}$
times larger than the node (with the default value $\eta_\mathrm{SB} = 1.2$).
Additional MACs specific to the physical modules are implemented by those modules
(see \S\ref{sec:gravity}).

The tree-walk is the most time consuming part of the tree-solver. Typically it 
takes more than 90\% of the computational time spent by the whole tree-solver. 
We stress that the tree-walk does not include any communication; the tree is 
traversed in parallel independently on each processor for all the grid cells 
in the spatial domain of that processor. The tree-solver exhibits very good 
scaling, with speed-up increasing almost linearly up to at least 1500 CPU cores 
(see \S\ref{sec:scaling}).

\subsection{Gravity module}
\label{sec:gravity}


This module calculates the gravitational potential and/or the gravitational
acceleration. We use the same approach as \citet{Springel2005} and store only
monopole moments in the tree, because this substantially reduces memory
requirements and communication costs. Since masses and centres of mass are
already stored on the tree-nodes by the tree-solver, the gravity module does not
contribute any extra quantities to the tree. 

In \S\ref{sec_MAC} we describe three data-dependent MACs which can be used
instead of the geometric MACs of the tree-solver: Maximum Partial Error (MPE),
Approximate Partial Error (APE) and the (experimental) implementation of the
SumSquare MAC. Furthermore, the code features three different types of gravity
boundary conditions. These are isolated (see \S\ref{sec_BND_isolated}), fully
periodic (\S\ref{sec_BND_periodic}), and mixed boundary conditions
(\S\ref{sec_BND_mixed}). Finally in \S\ref{sec:abu}, we describe a
technique called the {\em Adaptive Block Update} to save computational time
by re-using the solution from previous time-step when possible.

\subsubsection{Data-dependent MACs} \label{sec_MAC}

A general weakness of the purely geometric MACs is that they do not take into
account the amount and internal distribution of mass in a node. This can
make the code inefficient if the density is highly non-uniform. For example, if
the code calculates the gravitational potential of the multi-phase interstellar
medium, the contribution from nodes in the hot rarefied gas is very small, but
it is calculated with the same opening angle as the much more important
contribution from nodes in dense molecular cores.

{\sc MPE MAC} (Maximum Partial Error): To compensate for the above problem, SW94
propose a MAC based on evaluating the maximum possible error in the contribution
to the gravitational acceleration at the target point, ${\bf r}$, that could
derive from calculating it using the multipole expansion of the node up to
order $p$ (instead of adding directly the contributions from
all the constituent grid cells)
\begin{eqnarray}
\label{eq:mpe}
\Delta a_{(p)}^{\rm max} = \frac{G}{d^2} \frac{1}{(1-b_\mathrm{max}/d)^2}
\left\{ (p\!+\!2) \left(\frac{B_{(p+1)}}{d^{p+1}}\right)\!-\!
(p\!+\!1)\left( \frac{B_{(p+2)}}{d^{p+2}} \right) \right\}\!,&&\\
B_{(p)} = \sum_{i} |m_{i}| |\mathbf{r}_{i} - \mathbf{r}_{\rm a}|^{p}.\hspace{4.8cm}
\end{eqnarray}
Here, ${\bf r}_{\rm a}$ is the mass centre of the node; $d\!\equiv\!|{\bf
r}\!-\!{\bf r}_{\rm a}|$ is the distance from ${\bf r}_{\rm a}$ to the target
point; $b_{\rm max}$ is the distance from ${\bf r}_{\rm a}$ to the furthest
point in the node; $B_{(p)}$ is the $p^{\rm th}$-order multipole moment,
obtained by summing contributions from all the grid cells $i$ in
the node; $m_i$ and ${\bf r}_i$ are the masses and positions of these grid cells. 
The node is then accepted only if $\Delta a_{(p)}^{\rm max}$ is smaller than
some specified maximum allowable acceleration error. This threshold can either be 
set by the user as a constant value, $a_\mathrm{lim}$, in the
physical units used by the simulation 
\begin{equation}
\label{eq:abs_err}
\Delta a_{(p)}^{\rm max} < a_\mathrm{lim} \ ,
\end{equation}
or it can be set as a relative value, $\epsilon_\mathrm{lim}$, with respect to
the acceleration from the previous time-step $a_\mathrm{old}$
\begin{equation}
\label{eq:rel_err}
\Delta a_{(p)}^{\rm max} < \epsilon_\mathrm{lim} a_\mathrm{old} \ .
\end{equation}

{\sc APE MAC} (Approximate Partial Error): An alternative way to estimate the
partial error of a node contribution was suggested by \citet{Springel2001}. It
takes into account the node total mass, but it ignores the internal node mass
distribution. It is therefore faster, but less accurate. Using multipole moments
up to order $p$, the error of the gravitational acceleration is of order
the contribution from the $(\!p\!+\!1\!)^{\rm th}$ multipole moment
\begin{equation}
\label{eq:ape}
\Delta a_{(p)}^{\rm max} \simeq \frac{GM}{d^2} \left( \frac{h}{d} \right)^{p+1}\,,
\end{equation}
where $M$ is the mass in the node and $p=1$ in our case since we only store
monopole moments. Similar to the MPE MAC, the APE error limit can be either
set absolutely as $a_\mathrm{lim}$ (Equation~\ref{eq:abs_err}), or relatively
through $\epsilon_\mathrm{lim}$ (Equation~\ref{eq:rel_err}).

{\sc SumSquare MAC}: 
SW94 argue that it is unsafe to constrain the error using the contribution of a
single node only, since it is not known {\it a priori} how these contributions
combine. They suggest an alternative procedure, which limits the error in the
total acceleration at the target point; one variant of this procedure is the
SumSquare MAC which sums up squares of $a_{(p)}^{\rm max}$ given by
Equation~(\ref{eq:mpe}) over all nodes considered for the calculation of the
potential/acceleration at a given target point. In this way, the SumSquare MAC
controls the total error in acceleration resulting from the contribution of all
tree-nodes. This MAC requires a special tree-walk which does not proceed in the
depth-first manner. Instead it uses a priority queue, which on-the-fly reorders a
list of nodes waiting for evaluation according to the estimated error resulting
from their contribution. This feature is still experimental in our
implementation, nevertheless we evaluate its accuracy and performance and
compare it to other MACs in \S\ref{sec:macs}.


\subsubsection{Isolated boundary conditions}\label{sec_BND_isolated}

In the case of isolated boundary conditions (BCs), the gravitational
potential in a target point given by position vector $\mathbf{r}$ is 
\begin{equation}
\Phi(\mathbf{r}) = - \sum_{a=1}^{N} \frac{GM_a}{|\mathbf{r}-\mathbf{r_a}|}
\label{eq:pot}
\end{equation}
where index $a$ runs over all nodes accepted by the MAC during the tree-walk, 
$M_a$ and $\mathbf{r_a}$ are the node mass and position. The gravitational
acceleration is then obtained either by differentiating the potential
numerically, or it is calculated, as
\begin{equation}
\mathbf{a}(\mathbf{r}) = - \sum_{a=1}^{N}
\frac{GM_a(\mathbf{r}-\mathbf{r_a})}{|\mathbf{r}-\mathbf{r_a}|^3}\,.
\label{eq:accel}
\end{equation}
The first approach needs less memory and is slightly faster. The second 
approach results in less noise, because numerical differentiation is not 
needed.

\subsubsection{Periodic boundary conditions}\label{sec_BND_periodic}

In the case of periodic boundary conditions in all three directions, the 
gravitational potential is determined by the Ewald method 
\citep{Ewald1921,Klessen1997}, which is designed to mitigate the very 
slow convergence in case one evaluates contributions to the potential,
essentially $1/d$ where $d=|\mathbf{r}-\mathbf{r_a}|$, over an infinite number
of periodic copies, by brute force. This is achieved by splitting it into two
parts 
\begin{eqnarray}
1/d&=&\frac{\erfc(\alpha d)}{d} + \frac{\erf(\alpha d)}{d}
\end{eqnarray}
and summing the term $\erf(\alpha d)/d$ in Fourier space; $\alpha$ is an
arbitrary constant controlling the number of nearby and distant terms which have
to be taken into consideration. In this section, we present formulae only
for the potential. The expressions for acceleration are straightforward to
derive, and we list them in appendix \ref{ap:accEwald}.

The computational domain is assumed to be a rectangular cuboid, with sides
$L_x$, $L_y=bL_x$ and $L_z=cL_x$ where $b$ and $c$ are arbitrary real numbers.
The gravitational potential $\Phi$ at the target point, $\mathbf{r}$, is then 
\begin{eqnarray}
\Phi (\mathbf{r})=-G \sum_{a=1}^N M_a \left(\phi_S(\mathbf{r}-\mathbf{r_a}) 
+ \phi_L(\mathbf{r}-\mathbf{r_a})\right)\hspace{2.2cm}\\
= -G\!\sum_{a=1}^N M_a \left\{ \sum_{i_1,i_2,i_3}\!\frac{\erfc(\alpha |\mathbf{r}
\!-\!\mathbf{r_a}\!-\!i_1\mathbf{e_x}L_x\!-\!i_2\mathbf{e_y}bL_x\!-\!i_3\mathbf{e_z}cL_x|)}
{|\mathbf{r}\!-\!\mathbf{r_a}\!-\!i_1\mathbf{e_x}L_x\!-\!i_2\mathbf{e_y}bL_x\!-\!i_3\mathbf{e_z}cL_x|}\right.\nonumber\\
\left.+\frac{1}{bcL_x^3}\sum_{k_1,k_2,k_3,|k|\neq 0}\frac{4 \pi}{k^2}
\exp(-\frac{k^2}{4\alpha^2})\cos(\mathbf{k}\cdot(\mathbf{r}\!-\!\mathbf{r_a}))\right\}.
\label{ewpotential}
\end{eqnarray}
Here, the first inner sum corresponds to short-range contributions,
$\phi_S(\mathbf{r}-\mathbf{r_a})$, from the nearest domains in physical space,
and the second sum constitutes long-range contributions,
$\phi_L(\mathbf{r}-\mathbf{r_a})$. The outer sum runs over all accepted nodes in
the computational domain $\;M_a$ is the mass of a node, and ${\bf r}_{\rm a}$ is
its centre of mass\footnote{Note that the corresponding formula in \citet[][;
their Equation~(6)]{Klessen1997} has an incorrect sign before the
$\phi_L(\mathbf{r}-\mathbf{r_a})$ term.}. Indices $i_1$, $i_2$, $i_3$ are
integer numbers; $\mathbf{e_x}$, $\mathbf{e_y}$, $\mathbf{e_z}$ are unit vectors
in the corresponding directions; and $\mathbf{k}$ is a wavevector with
components $k_1=2 \pi l_1 /L_x$, $k_2=2 \pi l_2 /b L_x$, $k_3=2 \pi l_3  /c
L_x$, where $l_1$, $l_2$, $l_3$ are integer numbers. By virtue of the Ewald
method, both inner sums converge very fast. We follow \citet{Hernquist1991} in
setting
\begin{eqnarray}
i_1^2+(b i_2)^2+(c i_3)^2 & \leq & 15 \\
l_1^2+(l_2/b)^2+(l_3/c)^2 & \leq & 10
\label{ewcond}
\end{eqnarray}
and $\alpha = 2/L_x$.

\begin{figure}
\includegraphics[width=\columnwidth]{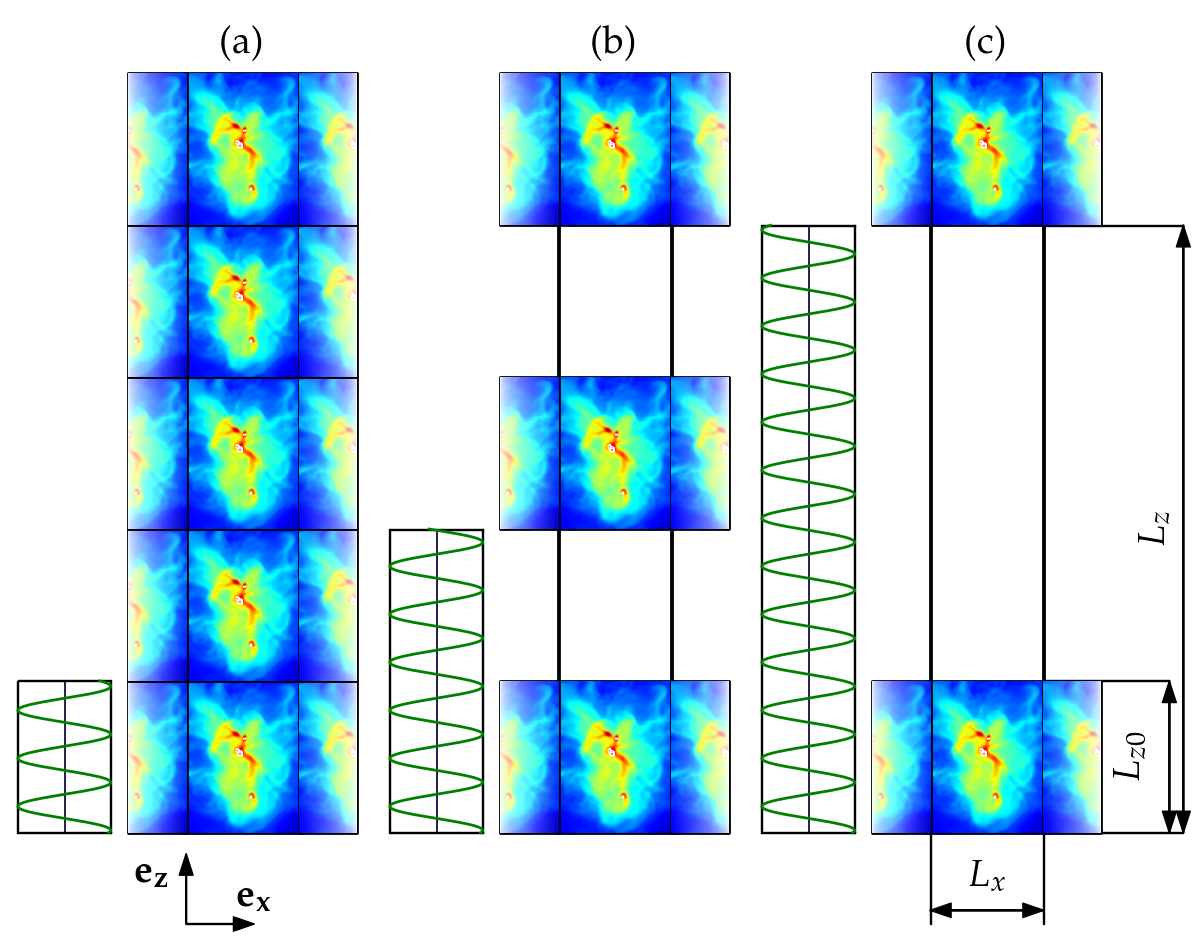}
\caption{An illustration of the limiting process which transforms a
configuration with periodic BCs to a configuration with mixed BCs. The
computational domain and its periodic copies are shown on slices of constant
$y$, the orientation of unit vectors $\mathbf{e_x}$ and $\mathbf{e_z}$ is
indicated at the bottom left. 
From left to right: (a) Configuration with periodic BCs (i.e. $n = 1$); 
(b) The material inside the periodic copies is displaced by distance $L_{z0}$ in
direction $\mathbf{e_z}$, and the density in the computational domain at $L_{z0}
< z < 2 L_{z0}$ is set to zero (i.e. $n = 2$); 
(c) The material in the periodic copies is displaced further ($n = 4$). The box
to the left of the computational domain shows the shortest wavelength in the
direction $\mathbf{e_z}$ fulfilling condition~(\ref{ewcond}).
The number of horizontal oscillations is proportional to the value of index $l_3$ for given $n$.}
\label{fig:ewald}
\end{figure}

\subsubsection{Mixed boundary conditions}\label{sec_BND_mixed}

We generalise the Ewald method, which was developed for computational domains
with periodic BCs in all spatial directions, to computational domains with mixed
BCs. In three dimensional space, mixed BCs can be of two types: periodic BCs in
two directions (without loss of generality we choose $x$- and $y$-directions),
and isolated BCs in the third ($z$-)direction; and periodic BCs in one direction
(we choose $x$), and isolated BCs in the other two directions. We abbreviate the
former case of mixed BCs as 2P1I, and the latter case as 1P2I. Configuration
2P1I has planar symmetry with axis $\mathbf{e_z}$, while configuration 1P2I has
an axial symmetry along axis $\mathbf{e_x}$. These configurations might be
convenient for studying systems with the symmetry (i.e. layers or filaments).
We note that directions that can be defined as periodic are given by
computational domain boundaries and thus they can only be parallel with one or
more of the Cartesian coordinate axes.

We find the expression for $\Phi(\mathbf{r})$ for mixed BCs of 2P1I type by
taking a limit of \req{ewpotential}. Consider a computational domain with
side-lengths $L_x$, $L_y\!=\!bL_x$, $L_z\!=\!cL_x$ and with periodic boundary
conditions in all three directions, for which the gravitational potential is
given by \req{ewpotential}. Next we shift periodic copies of this domain in the
$z$-direction so that the periodicity in the $z$-direction is $n$ times larger,
i.e. $L_z\!=\!n L_{z0}$, where $n$ is an integer number and $L_{z0}$ is the
extent in the $z$-direction of the original computational domain
(Figure~\ref{fig:ewald}). Since the copies are shifted and not stretched, the
mass distribution between $z=0$ and $z=L_{z0}$ is unaltered, and the density is
zero between $z=L_{z0}$ and $z\!=\!n L_{z0}$, leaving all mass concentrated in
plane-parallel layers of thickness $z=L_{z0}$ and with normals pointing in
direction $\mathbf{e_z}$. As $n$ increases, the layers move away from one
another, but \req{ewpotential} still holds. In the limit $n\!\to\!\infty$, the
periodic copies of the computational domain are touching one another in $x$- and
$y$-directions, however, neighbouring layers in the $z$-direction are at
infinite distance and hence they do not contribute to the gravitational field in
the original computational domain. 

As $n$ increases, the short--range contributions are zero for all $i_3 \neq 0$,
because the argument of the complementary error function in \req{ewpotential}
tends to infinity. The long--range term $\phi_L(\mathbf{r}-\mathbf{r_a})$
in the limit $n \to \infty$ becomes
\begin{eqnarray}
\phi_L(\mathbf{r}-\mathbf{r_a}) & \! \! = & \! \! \frac{1}{\pi L_x b}
\sum_{l_1,l_2} \exp\left[-\frac{\pi^2}{\alpha^2 L_x^2} 
\left(l_1^2+\left( l_2/b \right)^2\right)\right] \times \nonumber \\ 
&&\lim_{n \to \infty} \frac{1}{n} 
\sum_{l_3} \frac{\exp[-\pi^2 l_3^2/(\alpha c L_x n)^2]}
{c \, \left[l_1^2+\left(l_2/b \right)^2+\left(l_3/(c \, n)\right)^2 \right]}  \times \nonumber \\
&&\cos\left(\frac{2\pi l_1(x-x_a)}{L_x}+\frac{2\pi l_2(y-y_a)}{b L_x}\right. \nonumber \\
&&\left.+ \frac{2\pi l_3(z-z_a)}{c L_x n}\right).
\label{limitlong}
\end{eqnarray}
The condition (\ref{ewcond}), which is now $l_1^2+(l_2/b)^2+(l_3/cn)^2  \leq  10$
requires us to conserve resolution in the $z$-direction in Fourier
space, i.e. to increase the range of $l_3$ with $n$ linearly (see Figure~\ref{fig:ewald}).
Note that $2\pi (z-z_a)/(c L_x)$ is independent of
$n$, because we restrict all mass in the computational domain to interval $(0,L_{z0})$, 
(i.e. $|z-z_a| \leq c L_x = L_{z0}$ for any target point at $\mathbf{r}$ and node at $\mathbf{r_a}$).
Bearing this in mind, the term after the limit sign in Equation~(\ref{limitlong}) 
corresponds to a Riemann sum over interval $(-\sqrt{10},\sqrt{10})$ with equally spaced
partitions of size $1/nc$. Using the identity
$\cos(A+B)=\cos(A)\cos(B)-\sin(A)\sin(B)$ where $B=2\pi l_3(z-z_a)/(c L_x n)$,
the limit becomes
\begin{equation}
\cos \left(\frac{2\pi l_1(x-x_a)}{L_x}+\frac{2\pi l_2(y-y_a)}{L_x b}\right) I(l_1,l_2,z-z_a),
\end{equation}
where
\begin{equation}
I(l_1,l_2,z-z_a) = 2\int_0^\infty \frac{\exp(-\zeta u^2) \cos(\gamma u)}{l_1^2+(l_2/b)^2+u^2} \du.
\label{intdef}
\end{equation}
To keep the notation compact, we introduce $\gamma = 2 \pi(z-z_a)/L_x$ and
$\zeta = \pi^2/(\alpha L_x)^2$. In order to evaluate the integral analytically, 
we extend the interval of integration to infinity (this extension means that we
evaluate the sum even slightly more accurately than by condition \ref{ewcond})
If $|l_1|+|l_2| 
\neq 0$, we have
\begin{eqnarray}
I(l_1,l_2,z-z_a) &  = &  \frac{\pi \exp [-\gamma^2/(4 \zeta)]}{2\sqrtij}
\Bigg\{ \erfcx \left[ \frac{\zeta \sqrtij - \gamma/2}{\sqrt{\zeta}}  \right] \nonumber \\
&&+ \erfcx \left[  \frac{\zeta \sqrtij + \gamma/2}{\sqrt{\zeta}}    \right] \Bigg\},
\label{intij}
\end{eqnarray}
where $\erfcx(A)=\exp(A^2) \erfc(A)$. When $l_1 = l_2 = 0$, integral
(\ref{intdef}) is infinite, but this property can be circumvented. With the help
of $\cos(\gamma u)=1-2\sin^2(\gamma u/2)$ we get two integrals corresponding to
these two terms. The former one is infinite, but independent of the spatial
coordinates and we set it to zero. The latter one can easily be integrated
\begin{eqnarray}
I(0,0,z-z_a)  = -\pi \left\{ \gamma \erf(\frac{\gamma}{2\sqrt{\zeta}}) + 2 \sqrt{\frac{\zeta}{\pi}}
\exp(-\gamma^2/4\zeta)  \right\} + 2 \sqrt{\pi \zeta}.
\label{int00}
\end{eqnarray}

Now we can write the potential as
\footnote{In this section, we emphasise the way how the equations are derived.
For an implementation to a code, the form of \req{intij} possesses problems for numerical evaluation. 
We recommend to implement the potential in the form of
\begin{eqnarray}
\phi_L (\mathbf{r}-\mathbf{r_a}) & = & \frac{1}{\pi L_x b} \sum_{l_1,l_2,l_1^2+(l_2/b)^2 \leq 10} 
\cos \left(\frac{2\pi l_1(x-x_a)}{L_x}+\frac{2\pi l_2(y-y_a)}{L_x b}\right) \times  \nonumber \\
&& \widetilde{I}(l_1,l_2,z-z_a),
\label{potential2p1icode}
\end{eqnarray}
where function $\widetilde{I}(l_1,l_2,z-z_a)$ is defined by \req{intijscaled}.
}
\begin{eqnarray}
\Phi (\mathbf{r}) & = & -G \sum_{a=1}^N m_a \Big\{ \sum_{i_1,i_2,i_1^2+(b i_2)^2 \leq 10} 
\frac{\erfc(\alpha | \mathbf{r} - \mathbf{r_a} - i_1\mathbf{e_x}L_x -i_2\mathbf{e_y}b L_x|)}
{| \mathbf{r} - \mathbf{r_a} - i_1\mathbf{e_x}L_x -i_2\mathbf{e_y}b L_x|} \nonumber \\ 
&& +\frac{1}{\pi L_x b} 
\sum_{l_1,l_2,l_1^2+(l_2/b)^2 \leq 10} \exp(-\zeta( l_1^2+ (l_2/b)^2)) \times \nonumber \\  
&&\cos \left(\frac{2\pi l_1(x-x_a)}{L_x}+\frac{2\pi l_2(y-y_a)}{L_x b}\right) I(l_1,l_2,z-z_a) \Big\}.
\label{potential2p1i}
\end{eqnarray}
Note that the ratio $c$ is not contained in $\Phi (\mathbf{r})$ as we may
expect, because it is of no physical significance when the BCs are isolated in
this direction.

The modification of the Ewald method for a computational domain with mixed BCs
of type 1P2I can be derived in a similar way to the previous case. However, the
integration is more demanding here, because the result of the limiting process is a
double integral (we integrate \req{intdef} along $l_2/b$ instead of
equations~(\ref{intij}) and (\ref{int00})). Applying a substitution which corresponds
to a rotation, this integral can be transformed into  a 1D integral, but we have not been
able to express it in a closed form. In this case (1P2I), we arrive at
\begin{eqnarray}
\Phi (\mathbf{r}) & = & -G \sum_{a=1}^N m_a \Big\{ \sum_{i_1,i_1^2 \leq 10} 
\frac{\erfc(\alpha | \mathbf{r} - \mathbf{r_a} - i_1\mathbf{e_x}L_x|)}
{| \mathbf{r} - \mathbf{r_a} - i_1\mathbf{e_x}L_x |} \nonumber \\ 
&& +\frac{2}{L_x} 
\sum_{l_1,l_1^2 \leq 10} \exp(-\zeta l_1^2)  \cos \left(\frac{2\pi
l_1(x-x_a)}{L_x}\right)  \times \nonumber \\
&& K(l_1,y-y_a,z-z_a) \Big\},
\label{potential1p2i}
\end{eqnarray}
where function $K(l_1,y-y_a,z-z_a)$ is given by
\begin{equation}
K(l_1,y-y_a,z-z_a) = \int_0^{\infty} \frac{ J_0 (\eta q) \exp(-\zeta q^2)}{l_1^2+q^2} q \; \dq,
\label{int00p1}
\end{equation}
and $\eta = 2\pi \sqrt{(y-y_a)^2+(z-z_a)^2}/L_x$. Function $J_0$ is the Bessel
function of the first kind and zeroth order.

Formulae for accelerations corresponding to potentials \req{ewpotential}, \req{potential2p1i}
and \req{potential1p2i} are listed in appendix \ref{ap:accEwald}.

\subsubsection{Look-up table for the Ewald array}
\label{sec:ewald:lookup}

Since the explicit evaluation of $\phi_S(\mathbf{r}-\mathbf{r_a})$ and
$\phi_L(\mathbf{r}-\mathbf{r_a})$ at each time-step would be prohibitively time
consuming, these functions are pre-calculated before the first hydrodynamical time step, 
and their values are stored in a look-up table. 
We experiment with two approaches to approximate the above functions from the look-up table 
at the time when the gravitational potential is evaluated.

In the first approach, the function 
$\phi(\mathbf{r} -\mathbf{r_a})=\phi_S(\mathbf{r}-\mathbf{r_a}) +\phi_L(\mathbf{r}-\mathbf{r_a})$ 
is precalculated on a set of 
nested grids, and particular values are then found by trilinear interpolation on these grids. 
Coverage of the grids increases towards the singularity at the origin 
($|\mathbf{r}\!-\!\mathbf{r_a}|\rightarrow 0$). The
gravitational potential at target point ${\bf r}$ is then calculated as
\begin{equation}
\Phi (\mathbf{r}) = -\sum_{a=1}^{N} G M_a \phi(\mathbf{r} - \mathbf{r_a}).
\label{eq:evfld}
\end{equation}

In the second approach, we avoid the singularity of $\phi(\mathbf{r} - \mathbf{r_a})$ 
by subtracting the term $1/|\mathbf{r}\!-\!\mathbf{r_a}|$ from $\phi(\mathbf{r} - \mathbf{r_a})$. 
This enables us to use only one interpolating grid with uniform coverage for the whole computational domain. 
Moreover, for mixed BCs, $\phi(\mathbf{r} - \mathbf{r_a})$ can be approximated at some parts 
of the computational domain by analytic functions.
The function $\phi(\mathbf{r} - \mathbf{r_a})$ converges to $2 \pi |z - z_a|/(b L_x^2)$ with increasing $(z - z_a)/L_x$ 
for configuration 2P1I, and it 
converges to $2\mathrm{ln}(\sqrt{(y - y_a)^2 + (z-z_a)^2})/L_x$ 
with increasing $\sqrt{(y - y_a)^2 + (z-z_a)^2}/L_x$ for configuration 1P2I.
The convergence is exponential and the relative error in acceleration is always smaller than $10^{-4}$ if 
$(z - z_a) > 2 L_x$ and $\sqrt{(y - y_a)^2 +(z - z_a)^2} > 2 L_x$ for configuration 2P1I and 1P2I, respectively.
Accordingly, we use the analytic expression in these regions and pre-calculate $\phi(\mathbf{r} - \mathbf{r_a})$ 
only at the region where $(z - z_a) < 2 L_x$ or $\sqrt{(y - y_a)^2 +(z - z_a)^2} < 2 L_x$, so 
the grid covers only a fraction of the computational domain if the computational domain is elongated.
In combination with using only one interpolating grid this results in smaller demands on memory 
while it retains the same accuracy as in the first approach.

In the second approach, we pre-calculate not only $\phi(\mathbf{r} - \mathbf{r_a})$ but also its gradient.  
The actual value of $\phi(\mathbf{r} - \mathbf{r_a})$ at a given location is then 
estimated by a Taylor expansion to the first order. 
This is faster than the trilinear interpolation used in the first approach, and leads to 
a speed up in the Gravity module by a factor of $\simeq 1.4$ to $\simeq 1.9$ 
depending on the shape of the computational domain, the adopted BCs, and whether the
potential or acceleration is used. 
Thus the second approach appears to be superior to the first one.
In each approach, if gravitational accelerations rather than the potential are required, 
we adopt an analogous procedure for each of its Cartesian components.

Note that in a very elongated computational domain, the evaluation of
$\phi(\mathbf{r} - \mathbf{r_a})$ can be accelerated by adjusting 
the parameter $\alpha =2/L_x$. Since $\phi(\mathbf{r}-\mathbf{r_a})$ 
is pre-calculated, the choice of $\alpha$ is of little importance in our 
implementation and we do not discuss it further in this paper.


\subsubsection{Adaptive block update}
\label{sec:abu}

Often, it is not necessary to calculate the gravitational potential/acceleration
at each grid cell in each time-step. Since the {\sc FLASH} code uses a global
time-step controlled by the Courant-Friedrichs-Lewy (CFL) condition, there may be
large regions of the computational domain where the mass distribution almost
does not change during one time-step. In such regions, the gravitational
potential/acceleration from the previous time step may be accurate enough to be
used also in the current time-step. Therefore, to save the computational time, we
implement a technique called the {\em Adaptive Block Update (ABU)}. If
activated, the tree-walk is modified as follows. For each block, the tree-walk
is at first executed only for the eight corner grid cells of the current block. Then,
the gravitational potential or acceleration (or any other quantity calculated by
the tree-solver, e.g. the optical depth) in those eight grid cells is compared
to the values from the previous time-step. If all the differences are smaller
than the required accuracy (given e.g. by Equation~\ref{eq:abs_err} or
\ref{eq:rel_err}), the previous time-step values are adopted for all grid cells
of the block.

For some applications, the eight test cells in the block corners may not be
sufficient. For instance, if the gas changes its configuration in a spherically
symmetric way within a block, the gravitational acceleration at the block corners
does not change, even though the acceleration may change substantially in the
block interior. Such situation is more probable if larger blocks than default $8^3$
cells are used. Therefore, it is easily possible to add more test cells by editing
array \texttt{gr\_bhTestCells} in file \texttt{gr\_bhData.F90}, where test cells
are listed using cell indices within a block, i.e. in a form (1,1,1),
(1,1,8)\dots (8,8,8).

ABU can save a substantial amount of the computational time, however, on large
numbers of processors it works well only if a proper {\em load balancing} among
processors is ensured, i.e. each processor should be assigned with a task of
approximately the same computational cost. {\sc FLASH} is parallelized using a
domain decomposition scheme and individual blocks are distributed among
processors using the space filling Morton curve \citep[see][for
details]{Fryxell2000}. Each processor receives a number of blocks estimated so
that their total expected computational time measured by a {\em workload weight}
is approximately the same as the one for the other processors. By default, {\sc
FLASH} assumes that processing each leaf-block takes approximately the same amount of
time to compute, and it assigns workload weight $2$ to each leaf-block (because
it includes active grid cells) and workload weights $1$ to all other blocks
(they are used only for interpolations between different AMR levels).

The assumption of the same workload per leaf-block cannot be used with ABU,
because if the full tree-walk is executed for a given block less often, the
average computational time spent on it is substantially lower in comparison with
more frequently updated blocks. It is generally hard to predict whether a given
block will be fully updated in the next time-step or not without additional
information about the calculated problem. Therefore, we implement a simple block
workload estimate that leads in most cases to better performance than using the
uniform workload, even though it may not be optimal. It is based on the
assumption that the probability that the block will be updated is proportional
to the amount of work done on the block during several previous time-steps. This
assumption is motivated by considering that a typical simulation includes on one
hand regions where the density and the acceleration change rapidly (e.g. close
to fast moving dense massive objects), and on the other hand, regions where the
acceleration changes slowly (e.g. large volumes filled with hot rarefied gas).
Consequently, the past workload of a given block provides an approximate
estimate its current workload. However, this information is valid only until the
density field evolves enough to change the above property of the region. The
time at which this happens can be approximately estimated as the gas crossing
time of a singe block. Due to the CFL condition, the corresponding number of
time-steps is approximately a number of grid cells in a block along one
direction. Specifically, the block workload estimate works as follows. For each
leaf-block, a total number of node contributions during the tree-walk to all its
grid cells, $N_\mathrm{int}$, is determined. Then, the workload weight,
$W_b^{(n)}$, of that block is calculated as
\begin{equation}
W_b^{(n)} = W_b^{(n-1)} \exp\left(-\frac{1}{\tau_\mathrm{wl}}\right) +
\left[1-\exp\left(-\frac{1}{\tau_\mathrm{wl}}\right)\right]
\left(2+\omega_\mathrm{wl}\frac{N_\mathrm{int}}{N_\mathrm{max}}\right)
\end{equation}
where $W_b^{(n-1)}$ is the workload weight from the previous time-step,
$\tau_\mathrm{wl}$ is a characteristic number of time-steps on which the
workload changes, $\omega_\mathrm{wl}$ is a dimensionless number limiting the
maximum workload weight, and $N_\mathrm{max}$ is the maximum
$N_\mathrm{int}$ taken over all leaf-blocks in the simulation. In this way, the
block workload weight depends on its tree-solver computational cost during the last
several ($\sim\tau_\mathrm{wl}$) time-steps and is between $2$ (zero cost) and
$2+\omega_\mathrm{wl}$ (maximum cost). By default, we set two global
parameters $\tau_\mathrm{wl} = 10$ and $\omega_\mathrm{wl} = 8$. The workload
weight of non-leaf blocks remains equal to $1$.


\subsection{OpticalDepth module}
\label{sec:treecol}


The {\sc OpticalDepth} module is used to evaluate the simplified solution to the
radiative transfer equation
\begin{equation}
I_{\nu} = I_{\nu,0} \;e^{-\tau_{\nu}},
\end{equation}
where $I_{\nu}$ is the specific intensity at frequency $\nu$, $I_{\nu, 0}$ is
the specific intensity at the source location, and $\tau_{\nu}$ is the optical
depth along a given path through the computational domain at frequency $\nu$. In
this form, the problem of evaluating what radiation intensity reaches a given
point in the computational domain, i.e. a given target point, is reduced to
computing the optical depth in between a radiation source and the target point.
The optical depth is proportional to the absorption cross-section and the column
density along the path.

Hence, the {\sc OpticalDepth} module calculates the total and/or specific column
densities (e.g. of molecular hydrogen) for each cell in the computational
domain, and can therefore be used to compute the local attenuation of an
arbitrary external radiation field. The implementation presented here follows
the idea of the {\sc Treecol} method \citep{Clark2012a}, which has been
implemented in the {\sc Gadget} code \citep{Springel2001}. It has been
established as a fast but accurate enough approximative radiative transfer
scheme to treat the (self-)shielding of molecules --on-the-fly -- in simulations
of molecular cloud formation \citep[e.g.][]{Clark2014}. Recently, the method has
also been applied in larger-scale simulations of Milky-Way like galaxies
\citep{Smith2014} with the {\sc Arepo} code \citep{Springel2010}. The
implementation presented here has been successfully used in several recent works
on the evolution of the multi-phase ISM in galactic discs \citep{Walch2015,
Girichidis2016, Gatto2017, Peters2017}.

In principle, the {\sc OpticalDepth} module adds another dimension to the
accumulation of the node masses during the tree-walk. For each grid cell, the
module constructs a {\sc Healpix} sphere \citep{Gorski2005} with a given number
of pixels, $N_{_{\rm PIX}}$, each representing a sphere surface element
with index $i_{_{\rm PIX}}$ corresponding to polar and azimuth angles $\theta$
and $\phi$, respectively. This temporary map is filled while walking the tree,
as only the tree-nodes in the line of sight of a given pixel contribute to it,
and are added accordingly. At the end of the tree-walk, one has acquired a
column density map of a given quantity, e.g. total mass. 

Since the tree-walk in {\sc FLASH} is executed on a block-by-block basis, the
additional memory requirement for the local pixel maps is $2^{l_\mathrm{bt}}
\times N_{_{\rm PIX}} \times l_\mathrm{q}$, where $l_\mathrm{q}$ is the number
of quantities that are mapped and stored.  For this paper, we map
$l_\mathrm{q}=3$ variables: (1) the total mass giving the total hydrogen column
density, $\NHipix$; (2) the H$_2$ column of molecular hydrogen,
which is used to compute its self-shielding and which contributes to the
shielding of CO; and (3) the CO column of carbon-monoxide, which is necessary to
compute the self-shielding of CO. We store three separate maps because we
actually follow the relative mass fractions of multiple species in the
simulation using the {\sc FLASH Multispecies} module. After the tree-walk for a
given block has finished, the local maps are erased and the arrays can be
re-used for the next block. This approach is only possible because the tree-walk
is computed locally on each processor (see \S\ref{sec:treesolver}).

When using the {\sc OpticalDepth} module, there are {\it two major
modifications} with respect to the usual tree-walk (as described above). First,
the intersection of a given tree-node with the line of sight of each pixel has
to be evaluated during the tree-walk. Second, at the end of the tree-walk for a
given block, the acquired column density maps have to be evaluated for each
cell. 

{\sc Node-Ray intersection:} The mapping of tree-nodes onto the individual
pixels represents the core of all additional numerical operations that have to
be carried out when running {\sc OpticalDepth} in addition to the gravity
calculation. It has to be computationally efficient in order to minimise
additional costs. At this point, we do not follow the implementation of
\citet{Clark2012a}, who make a number of assumptions about the shape of the
nodes and their projection onto the pixels, which are necessary to reduce the
computational cost. Instead, we pre-compute the number of intersecting {\sc
Healpix} rays and their respective, relative weight for a large set of nodes at
different angular positions ($\theta$, $\phi$) and different angular sizes
$\psi$. These values are stored in a look-up table, which is accessed during
the tree-walk. In this way, the mapping of the nodes is highly efficient. Since
$\theta$, $\phi$, and $\psi$ are known, we can easily compute the contribution
of a node to all intersecting pixels by simply multiplying the mass (or any
other quantity that should be mapped) of the node with the corresponding weight
for each pixel and adding this contribution to the pixel map. For better
accuracy, we over-sample the {\sc Healpix} tessellation and construct the table
for four times more rays than actually used in the simulation. 

{\sc Radiative heating and molecule formation:} The information that is obtained
by the {\sc OpticalDepth} module is necessary to compute the local heating rates
and the formation and dissociation rates of H$_2$ and CO. At the end of the
tree-walk for a given block, the {\it mean} physical quantities needed by the
{\sc Chemistry} module calculating the interaction of the radiation with the gas
are determined. For instance, the mean visual extinction in a given grid cell is 
\begin{equation}
A_\mathrm{V} = -\frac{1}{2.5}\ln\left[\frac{1}{N_{_{\rm PIX}} }
\sum_{i_\mathrm{PIX}=1}^{N_{_{\rm PIX}} }
\exp\left(-2.5 \frac{\NHipix}{1.87\times 10^{21}\,\mathrm{cm}^{-2}}\right)\right]
\label{eq:av}
\end{equation}
where the constant $1.87\times 10^{21}\,\mathrm{cm}^{-2}$ comes from the
standard relation between the hydrogen column density, $\NHipix$, and the visual
extinction in a given direction \citep{Draine1996}. The weighted mean is calculated in
this fashion, because the photodissociation rates of molecules such as CO and
the photoelectric heating rate of the gas all depend on exponential functions of
the visual extinction \citep[see][for details]{Clark2012b}.
Additionally, the shielding coefficients, $f_\mathrm{shield, H_2}$ and
$f_\mathrm{shield, CO}$ \citep{Glover2007, Glover2010}, as well as the dust
attenuation, $\chi_\mathrm{dust}$ \citep{GloverClark2012a, Clark2012b}, are
computed by averaging over the {\sc Healpix} maps in a similar way. These quantities
are stored as globally accessible variables and can be used by other modules. In
particular, we access them in the {\sc Chemistry} module, which locally (in
every cell) evaluates a small chemical network \citep{Glover2010} on the basis
of its current density and internal energy and re-computes the relative mass
fractions of the different chemical species. The evaluation of the chemical
network is operator split and employs the {\sc Dvode} solver \citep{Brown1989}
to solve a system of coupled ODEs that describes the chemically reactive flow
for the given species, i.e. their creation and destruction within a given time
step. Here, we explicitly follow the evolution of five species, i.e. the
different forms of hydrogen (ionised, H$^+$, atomic, H, and molecular, H$_2$) as
well as ionised carbon (C$^+$) and carbon-monoxide (CO). Details about the
chemical network, e.g. the considered reactions and the employed rate
coefficients in the current implementation can be found in \citet{Glover2010}
and \citet{Walch2015}.
 
{\sc Parameters:} The main parameters controlling both the accuracy and the
speed of the calculation are the number of pixels per map $N_{_{\rm PIX}}$, and
the opening angle, $\theta_\mathrm{lim}$, with which the tree is walked (see
Equation~(\ref{eq:bhmac})). Both should be varied at the same time. A high
number of $N_{_{\rm PIX}}$ used with a relatively large opening angle will not
improve the directional information since the nodes that are mapped into each
solid angle will not be opened and thus, a spatial resolution that is sufficient
for a fine-grained map cannot not be achieved. Therefore we vary both $N_{_{\rm
PIX}}$ and $\theta_\mathrm{lim}$ at the same time.

The number of {\sc Healpix} pixels is directly related to the solid angle of
each element on the unit sphere
\begin{equation}
\Omega_{_{\rm PIX}} = \frac{4 \pi}{ N_{_{\rm PIX}} } [sr].
\end{equation}
Tests in \S\ref{sec:ODtest:cd} show, in agreement with \citet{Clark2012a},
that the code efficiency is optimal if $\theta_\mathrm{lim}$ is approximately
the same as the angular size {\sc Healpix} elements, i.e.
\begin{equation}
\theta_\mathrm{lim} = \sqrt{\Omega_{_{\rm PIX}} }.
\end{equation}
Therefore, for $N_{_{\rm PIX}}=12$, 48, 192 pixels we recommend to use
$\theta_\mathrm{lim}\approx 1.0$, $0.5$, $0.25$.


\section{Accuracy and performance}
\label{sec:accperf}

Since more computational time is needed to reach higher accuracy when solving
numerical problems, accuracy and performance are connected and therefore, these
two properties should always be evaluated at the same time. However, they are
often highly dependent on the specific type of the problem and finding a test
that allows one to objectively measure both accuracy and performance is hard.
Another complication is that the tree-solver saves time by using the information
from the previous time-step (if ABU is switched on), and thus any realistic
estimate of the performance must be measured by running a simulation in which
the mass moves in a similar way as in real applications and by integrating the
computational time over a number of time-steps. Unfortunately, such
simulations are unavoidably too complex to have an analytic solution against
which the accuracy could be easily evaluated.

Therefore we perform two types of tests: static tests that measure accuracy
using simple problems and dynamic tests that evaluate accuracy and performance
together. The static tests need substantially less CPU time and thus allow for a
higher number of parameter sets to be tested. Furthermore, analytic or
semi-analytic solutions are known and the results can be compared to them. On
the other hand, the dynamic tests represent more complex simulations which are
more similar to problems that one would actually want to solve with the
presented code. They also show how well the tree-solver is coupled with the
hydrodynamic evolution (where we use the standard PPM Riemann solver of the {\sc
Flash} code) and how the error accumulates during the evolution. In this
section, we describe four static and two dynamic tests of the {\sc Gravity}
module and one test of the {\sc OpticalDepth} module.

When possible, i.e. for fully periodic of fully isolated boundary conditions, we
compare the results obtained with the new tree-solver to the results obtained
with the default multi-grid Poisson solver of FLASH \citep{Ricker2008}. The
multi-grid solver is an iterative solver and the accuracy is controlled by
checking the convergence of the L2 norm of the Poisson equation residual
$R(\mathbf{r}) \equiv 4\pi G\rho(\mathbf{r}) - \nabla\Phi(\mathbf{r})$. The
iteration process is stopped when $||R_{n}||/||R_{n-1}|| <
\epsilon_\mathrm{mg,lim}$ where $||R_n||$ is the residual norm in the $n$-th
iteration and $\epsilon_\mathrm{mg,lim}$ is the limit set by user. If isolated
boundary conditions are used, the gravitational potential at the boundary is
calculated by a multipole Poisson solver expanding the density and potential
field into a series up to a multipole of order $m_\mathrm{mp}$. By default
$m_\mathrm{mp}=0$ in {\sc Flash} version 4.4. However, using this value we found
unexpectedly high errors close the boundaries (see test \S\ref{sec:bes} and
Figures~\ref{fig:bes:1d} and \ref{fig:bes:2d}), and therefore we use
$m_\mathrm{mp} = 15$ (the highest value allowed for technical reasons) in most
tests because it yields the smallest error.

In general, the calculated gravitational acceleration deviates from the exact
analytical solution due to two effects. The first one is the inherent inaccuracy
of the gravity solver (either the tree gravity solver or the multi-grid solver),
the second one is caused by an imperfect discretisation of the density field on
the grid. Since we are mainly interested in evaluating the first effect, we
measure the error by comparing the calculated accelerations to the {\em
reference solution} obtained by direct "$N^2$" summation of all interactions of
each grid cell with all the other grid cells in the computational domain. We
additionally give the difference between the analytical and the
"$N^2$-integrated" acceleration when possible.

We define the relative error $e_{\rm a}$ of the gravitational acceleration ${\bf
a}$ at the point ${\bf r}$ as
\begin{equation}
e_{\rm a}({\bf r}) \equiv \frac{| {\bf a}({\bf r}) - {\bf a}_{\rm ref}({\bf r}) |}{a_{\rm ref, max}},
\label{eq:ea}
\end{equation}
where ${\bf a}_{\rm ref}$ is the acceleration of the reference solution and
$a_{\rm ref, max}$ is its maximum taken over the whole computational domain.

In most of the gravity module tests, we control the error by setting the
absolute limit $a_{\rm lim}$ on the acceleration, which is calculated from the
initial maximum acceleration in the computational domain, $a_{\rm max}$, as
$a_{\rm lim} = \varepsilon_\mathrm{lim} \times a_{\rm max}$; typically,
$\varepsilon_\mathrm{lim} = 10^{-2}$ or $10^{-3}$. The difference\footnote{Note
that $\varepsilon_\mathrm{lim}$ is only a device to set $a_\mathrm{lim}$ and it
differs from the code parameter $\epsilon_\mathrm{lim}$, which sets the limit on
the acceleration error "on-the-fly" with respect to the previous time-step
acceleration.} between using the absolute or the relative error control is
discussed in \S\ref{sec:macs}.

Most of the tests were carried out on cluster Salomon of the Czech National
Supercomputing Centre IT4I \footnote{http://www.it4i.cz/?lang=en}. A few static
tests that do not need larger computational power have been run on a workstation
equipped with a $4$-core Intel Core i7-2600 processor.


\subsection{Static tests of gravity module}
\label{sec:static}


In order to test all combinations of the boundary conditions implemented in the
{\sc Gravity} module, we present four static tests. A marginally stable
Bonnor-Ebert sphere is used to test the code with isolated boundary conditions
(see \S\ref{sec:bes}) and a density field perturbed by a sine wave not aligned
with any coordinate axis is used to test setups with fully periodic boundary
conditions (\S\ref{sec:jeans}). For mixed boundary conditions, periodic in two
directions and isolated in a third one, or periodic in a single direction and
isolated in the remaining two, we use an isothermal layer in hydrostatic
equilibrium (\S\ref{sec:layer}) and an isothermal cylinder in hydrostatic
equilibrium, respectively (\S\ref{sec:fil}). Finally, in \S{\ref{sec:icyl}}, we
test how the code accuracy depends on the alignment or non-alignment of the gas
structures with the grid axes using a set of parallel cylinders lying in the
xy-plane inclined at various angles with respect to the x-axis.


\subsubsection{Bonnor-Ebert sphere}
\label{sec:bes}


\begin{figure*}
\includegraphics[width=\textwidth]{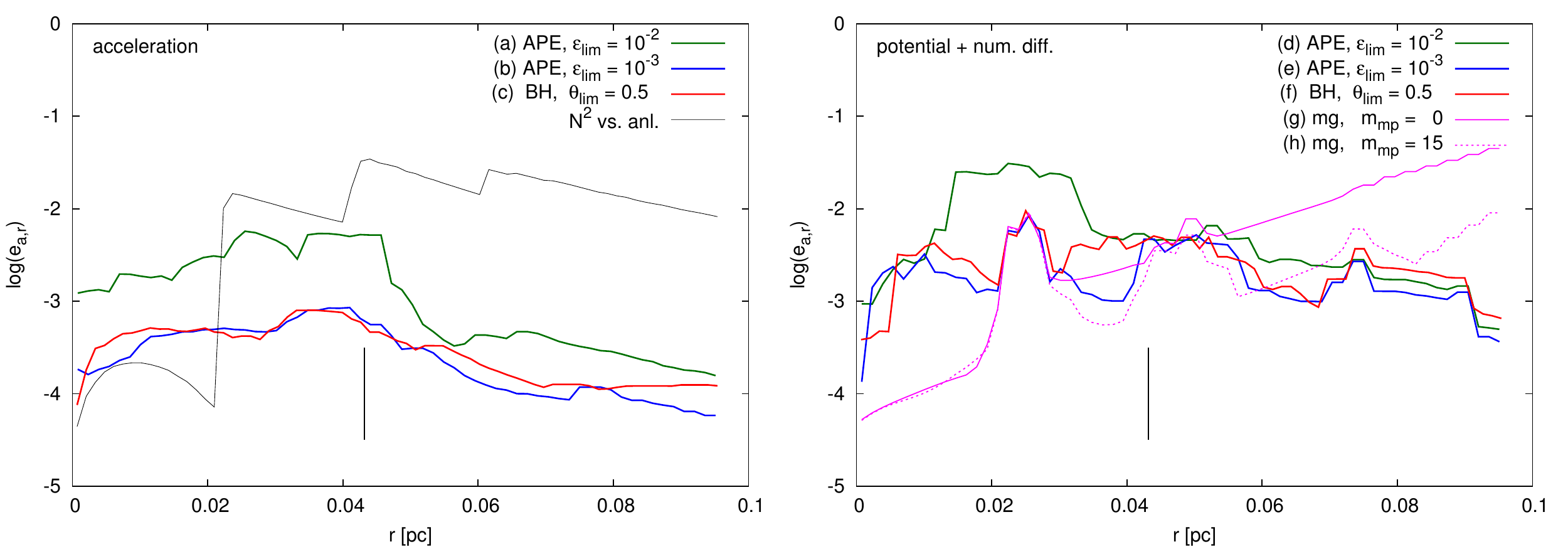}
\caption{Error in the gravitational acceleration for the Bonnor-Ebert sphere as
a function of radius. At a given radius, $r$, the error $e_{a,r}$ is calculated as
a maximum over all angular directions $\phi$ and $\theta$. The vertical black
line shows the BE sphere edge. The solid black line shows the difference between
the acceleration obtained analytically and the reference solution calculated
using the $N^2$ summation. {Left panel:} shows tests where the acceleration was
calculated directly using Equation~(\ref{eq:accel}), the green, blue and red
lines show errors of runs (a), (b) and (c), respectively, with parameters given
in Table~\ref{tab:stat:bes}. {\bf Right panel:} displays tests where the
tree-solver calculates the gravitational potential using Equation~(\ref{eq:pot})
and the acceleration is obtained by numerical differentiation. The green, blue
and red line denote models (d), (e) and (f). The magenta lines show tests
calculated with the multi-grid solver using $m_\mathrm{mp} = 0$ (dashed) and
$m_\mathrm{mp} = 15$ (dotted), respectively. }
\label{fig:bes:1d}
\end{figure*}

\begin{figure}
\includegraphics[width=\columnwidth]{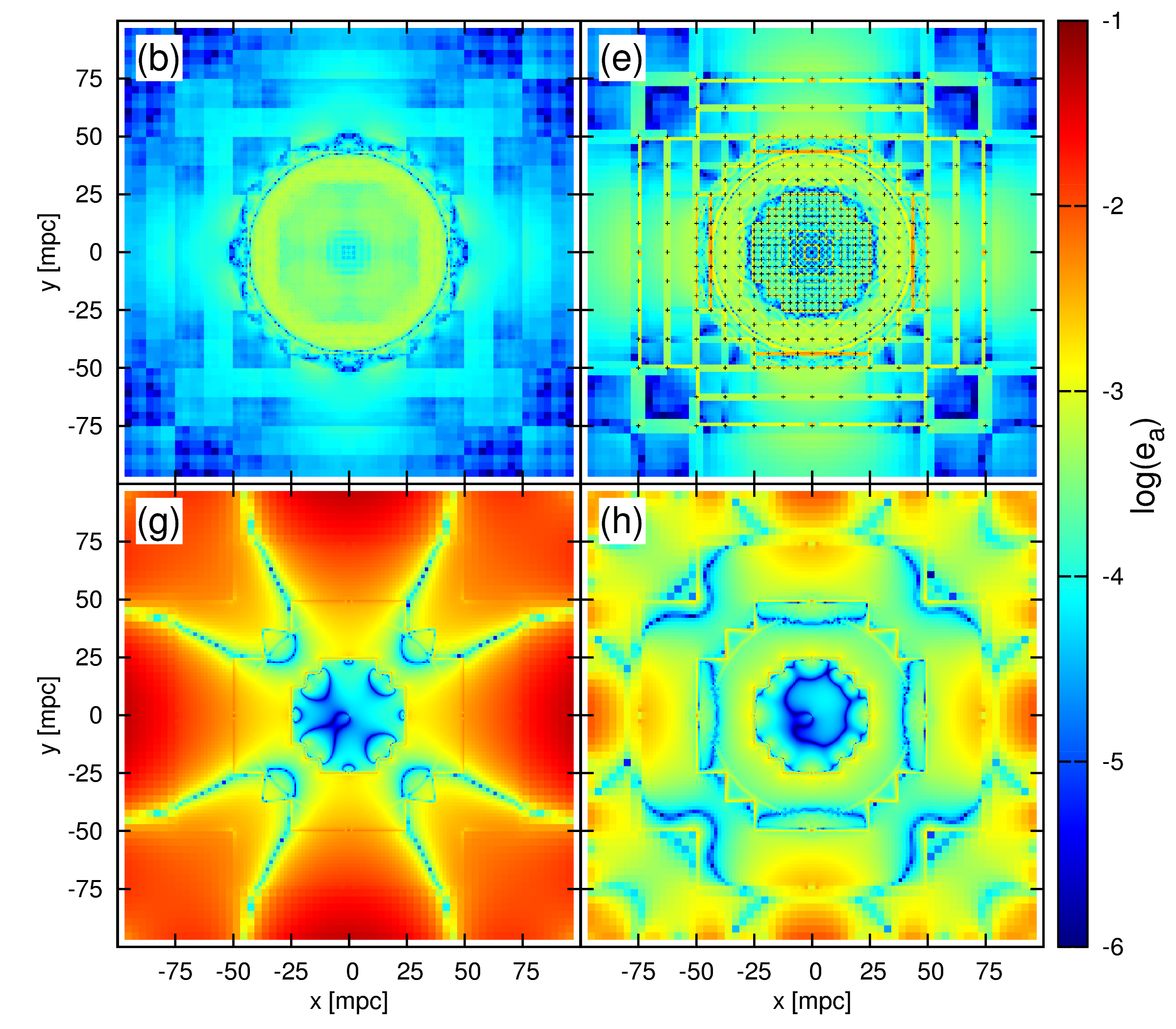}
\caption{Error in the gravitational acceleration, $e_a$, displayed in the $z=0$
plane for the Bonnor-Ebert sphere test. The four panels show four selected runs
with parameters given in Table~\ref{tab:stat:bes}: {\bf top left} corresponds to
model (b) using the tree-solver calculating the grav. acceleration directly;
{\bf top right} shows model (e) where the tree-solver calculated the potential;
{\bf bottom left} is model (g) calculated using the multi-grid solver with
$m_\mathrm{mp} = 0$; and {\bf bottom right} is model (h) calculated using the
multi-grid solver with $m_\mathrm{mp} = 15$. The grid geometry (borders of $8^3$
blocks) is shown in the top right panel.}
\label{fig:bes:2d}
\end{figure}

\begin{table}
\caption{Results of the marginally stable Bonnor-Ebert sphere test.}
\label{tab:stat:bes}
\begin{center}
\begin{tabular}{lllcllclc}
\hline
mod. & solver & quan. & MAC & $\varepsilon_\mathrm{lim}$ & $\theta_\mathrm{lim}$ 
& $m_\mathrm{mp}$ & $e_{a,\mathrm{max}}$ & $t_\mathrm{grv}$\\
\hline
(a) &tree & accel. & APE & $10^{-3}$ & -   & -  & 0.0009  & 83  \\
(b) &tree & accel. & APE & $10^{-2}$ & -   & -  & 0.0057  & 35  \\
(c) &tree & accel. & BH  &  -        & 0.5 & -  & 0.0008  & 110 \\
(d) &tree & pot.   & APE & $10^{-3}$ & -   & -  & 0.0085  & 80  \\
(e) &tree & pot.   & APE & $10^{-2}$ & -   & -  & 0.031   & 38  \\
(f) &tree & pot.   & BH  &  -        & 0.5 & -  & 0.0095  & 106 \\
(g) &mg   & pot.   & -   &  -        & -   & 0  & 0.058   & 21  \\
(h) &mg   & pot.   & -   &  -        & -   & 15 & 0.077   & 20  \\
\hline
\end{tabular}
\end{center}
\begin{flushleft}
We give the model name in column 1. 
The following columns are:
\begin{itemize}
\item solver: indicates whether the tree-solver or the multi-grid solver (mg) is used
\item quan.: quantity calculated by the gravity solver (acceleration or
potential which is then differentiated)
\item MAC: Multipole Acceptance Criterion (Barnes-Hut or Approximate Partial Error)
\item $\varepsilon_\mathrm{lim}$: requested accuracy of the solver as given by
Equation~(\ref{eq:abs_err}) ($a_\mathrm{lim} = \varepsilon_\mathrm{lim}\times
a_\mathrm{max}$ where $a_\mathrm{max}$ is the maximum gravitational acceleration
in the computational domain)
\item $\theta_\mathrm{lim}$: maximum opening angle when the Barnes-Hut MAC is used
\item $e_{a,\mathrm{max}}$: maximum relative error in the computational domain given by
Equation~(\ref{eq:ea})
\item $t_\mathrm{grv}$: time (in seconds) to calculate a single time-step on 8 cores
\end{itemize}
\end{flushleft}
\end{table}

We calculate the radial gravitational acceleration of a marginally stable
Bonnor-Ebert sphere \citep[][BES]{Ebert1955, Bonnor1956} with mass
$M_\mathrm{BE} = 1$~M$_\odot$, temperature $T_\mathrm{BE} = 10$~K and
dimensionless radius $\xi = 6$. The resulting BES radius is $R_\mathrm{BE} =
0.043$~pc and the central density is $\rho_0 = 1.0\times 10^{-18}$~g\,cm$^{-3}$.
The sphere is embedded in a warm rarefied medium with temperature
$T_\mathrm{amb} = 10^4$~K and density $\rho_\mathrm{amb} = 8.5\times
10^{-23}$~g\,cm$^{-3}$, which ensures that the gas pressure across the BES edge
is continuous. We use an AMR grid controlled by the Jeans criterion -- the Jeans
length has to be resolved by at least by 64 cells and at most by 128 cells. It
results in an effective resolution of $512^3$ in the centre of the BES.

Figure \ref{fig:bes:1d} shows the relative error in the gravitational
acceleration, $e_{a,r}$, as a function of radial coordinate, $r$, and Table
\ref{tab:stat:bes} lists all models, their maximum relative error,
$e_{a,\mathrm{max}}$, and the time to calculate one time step, $t_\mathrm{grv}$.
We compare the solutions calculated with the tree gravity solver using the
geometric (BH) MAC with $\theta_\mathrm{lim} = 0.5$ (red curves) to the ones
calculated using the APE MAC with $\varepsilon_\mathrm{lim} =10^{-2}$ (green
lines) and $\varepsilon_\mathrm{lim} =10^{-3}$ (blue lines), respectively. The
APE MAC and $\varepsilon_\mathrm{lim} = 10^{-3}$ as well as the geometric MAC
with $\theta_\mathrm{lim} = 0.5$ always give a maximum relative error which is
smaller than $0.1\%$. In case of the APE MAC and $\varepsilon_\mathrm{lim} =
10^{-2}$, the maximum relative error reaches $\sim 1\%$. Note that the error due
to the discretisation of the density field is also of the order of 1\% (black
line; the jumps are due to changes in the refinement level in the AMR grid).

With the tree gravity solver, the user may choose to directly compute the
gravitational accelerations (left panel of Figure~\ref{fig:bes:1d}) or to
calculate them by numerical differentiation of the gravitational potential
(right panel of Figure~\ref{fig:bes:1d}). Usually, the latter is the standard
practice in grid-based 3D simulations, also because only one field variable, the
potential, has to be stored instead of three, the accelerations in three spatial
directions. However, for the tree-solver we generally find that the error in the
gravitational accelerations is significantly smaller (about a factor of $10$ in
the test presented here) if they are computed directly. This is independent of
the used MAC. 

For comparison, we also show the results obtained with the multi-grid solver
(magenta lines) using $\varepsilon_\mathrm{mg,lim}=10^{-6}$ and
$m_\mathrm{mp}=0$ (solid lines) or $m_\mathrm{mp}=15$ (dotted lines),
respectively. Although the mass distribution is spherically symmetric, the order
of the multipole expansion of the boundary condition affects the accuracy of the
multi-grid solver relatively far away from boundaries, even inside the BES. The
error of the multi-grid solver is very low in the central region, it reaches
$\sim 1\%$ in regions where the refinement level changes (due to numerical
differentiation of the potential), and increases to relatively high values at
the border of the computational domain ($\sim 1\%$ for $m_\mathrm{mp} = 15$ and
$\sim 5\%$ for $m_\mathrm{mp} = 0$), due to inaccuracy of the boundary
conditions calculated by the multipole solver. We note that a direct calculation
of the gravitational acceleration is not possible with the multi-grid solver.

The distribution of the relative error ${\bf e}_a$ in the $z=0$ plane through
the centre of the BES is depicted in Figure~\ref{fig:bes:2d}. The results show
that the acceleration obtained with the tree gravity solver using the APE MAC
with $\varepsilon_\mathrm{lim} = 10^{-2}$ has a substantially smaller error if
it is calculated directly (top left panel; see Table \ref{tab:stat:bes} model
(b)) instead of by numerical differentiation of the potential (top right panel;
model (e)). The bottom panels show the results for the multi-grid solver with
$m_\mathrm{mp} = 0$ (model (g)) and $m_\mathrm{mp} = 15$ (model (h)),
respectively. The default setting of $m_\mathrm{mp} = 0$ gives errors of $\sim$
5\% near the domain boundaries due to the low accuracy of the multi-pole solver.
This error propagates into a large fraction of the computational domain.


\subsubsection{Sine-wave perturbation (Jeans test)}
\label{sec:jeans}


\begin{figure}
\includegraphics[width=\columnwidth]{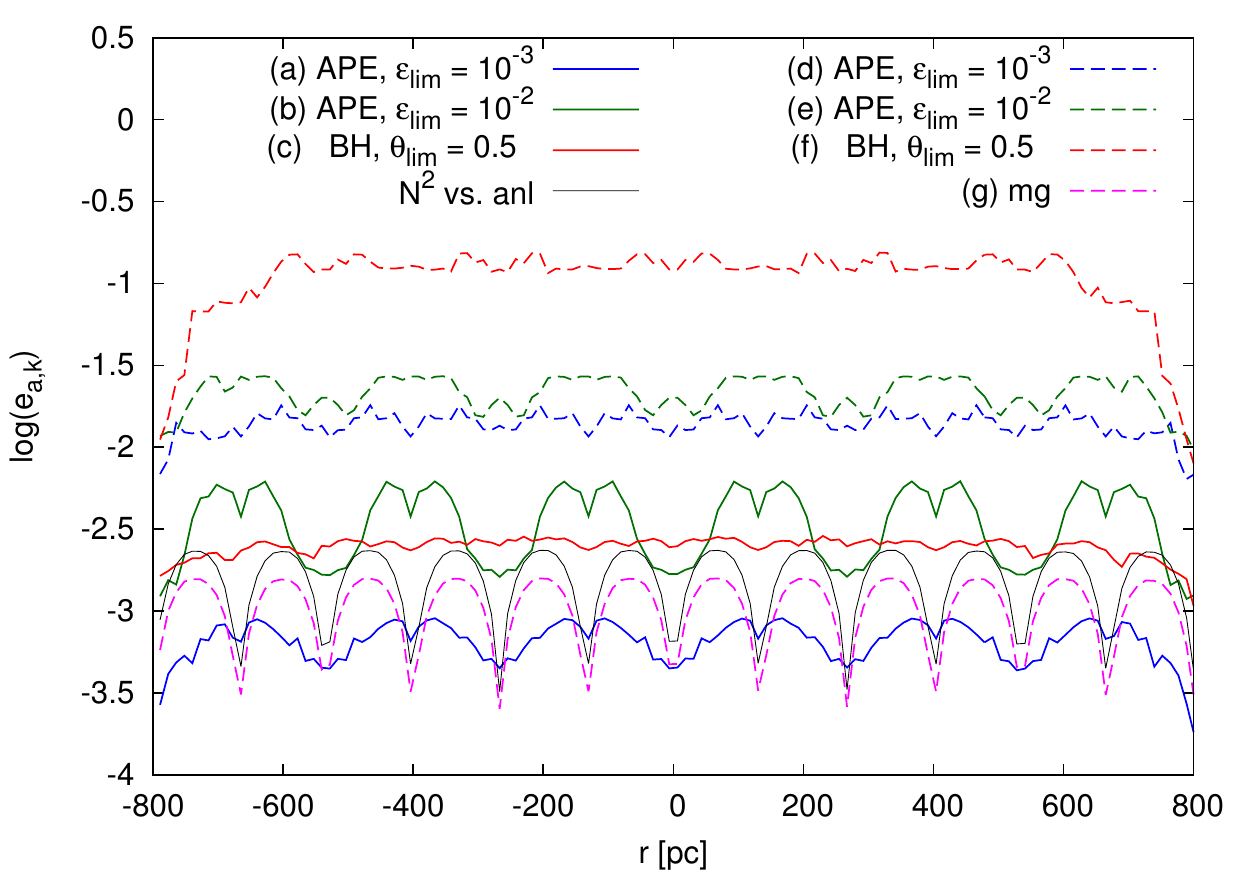}
\caption{Maximum relative error of the gravitational acceleration for the Jeans
test. Solid lines show acceleration calculated directly, while dashed lines show
acceleration calculated by numerically differentiating the potential. The
acceptance criteria are the same as in Figure~\ref{fig:bes:1d}.
}
\label{fig:jeans}
\end{figure}

In a computational domain with fully periodic boundary conditions we calculate
the gravitational acceleration of a smooth density field with a harmonic
perturbation,
\begin{equation}
\rho(\mathbf{r}) = \rho_0 + \rho_1 \cos(\mathbf{k}\cdot\mathbf{r}),
\end{equation}
where $\rho_0 = 1.66\times 10^{-24}$ is the mean density and $\rho_1 =
0.99\rho_0$ is the amplitude of the perturbation. The computational domain is a
cube of size $500$~pc with $128$ grid cells in each direction. The wave-vector
$\mathbf{k} = 6\pi(3,2,1)/L$ was chosen such that it is not aligned with any of
the coordinate axes. The gravitational acceleration can be obtained analytically
with the help of the Jeans swindle \citep{Jeans1902,Kiessling1999}
\begin{equation}
\mathbf{g}(\mathbf{r}) = -4\pi G\rho_1\frac{\mathbf{k}}{k^2}\sin(\mathbf{k}\cdot\mathbf{r})
\ .
\end{equation}

Figure~\ref{fig:jeans} shows the maximum relative error $e_{a,k}$ as a function of
the position $x_k$ on a line parallel to the perturbation wave-vector
$\mathbf{k}$. The maximum error is computed from all points projected to a given
position on the line. It can be seen that the error of the multi-grid solver
(magenta curve) is very small, almost the same as the difference between the
analytical solution and the reference solution (black line). This is because
without the need to calculate the boundary conditions separately, and on a
uniform grid, the FFT accelerated multi-grid method is extremely efficient.
Again, the results for the tree-solver simulations show that direct calculation
of the acceleration (solid curves) leads to a much lower error than the
calculation of the potential and subsequent differentiation (dashed lines). In
particular, the calculation of the potential with the geometric MAC that does
not take into account the different mass density in the tree-nodes leads to a
relative error greater than $10$\%. However, a direct calculation of the
acceleration gives very accurate results for both, the geometric MAC and the APE
MAC with $\varepsilon_\mathrm{lim} = 10^{-3}$. In Table \ref{tab:stat:jeans} we
list all models with their respective $e_{a,\mathrm{max}}$ and $t_\mathrm{grv}$.

\begin{table}
\caption{Results of the second static test: sine-wave perturbation.
The meaning of the columns is the same as in Table~\ref{tab:stat:bes}.}
\label{tab:stat:jeans}
\begin{center}
\begin{tabular}{l|l|l|l|c|c|l|r}
\hline
model & solver & quan. & MAC & $\varepsilon_\mathrm{lim}$ & $\theta_\mathrm{lim}$ 
& $e_{a,\mathrm{max}}$ & $t_\mathrm{grv}$\\
\hline
(a) & tree & accel. & APE & $10^{-3}$ & -   &  0.0009 & 480 \\
(b) & tree & accel. & APE & $10^{-2}$ & -   &  0.0062 & 210 \\
(c) & tree & accel. & BH  &  -        & 0.5 &  0.0029 & 250 \\
(d) & tree & pot.   & APE & $10^{-3}$ & -   &  0.0180 & 330 \\
(e) & tree & pot.   & APE & $10^{-2}$ & -   &  0.0270 & 130 \\
(f) & tree & pot.   & BH  &  -        & 0.5 &  0.15   & 150 \\
(g) & mg   & pot.   & -   &  -        & -   &  0.0016 & 9   \\
\hline
\end{tabular}
\end{center}
\end{table}


\subsubsection{Isothermal layer in hydrostatic equilibrium}
\label{sec:layer}


\begin{figure}
\includegraphics[width=\columnwidth]{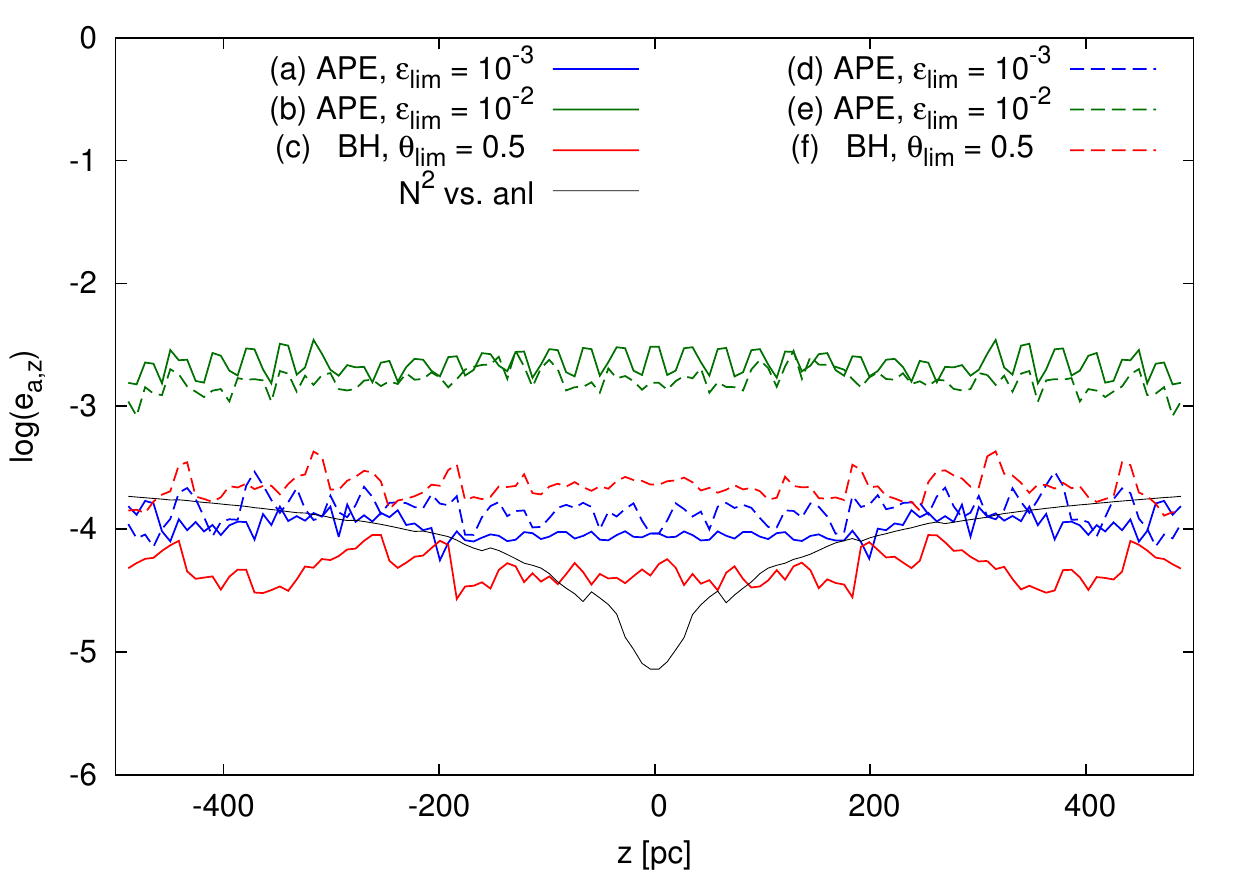}
\caption{Maximum relative error of the gravitational acceleration for the
isothermal layer. Meaning of line types is the same as in
Fig.~\ref{fig:bes:1d}.}
\label{fig:layer}
\end{figure}

In order to test the accuracy of the tree gravity module with mixed boundary
conditions (periodic in two directions and isolated in the third one), we
calculate the gravitational acceleration of an isothermal layer in hydrostatic
equilibrium. The vertical density distribution of the layer is \citep{Spitzer1942}
\begin{equation}
\rho(z) = \rho_0 \mathrm{sech}^{2}\left( \sqrt{ \frac{2\pi G\rho_0}{c_{s}^{2}} } z \right)
\end{equation}
where $\rho_0 = 1.6\times 10^{-24}$~g\,cm$^{-3}$ is the mid-plane density and
$c_{s} = 11.7$~km\,s$^{-1}$ is the isothermal sound speed. The corresponding
vertical component of the gravitational acceleration is
\begin{equation}
g_{z}(z) = 2\sqrt{2\pi G\rho_0 c_s^2} \tanh 
\left( \sqrt{ \frac{2\pi G\rho_0}{c_{s}^{2}} } z \right)
\ .
\end{equation}
The computational domain is a cube of side length $L = 1000$~pc and a uniform
resolution of $128$ grid cells in each direction.

Figure~\ref{fig:layer} shows the maximum relative error $e_{a,z}$ in the
acceleration as a function of the $z$-coordinate, where the maximum is taken
over all cells with the same $z$-coordinate. It can be seen that the error is
almost independent of $z$ and there is only a small difference between the cases
where the gravitational acceleration is calculated directly (solid lines) or
where it is obtained by differentiation of the potential (dashed lines). The
reason is that the density field in this test has relatively shallow gradients
(e.g. compared to the Jeans test discussed in the previous section) and
numerical differentiation leads to particularly severe errors for steep
gradients. We find the largest error for runs with APE MAC and
$\varepsilon_\mathrm{lim} = 10^{-2}$. All other runs have small errors, which
are comparable to the difference between the analytical and the reference
solution, resulting from the discretisation of the density field. The results
are summarised in Table \ref{tab:stat:layer}.

\begin{table}
\caption{Results of the second static test: isothermal layer in hydrostatic
equilibrium. The meaning of columns is the same as in Table~\ref{tab:stat:bes}.}
\label{tab:stat:layer}
\begin{center}
\begin{tabular}{l|l|l|l|c|c|l|r}
\hline
model & solver & quan. & MAC & $\varepsilon_\mathrm{lim}$ & $\theta_\mathrm{lim}$ 
& $e_{a,\mathrm{max}}$ & $t_\mathrm{grv}$\\
\hline
(a) & tree & accel. & APE & $10^{-3}$ & -   &  0.00017 & 170 \\
(b) & tree & accel. & APE & $10^{-2}$ & -   &  0.0035  & 106 \\
(c) & tree & accel. & BH  &  -        & 0.5 &  $9.0\times 10^{-5}$ & 180 \\
(d) & tree & pot.   & APE & $10^{-3}$ & -   &  0.00029 & 99  \\
(e) & tree & pot.   & APE & $10^{-2}$ & -   &  0.0028  & 45  \\
(f) & tree & pot.   & BH  &  -        & 0.5 &  0.00043 & 107 \\
\hline
\end{tabular}
\end{center}
\end{table}


\subsubsection{Isothermal cylinder in hydrostatic equilibrium}
\label{sec:fil}


\begin{figure}
\includegraphics[width=\columnwidth]{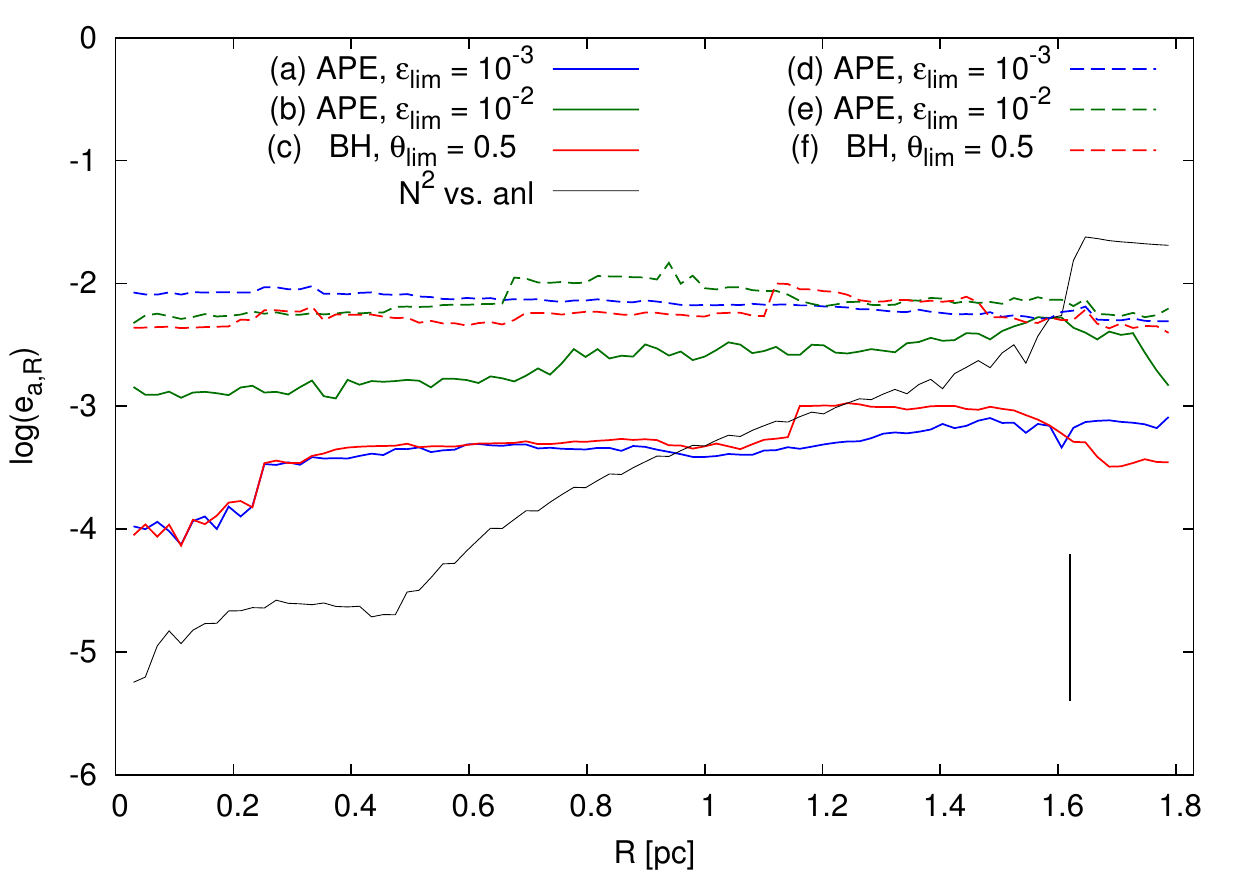}
\caption{Maximum relative error of the gravitational acceleration for the
isothermal cylinder. Meaning of line types is the same as in
Figure~\ref{fig:bes:1d}. The black vertical line denotes the edge of the cylinder.}
\label{fig:cyl}
\end{figure}

In the next static test, we evaluate the accuracy of the tree gravity module for
mixed boundary conditions, which are isolated in two directions and periodic in
the third one. We calculate the gravitational acceleration of an isothermal
cylinder in hydrostatic equilibrium. The long axis of the cylinder is parallel
to $x$-coordinate and the radius is given as $R = \sqrt{y^2+z^2}$. The density
distribution is \citep{Ostriker1964}
\begin{equation}
\rho(R) = \rho_0 \left(1+\frac{\pi G \rho_0 R^2}{2c_s^2}\right)^{-2}
\end{equation}
where $\rho_0 = 3.69\times 10^{-23}$~g\,cm$^{-3}$ is the central density and
$c_s = 0.2$~km\,s$^{-1}$ is the isothermal sound speed. The density distribution
is cut off at radius $R_\mathrm{cyl} = 1.62$~pc and embedded in an ambient gas
with $c_{s,\mathrm{amb}} = 10$~km\,s$^{-1}$ and the same pressure as the pressure
at the cylinder boundary. The corresponding gravitational acceleration is
\begin{equation}
\mathbf{g}(R) = 2 \pi G \rho_0 R \left(1+\frac{\pi G \rho_0 R^2}{2c_s^2}\right)^{-1}
\ .
\end{equation}
The computational domain has dimensions $3.6 \mathrm{pc}\times 1.8 \mathrm{pc}
\times 1.8 \mathrm{pc}$ and contains $256\times 128\times 128$ grid cells.

Figure~\ref{fig:cyl} shows the maximum relative error $e_{a,R}$ of the
gravitational acceleration in radial direction, where the maximum error is
calculated for all grid cells at the same distance $R$ to the cylinder axis. In
all runs, the error is a very weak function of $R$. If numerical differentiation
of the potential is used, it is the dominant source of the error, which is as
large as 1\% in these cases (see dashed lines). The results are summarised in
Table \ref{tab:stat:cyl}.

\begin{table}
\caption{Results of the fourth static test: isothermal cylinder in hydrostatic
equilibrium. The meaning of columns is the same as in Table~\ref{tab:stat:bes}.}
\label{tab:stat:cyl}
\begin{center}
\begin{tabular}{l|l|l|l|c|c|l|r}
\hline
model & solver & quan. & MAC & $\varepsilon_\mathrm{lim}$ & $\theta_\mathrm{lim}$ 
& $e_{a,\mathrm{max}}$ & $t_\mathrm{grv}$\\
\hline
(a) & tree & accel. & APE & $10^{-3}$ & -   & $0.00082$ & $270$ \\
(b) & tree & accel. & APE & $10^{-2}$ & -   & $0.0053$  & $110$ \\
(c) & tree & accel. & BH  &  -        & 0.5 & $0.0011$  & $280$\\
(d) & tree & pot.   & APE & $10^{-3}$ & -   & $0.0095$  & $180$\\
(e) & tree & pot.   & APE & $10^{-2}$ & -   & $0.015$   & $73$ \\
(f) & tree & pot.   & BH  &  -        & 0.5 & $0.010$   & $175$\\
\hline
\end{tabular}
\end{center}
\end{table}


\subsubsection{Inclined cylinders}
\label{sec:icyl}


\begin{figure}
\includegraphics[width=\columnwidth]{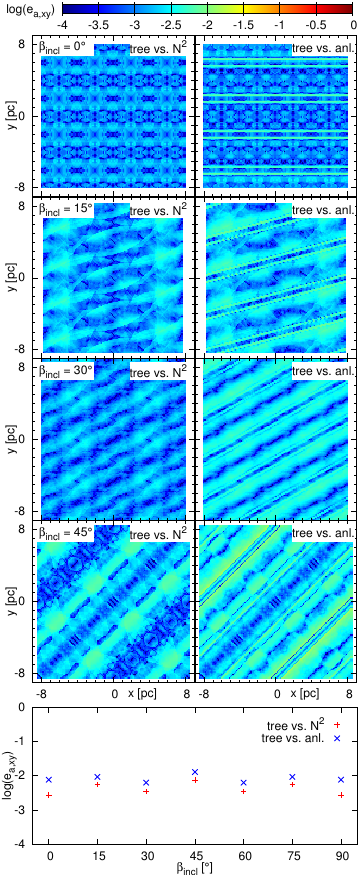}
\caption{Relative error of the gravitational acceleration in the xy-plane,
$e_\mathrm{a,xy}$, for the set of inclined cylinders. Left panels show the
logarithm of the error measured with respect to the direct $N^2$ integration,
right panels show the error with respect to analytically obtained accelerations.
Each of the top four rows show the calculation with different inclination angle
of the cylinders: $0$, $15$, $30$ and $45$ degrees from top to bottom. The panel
at the very bottom shows the logarithm of the maximum error in the acceleration
$e_\mathrm{a,xy}$ as a function of the cylinder inclination angle,
$\beta_\mathrm{incl}$.
}
\label{fig:icyl}
\end{figure}

In order to test whether the alignment of gas structures with the coordinate
axes has an impact on the code accuracy, i.e. whether the algorithm is sensitive
to any grid effects, we calculate gravitational field of the set of parallel
cylinders in the 2P1I geometry. The axes of all cylinders lie in the $xy$-plane
and they are inclined at angle $\beta_\mathrm{incl}$ with respect to the
$x$-axis. The computational domain has an extent $48$\,pc in the isolated
$z$-direction and approximately $16$\,pc in the periodic $x-$ and
$y-$directions. The exact extents in the latter two directions are chosen so
that the computational domain composes a periodic cell of the infinite plane of
cylinders, i.e. the cylinders connect contiguously to each other at the $x$ and
$y$ periodic boundaries. Each cylinder has the same radius and density profile
as the cylinder described in section \S\ref{sec:fil}, the distance between the
cylinder axes is $4$\,pc. We have calculated $7$ models with
$\beta_\mathrm{incl}$ increasing from $0\,\degr$ to $90\,\degr$ with a step
$15\,\degr$. For all models, the gravity tree solver was running with the BH MAC
and maximum opening angle $\theta_\mathrm{lim} = 0.5$.

Figure~\ref{fig:icyl} shows the relative error of the gravitational
acceleration, $e_\mathrm{a,xy}$, calculated in the $xy$-plane using
Equation~(\ref{eq:ea}). The reference acceleration, $\mathbf{a}_\mathrm{ref}$,
is either obtained numerically by the $N^2$-integration (four panels on the left
for $\beta_\mathrm{incl} = 0\degr - 45\degr$), or analytically by summing up
potential of $1000$ parallel cylinders (four panels on the right). The error
with respect to the $N^2$-integration is always smaller than $1\%$. The error
with respect to the analytical acceleration is of order $1\%$ and is always
slightly higher than the former error, as it includes contribution from the
imperfect discretisation of the density field reaching the highest values along
the cylinder edges where the density field has a discontinuity. The bottom panel
show the maximum $e_\mathrm{a,xy}$ as a function of $\beta_\mathrm{incl}$
demonstrating that the code accuracy is almost independent of the inclination of
the gaseous structures with respect to coordinate axes.


\subsection{Dynamic tests of gravity module}


We run two dynamic tests of the gravity module. The first one (described in
\S\ref{sec:evrard}) is a collapse of a cold adiabatic sphere suggested by
\citet{Evrard1988} and it tests how well the energy is conserved during the
gravitational collapse. The second one, describes the evolution of a turbulent
sphere (\S\ref{sec:turbsphere}). Both test the accuracy of the gravity module
and its coupling to the hydrodynamic solver.


\subsubsection{Evrard test}
\label{sec:evrard}


\begin{figure}
\includegraphics[width=\columnwidth]{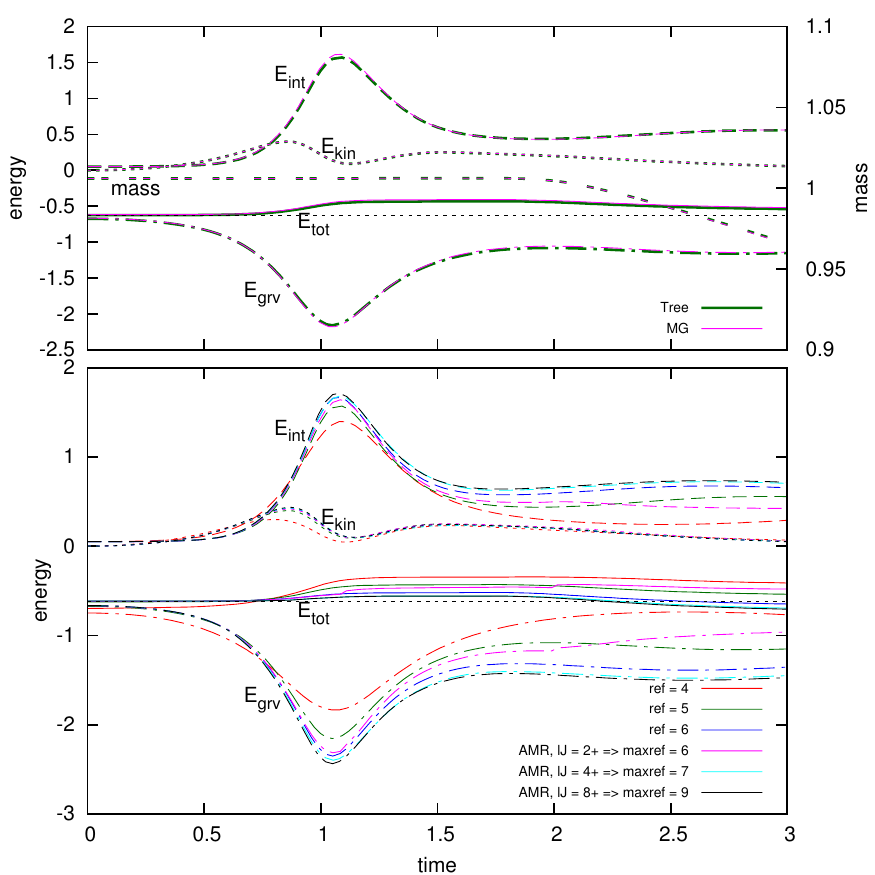}
\caption{Time evolution of the total mass and thermal, kinetic, gravitational and
total energy for the Evrard test. {\bf Top panel} compares calculation with the
tree gravity solver (green lines) and the multi-grid solver (magenta lines) at
the same grid with uniform resolution $128^3$. The two runs are almost
indistinguishable. {\bf Bottom panel} compares calculations with the tree-solver
at different resolution. The red, green and blue lines show calculations done on
a uniform grid with constant refinement levels $4$, $5$ and $6$, corresponding to
grid sizes $64^3$, $128^3$ and $256^3$, respectively. The magenta, cyan and
black lines show runs with the AMR grid where the resolution was set so that the
Jeans length is always resolved at least by $2$, $4$ and $8$ grid cells,
respectively. It resulted in the maximum refinement levels reached $6$, $7$ and
$9$, respectively.}
\label{fig:evrard}
\end{figure}

The Evrard test \citep{Evrard1988} describes the gravitational collapse and a
subsequent re-bounce of an adiabatic, initially cold sphere. It is often used to
verify energy conservation in SPH codes
\citep[e.g.][]{Springel2001,Wetzstein2009}, its application on grid-based
codes is unfortunately less common. The initial conditions consist of a gaseous
sphere of mass $M$, radius $R$ and density profile
\begin{equation}
\rho(r) = \frac{M}{2\pi R^2 r} \ .
\end{equation}
The initial, spatially constant temperature is set so that the internal energy
per unit mass is
\begin{equation}
u = 0.05 \frac{GM}{R} \ ,
\end{equation}
where $G$ is the gravitational constant. The standard values of the above
parameters, used also in this work, are $M = R = G = 1$.

In Figure~\ref{fig:evrard} we show the time evolution of the total mass as well
as the gravitational, kinetic, internal, and total energy. On the top panel, we
compare the results obtained with the tree gravity solver and the multi-grid
solver, both computed on a uniform grid of size $128^3$ corresponding to a
constant refinement level equal to $5$. The tree-solver run uses the Barnes-Hut
MAC with $\theta_\mathrm{lim} = 0.5$, the multi-grid run was calculated with the
default accuracy $\epsilon_\mathrm{mg,lim} = 10^{-6}$ and $m_\mathrm{mp} = 0$.
The two runs are practically indistinguishable, however, the total energy (that
should stay constant) rises by approximately $0.1$ during the period of maximum
compression. Since the distribution of the error in the gravitational
acceleration calculated by the two solvers is very different, the same results
indicate that the error in the energy conservation is not caused by the
calculation of the gravitational acceleration and that the acceleration errors
are below the sensitivity of this test.

The bottom panel of Figure~\ref{fig:evrard} compares runs calculated with the
tree-solver at different resolutions. It includes three runs with uniform grids
of sizes $64^3$, $128^3$ and $256^3$ (corresponding to constant refinement
levels of $4$, $5$, and $6$) and three runs calculated on adaptive grids, which
are refined such that the Jeans length is resolved by at least $2$, $4$, and $8$
grid cells, respectively.

We find that low resolution leads to a higher numerical dissipation and
artificial heating of the gas. Furthermore, lower resolution does not allow high
compression of the sphere centre leading to less pronounced peaks of the
internal and gravitational energies. Consequently, the results of this test show
that high resolution is needed only in the centre of the sphere where the
highest density is reached.


\subsubsection{Turbulent sphere}
\label{sec:turbsphere}


\begin{figure}
\includegraphics[width=\columnwidth]{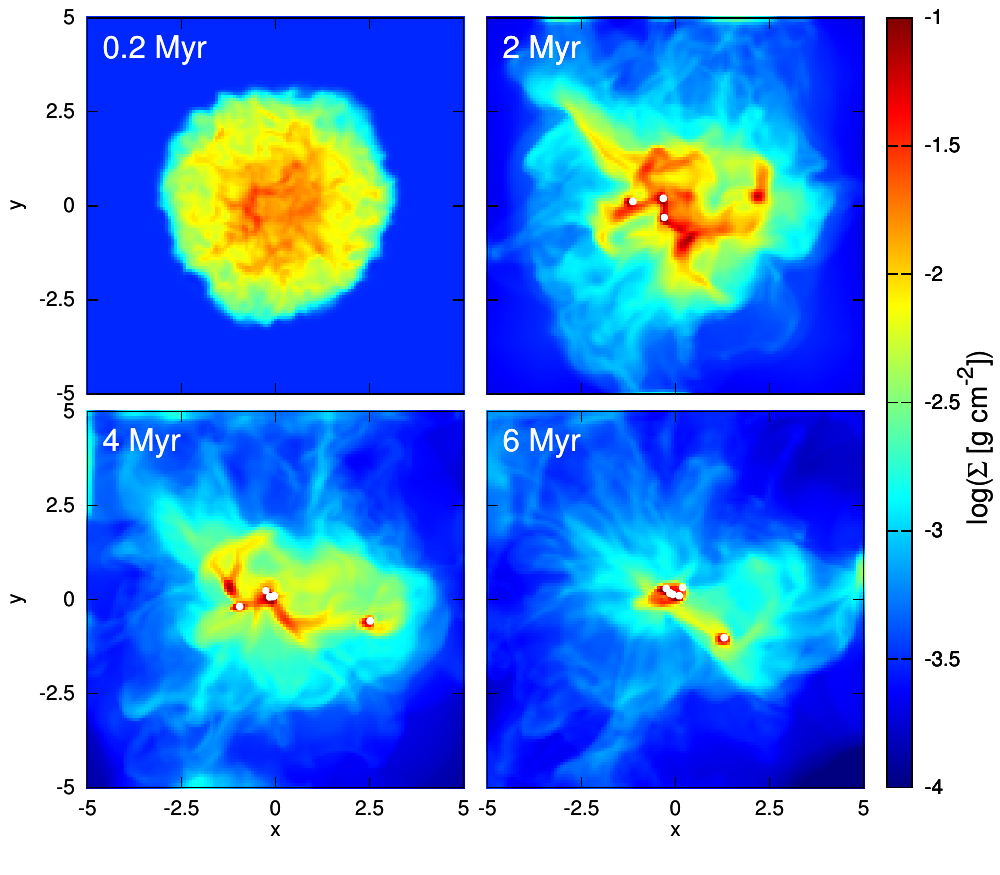}
\caption{Evolution of the mass column density in the turbulent sphere test
(shown run a). Individual panels show different stages of the evolution at
$0.2$, $2$, $4$ and $6$~Myr. Sink particles are shown as white circles.}
\label{fig:tsph:evol:2d}
\end{figure}

\begin{figure}
\includegraphics[width=\columnwidth]{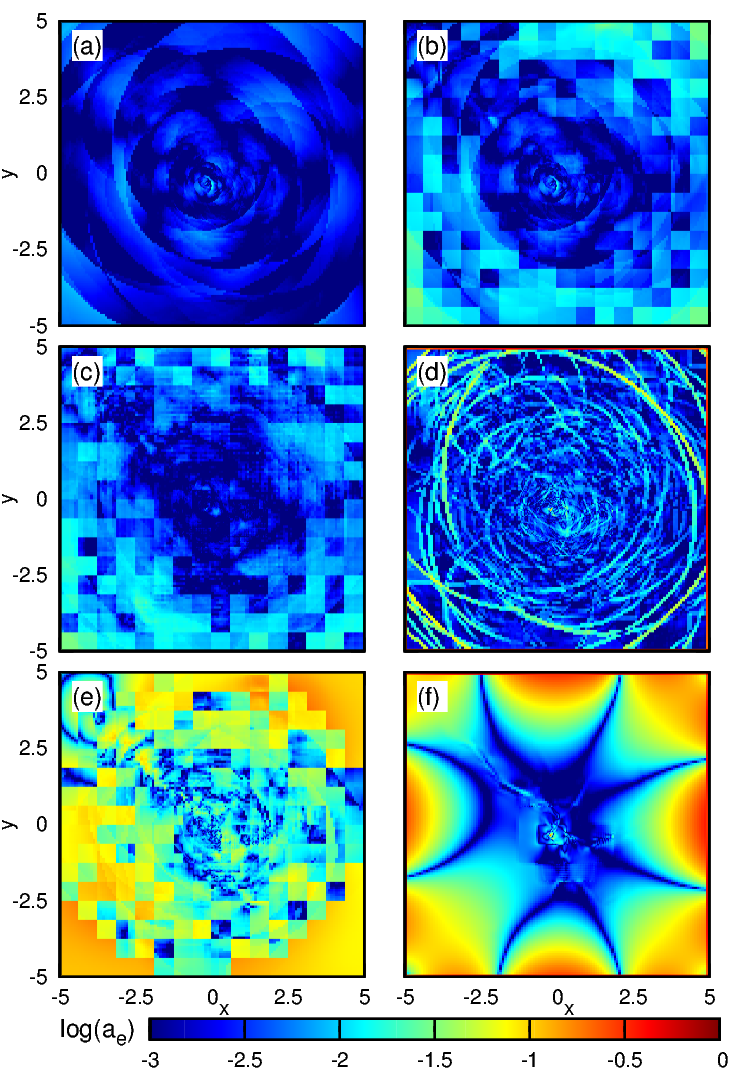}
\caption{Error in the gravitational acceleration in the $xy$-plane of the
turbulent sphere at $t = 2$\,Myr. The panels show: (a) tree gravity solver
calculating the acceleration with BH MAC, $\theta_\mathrm{lim} = 0.5$, and
adaptive block update (ABU) switched off, (b) tree gravity solver calculating
the acceleration with BH MAC, $\theta_\mathrm{lim} = 0.5$, and ABU on, (c) tree
gravity solver calculating the acceleration with APE MAC, $\varepsilon_{lim} =
10^{-2}$ and ABU on, (d) tree gravity solver calculating the potential with APE
MAC, $\varepsilon_{lim} = 10^{-2}$ and ABU on, (e) tree gravity solver
calculating the potential with APE MAC, $\varepsilon_{lim} = 10^{-1}$ and ABU
on, (f) multi-grid solver calculating the potential with $\epsilon_{mg,lim} =
10^{-6}$ and $m_\mathrm{mp} = 10$.}
\label{fig:tsph:gerr:2d}
\end{figure}

\begin{figure}
\includegraphics[width=\columnwidth]{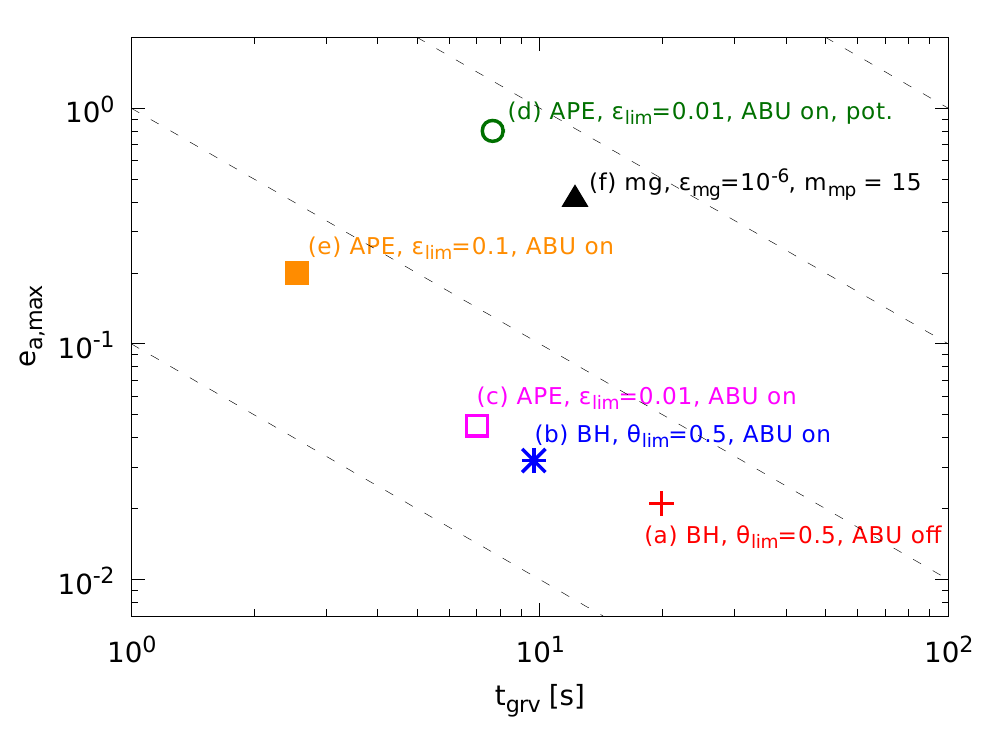}
\caption{Results of the turbulent sphere plotted in the plane of the gravity
calculation duration in seconds, $t_\mathrm{grv}$ (x-axis) versus the maximum
relative error in the gravitational acceleration, $e_{a,\mathrm{max}}$ (y-axis).
The error is determined at $t = 2$\,Myr and the maximum is taken over the whole
computational domain. The thin dashed lines are iso-lines of constant
$t_\mathrm{grv}\times e_{a\mathrm{max}}$ assessing the code efficiency. 
Parameters of the displayed runs are given in Table~\ref{tab:dyn:tsph}.
}
\label{fig:tsph:accperf}
\end{figure}

\begin{figure}
\includegraphics[width=\columnwidth]{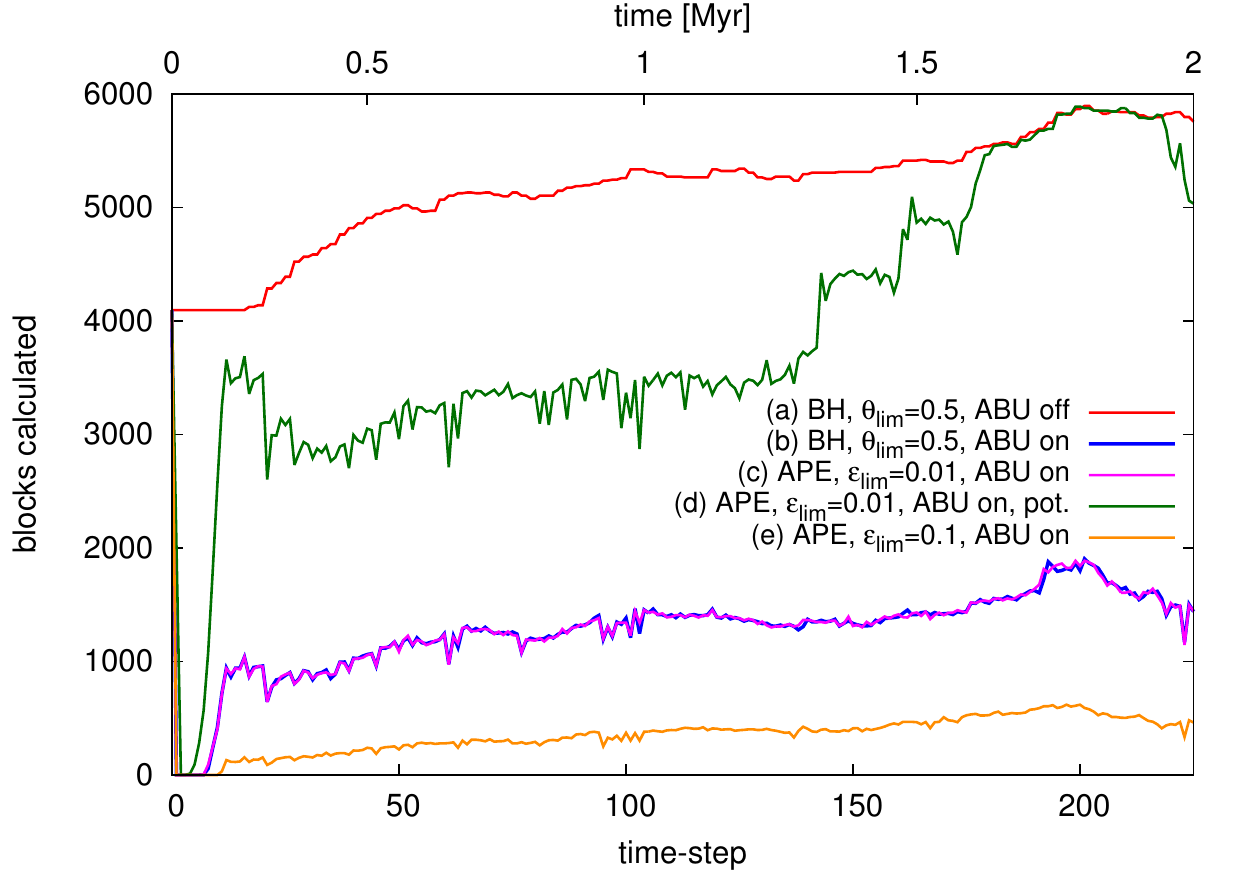}
\caption{Number of updated blocks, i.e. blocks for which the tree walk was
executed for all grid cells in a given time-step, as a function of time (top
x-axis) or time-step number (bottom x-axis). The figure shows first 2 Myr of the
evolution of the turbulent sphere test. Individual curves represent models (a)
-- (e) as given in the legend (see also Table~\ref{tab:dyn:tsph} for model
parameters). Note that blue and magenta lines (models (b) and (c)) are on the
top of each other.
}
\label{fig:tsph:blkstat}
\end{figure}

The turbulent sphere represents a proto-typical star formation test. We set up a
turbulent, isothermal sphere with a total mass of $10^3\;{\rm M}_{\odot}$,
radius $3$\,pc, and temperature $10$\,K. The initial density profile is Gaussian
with a central density of $\rho_{0}=1.1\times 10^{-21} \;{\rm g\;cm}^{-3}$ and
the density at the edge is $\rho_0/3$. It is embedded in a cubic box with side
length $L=10$ pc, which is filled with a rarefied ambient medium of density
$\rho_{\rm amb} = 10^{-23} \;{\rm g\;cm}^{-3}$ and temperature $100$\,K. We add
an initial turbulent velocity field to the sphere with a Kolmogorov spectrum on all
modes with wave numbers between $k_{\rm min} = 2\times (2\pi/L)$ and $k_{\rm max}
= 32\times (2\pi/L)$. The magnitude of velocity perturbations is scaled so that
the total kinetic energy is $0.7$ times the absolute value of the total
potential energy.

The sphere is evolved under the influence of self-gravity and hydrodynamics, and
since it is gravitationally bound it collapses towards the center and forms
stars. We use isolated gravity boundary conditions, while the hydrodynamic
boundary conditions are set to "outflow". The spatial resolution on the base
grid is $128^3$ (refinement level $5$) and with AMR we allow for a maximum
effective resolution of $1024^3$ (refinement level $8$). All calculations were
carried out on the IT4I/Salomon supercomputer running on 96 processor cores.

To model the star formation process, we introduce sink particles according to a
Jeans criterion if the gas density is above a threshold density of $\rho_{\rm
thres}=10^{-18}$\,g\,cm$^{-3}$ and other criteria are fulfilled \citep[see][for
a description of the sink particles in FLASH]{Federrath2010}. All sink particles
live on the maximum allowed refinement level and the gravitational forces among
all sink particles and between the particles and the gas are computed by direct
summation. They are evolved using a Leapfrog integrator.

In Figure~\ref{fig:tsph:evol:2d} we show the evolving column density in the
$xy$-plane at times $0.2$, $2.0$, $4.0$, and $6.0$\,Myr. Although the simulation
is interesting in itself, we only focus on the error in the resulting
gravitational acceleration. Therefore, we compute the same initial conditions
six times with different gravity settings and measure the resulting error of the
gravitational acceleration, where the supposedly accurate result compared to
which we calculate the error is obtained using $N^2$ integration. The results of
our analysis are shown in Figure~\ref{fig:tsph:gerr:2d}, which depicts the error
in the $xy$-plane at $t=2$ Myr. The maximum and average errors
$e_{a,\mathrm{max}}$ and $e_{a,\mathrm{avg}}$, respectively, and mean times per
gravity and hydrodynamic time-step computations $t_\mathrm{grv}$ and
$t_\mathrm{hydro}$, respectively, are given in Table~\ref{tab:dyn:tsph}. The
runs are also shown in the $t_\mathrm{grv}$-$e_{a,\mathrm{max}}$ plane in
Figure~\ref{fig:tsph:accperf}.

The top two panels of Figure~\ref{fig:tsph:gerr:2d} show calculations with the
tree-solver calculating directly the gravitational acceleration using the
geometric MAC with $\theta_\mathrm{lim} = 0.5$. The left panel
(\ref{fig:tsph:gerr:2d}a) was calculated without the Adaptive Block Update (ABU
off). The relative error is very small everywhere, with sudden changes at
constant distances from massive concentrations of gas, resulting from switching
tree-node sizes as prescribed by the geometric MAC criterion. The maximum error
is approximately $2$\,\%, the average error is even an order of magnitude
smaller. One iteration of the tree-solver took approximately $20$ seconds, i.e.
it was the slowest run. The right panel (\ref{fig:tsph:gerr:2d}b) shows the same
calculation, but the ABU was switched on in this case. The relative error
exhibits a rectangular pattern, because some blocks, in particular in the outer
regions, were not updated in a given time-step and the error in them is larger.
The maximum error is approximately $3$\,\% , i.e. $1.5$ times more than in the
run with ABU off, and at the same time, the ABU makes the calculation
approximately two times faster.

Panel (\ref{fig:tsph:gerr:2d}c) shows a run with the tree-solver using the APE
MAC with $\varepsilon_\mathrm{lim} = 10^{-2}$. The results are very similar to
the one in run (\ref{fig:tsph:gerr:2d}b), with a maximum relative error of
approximately $4.5$\,\% ($\sim 1.5$ larger) and the mean time per gravity
time-step is $7$ seconds (slightly smaller). Panel (\ref{fig:tsph:gerr:2d}d)
shows the run with the same tree-solver parameters, but instead of
calculating the acceleration directly, the tree-solver calculates the potential
and differentiates it numerically. The relative error exhibits a similar pattern
to run (a), however, instead of sudden changes it includes high peaks of the
error resulting from a numerical differentiation. Even though the mean error is
comparable to runs (b) and (c), the maximum error is much higher, reaching
$80$\,\%. The time of the gravity calculation is slightly higher than in run
(c), even though calculating the potential is cheaper than the acceleration for
a single target cell. It is because a higher number of blocks must be updated in
each time-step due to the higher error.

Panel (\ref{fig:tsph:gerr:2d}e) includes the run with a reduced accuracy of
$\varepsilon_\mathrm{lim} = 10^{-1}$ made to test the limits of the tree-solver
usability. The relative error is high, in particular in the outer regions where
the blocks are updated less often, reaching a maximum of $20$\,\%, however, it
is still a factor of $2$ smaller than the error at the boundaries of the
computational domain found in the run with the multi-grid solver (see below). On
the other hand, the calculation is very fast with a mean time per gravity
time-step of $3.5$ seconds, which is $\sim 70$\,\% of the time needed by the
hydro solver. 

Panel (\ref{fig:tsph:gerr:2d}f) displays the calculation with the multi-grid
solver and as in \S\ref{sec:bes} it shows that the largest error (reaching $\sim
40$\,\%) is along the boundaries of the computational domain where the potential
is influenced by the boundary values obtained by the multipole expansion. The
error in the central region is of the order of several percent, comparable to
the runs in panels (\ref{fig:tsph:gerr:2d}b) and (\ref{fig:tsph:gerr:2d}c). The
run with the multi-grid solver is $20 - 30$\,\% slower.

In order to evaluate the efficiency of the ABU, we show in
Figure~\ref{fig:tsph:blkstat} a number of updated blocks in each time-step as a
function of time / time-step number. The red curve corresponds to model (a)
where ABU was switched off, i.e. it shows the number of all blocks in the
simulation. It grows from 4096 to almost 6000, as the AMR creates more blocks in
regions of high density formed by the gravitational collapse. The number of all
blocks is the same for all simulations, as they run the identical model. For
model (b) (blue curve), the number of updated blocks stays very small for the
first $\sim 10$ time-steps, because the initial time-step is very low and the
density and gravitational acceleration fields almost do not change. As soon as
the time-step reaches a value given by the CFL condition, the number of updated
blocks quickly rises up to $~1000$ and then it increases slowly to almost $2000$
at $2\,$Myr. Throughout the evolution, the number of updated blocks is
approximately three times lower than in model (a) with ABU off. As a result,
model (b) runs more than twice as fast as model (a) and the maximum error in the
acceleration is approximately $1.5$ times larger (see Table~\ref{tab:dyn:tsph}
and Figures~\ref{fig:tsph:gerr:2d} and \ref{fig:tsph:accperf}). For model (c),
the fraction of updated blocks is almost the same as for model (b). However, the
model runs $\sim 30\%$ faster as the APE MAC needs less interaction than the BH
MAC of model (b) and consequently, the maximum error is $\sim 1.5$ larger. Model
(e) with larger error limit updates less than 10\% of blocks in each time-step
and as a result it runs $8\times$ faster than model (a) and its maximum error is
almost $10\times$ larger. Model (d) calculating the potential instead of the
acceleration behaves in a different way. The number of updated blocks exceeds
$3000$ shortly after the start of the simulation, their fraction stays above
$50$\% and reaches $100$\% in the last quarter of the time. It is because the
numerical differentiation of the potential at the border between updated and
not-updated blocks tends to give high error in the acceleration. Therefore we do
not recommend the use of ABU together with calculating the potential.

Note that the efficiency of the ABU test is highly problem dependent. In this
regard, the used turbulent sphere setup is a relatively hard one, because the
sphere quickly forms dense filaments with large density gradients and they move
supersonically as the whole structure collapses (i.e. the time-step is given
mainly by the gas velocity, not the sound-speed). On the other hand, there are
still regions where the gravitational acceleration changes slowly, e.g. in the
computational domain corners, and these regions can be updated less often making
the ABU efficient. If the volume with fast moving dense objects is larger, the
ABU can be less efficient and vice verse.

\begin{table*}
\caption{Accuracy and performance of the turbulent sphere test.}
\label{tab:dyn:tsph}
\begin{center}
\begin{tabular}{c|l|l|l|l|c|c|c|c|c|c}
\hline
model & solver & quan. & MAC & ABU& $\varepsilon_\mathrm{lim}$ & $\theta_\mathrm{lim}$ &
$e_{a,\mathrm{max}}$  & $e_{a,\mathrm{avg}}$ & $t_\mathrm{grv}$ & $t_\mathrm{hydro}$\\
\hline
(a) & tree & accel. & BH  & off &  -        & $0.5$ & $0.021$ & $0.0020$ & $19.9$ & $7.5$ \\
(b) & tree & accel. & BH  & on  &  -        & $0.5$ & $0.032$ & $0.0061$ &  $9.7$ & $5.3$ \\
(c) & tree & accel. & APE & on  & $10^{-2}$ & -     & $0.045$ & $0.0056$ &  $7.0$ & $4.5$ \\
(d) & tree & pot.   & APE & on  & $10^{-2}$ & -     & $0.801$ & $0.0062$ &  $7.7$ & $4.0$ \\
(e) & tree & accel. & APE & on  & $10^{-1}$ & -     & $0.200$ & $0.0472$ &  $2.5$ & $4.8$ \\
(f) & mg   & pot.   & -   & -   &  -        & -     & $0.416$ & $0.0447$ & $12.2$ & $4.7$ \\
\hline
\end{tabular}
\end{center}
\begin{flushleft}
Column 1 gives the model name. The following columns list:
\begin{itemize}
\item solver: indicates whether the tree-solver or the multi-grid solver (mg) is used
\item quan.: quantity calculated by the gravity solver (acceleration or
potential which is then differentiated)
\item MAC: Multipole acceptance Criterion (Barnes-Hut or Approximate Partial Error)
\item ABU: Adaptive Block Update (on or off)
\item $\varepsilon_\mathrm{lim}$: requested accuracy of the solver as given by
Equation~(\ref{eq:abs_err}) ($a_\mathrm{lim} = \varepsilon_\mathrm{lim}\times
a_\mathrm{max}$ where $a_\mathrm{max}$ is the maximum gravitational acceleration
in the domain)
\item $\theta_\mathrm{lim}$: maximum opening angle when the Barnes-Hut MAC is used
\item $e_{a,\mathrm{max}}$: maximum relative error in the computational domain given by
Equation~(\ref{eq:ea}) measured at $t = 2$ Myr
\item $e_{a,\mathrm{avg}}$: average relative error in the computational domain given by
Equation~(\ref{eq:ea}) measured at $t = 2$ Myr
\item $t_\mathrm{grv}$: time per time-step (in seconds) to calculate the gravitational
acceleration on 96 cores
\item $t_\mathrm{hydro}$: time per time-step spent in the hydrodynamic solver on 96 cores
\end{itemize}
\end{flushleft}
\end{table*}


\subsection{Test of the OpticalDepth module}
\label{sec:ODtest}


In order to evaluate the accuracy of the {\sc OpticalDepth} module, we perform
two tests. For both of them, we repeat the calculation of the turbulent sphere
described in \S\ref{sec:turbsphere}, using an adiabatic equation of state with
$\gamma=5/3$ (instead of the isothermal equation of state used
previously). Additionally, we switch on the {\sc OpticalDepth} and {\sc
Chemistry} modules calculating the gas cooling and heating, and the mass
fractions of various species. The sphere is heated from the outside assuming a
typical interstellar radiation field (ISRF) of strength $G_0=1.7$ times the
Habing field. This causes the low ambient density gas to heat up to a few$\times
10^3$ K, while the interior of the sphere is cold and thus it collapses to form
stars as in runs with the isothermal equation of state in
\S\ref{sec:turbsphere}. A detailed description of the chemical network, the
heating \& cooling processes it includes, the dust temperature it calculates,
and how the {\sc OpticalDepth} module is coupled to it can be found in
\citet{Walch2015}. Here, we are only concerned with the workings of the {\sc
OpticalDepth} module and with the column density (or optical depth) it delivers. 

In the first test, we evaluate how accurately the {\sc OpticalDepth} module
determines the column density depending on the chosen angular resolution;
and in the second test, we compare the resulting optical depth with the optical
depth computed using the Monte Carlo radiative transfer code {\sc RADMC-3D}
\footnote{See
http://www.ita.uni-heidelberg.de/$\sim$dullemond/software/radmc-3d/}.


\subsubsection{Column density with increasing $N_{_{\rm PIX}}$}
\label{sec:ODtest:cd}


\begin{figure*}
\includegraphics[width=0.9\textwidth]{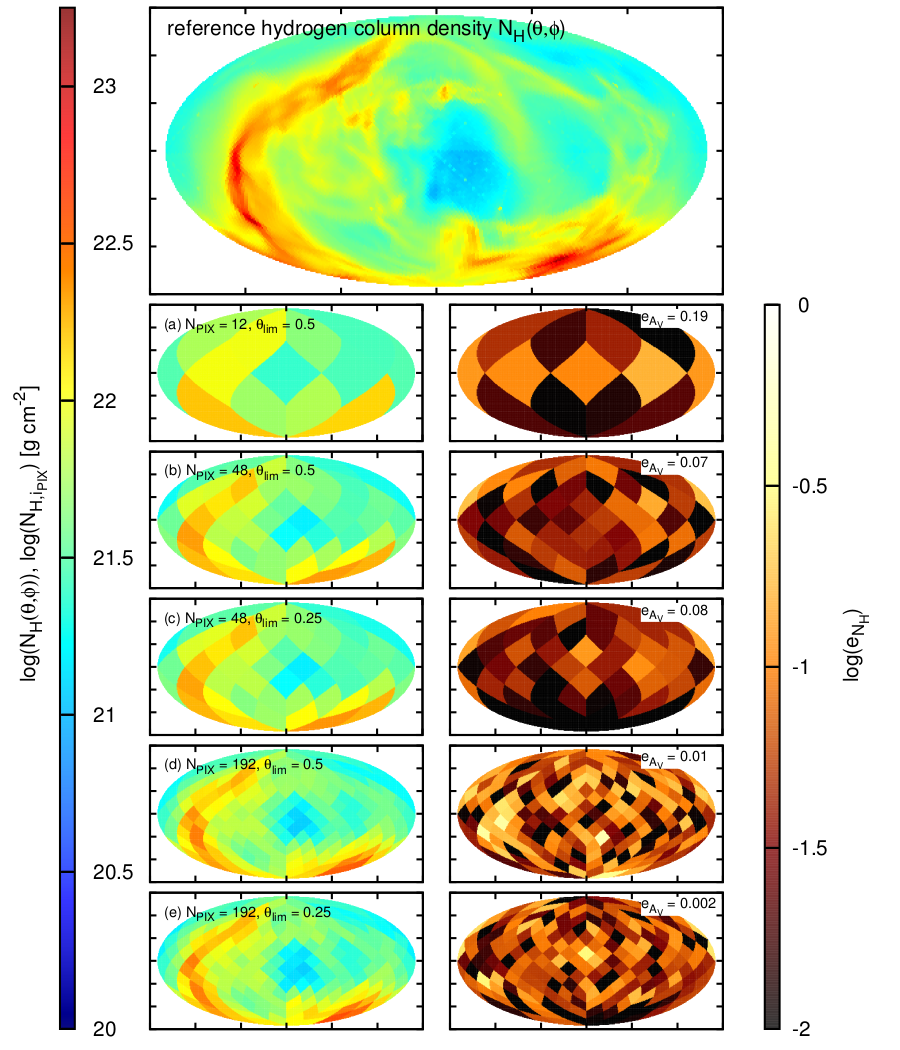}
\caption{Results of the {\sc OpticalDepth} module test evaluating the accuracy
of the hydrogen column density calculation as seen from the centre of the
turbulent sphere, as a function of the used angular resolution. {\bf Top panel:}
shows the reference hydrogen column density, $N_\mathrm{H}(\theta,\phi)$, displayed in
the Hammer projection. {\bf Left panels below:} show $\NHipix$ determined by the
{\sc OpticalDepth} module with $N_{_{\rm PIX}}=12, 48$, and $192$, and
$\theta_\mathrm{lim} = 0.5$ and $0.25$ as denoted in the top left corner of each
panel. {\bf Right panels below:} show the relative error in the column density,
$e_{\mathrm{N_{H}},i_{\rm PIX}}$, calculated using Equation~(\ref{eq:eNH}). The
relative error in the mean visual extinction, $e_\mathrm{A_{V}}$, is given in
top right corners of the panels.}
\label{fig:odtest}
\end{figure*}

\begin{table}
\caption{Results of the {\sc OpticalDepth} module test studying the dependency
of the column density accuracy on the resolution.}
\label{tab:ODtest:cd}
\begin{center}
\begin{tabular}{lrccccc}
\hline
model & $N_{_{\rm PIX}}$ & $\theta_\mathrm{lim}$ & $e_\mathrm{N_{H},max}$ &
$e_{A_\mathrm{V}}$ & $t_\mathrm{tree}$ & $N \mathrm{vs.} H$\\
\hline
(a) &  12 & 0.5  &  0.16  & 0.19  & 41  & $N < H$\\
(b) &  48 & 0.5  &  0.18  & 0.07  & 44  & $N \approx H$\\
(c) &  48 & 0.25 &  0.14  & 0.08  & 256 & $N < H$\\
(d) & 192 & 0.5  &  0.48  & 0.01  & 48  & $N > H$\\
(e) & 192 & 0.25 &  0.30  & 0.002 & 286 & $N \approx H$\\
\hline
\end{tabular}
\end{center}
\begin{flushleft}
We give the model name in column 1. 
The following columns are:
\end{flushleft}
\begin{itemize}
\item $N_{_{\rm PIX}}$: number {\sc Healpix} pixels corresponding to
the angular resolution
\item $\theta_\mathrm{lim}$: maximum opening angle (Barnes-Hut MAC is
used in all tests)
\item $e_\mathrm{N_{H},max}$: maximum relative error in the hydrogen
column density (Equation \ref{eq:eNH}); maximum is taken over all {\sc Healpix}
pixels
\item $e_{A_\mathrm{V}}$: relative error in the mean visual extinction
(Equation \ref{eq:eAV})
\item $t_\mathrm{tree}$: time per time-step (in seconds) spent in 
the tree-solver on $96$ cores
\item $N \mathrm{vs.} H$; indicates the relative size of tree nodes (N) and {\sc
Healpix} elements (H)
\end{itemize}
\end{table}

We perform a test similar to the one by \citet[][their \S3.2]{Clark2012a}, and
calculate the "sky map" of hydrogen column density, $N_\mathrm{H}$, as seen from
the centre of the computational domain, for the turbulent sphere simulation at
time $t = 2$\,Myr (see the top right panel of Figure~\ref{fig:tsph:evol:2d}).
The hydrogen column density determined by the {\sc OpticalDepth} module,
$\NHipix$, is compared to the `actual' reference hydrogen column density,
$N_\mathrm{H}(\theta,\phi)$, obtained using a direct integration over individual
grid cells of the simulation and very high {\sc Healpix} resolution $N_{_{\rm
PIX}} = 3072$. The angular resolution of the {\sc OpticalDepth} module is
controlled by two parameters: number of {\sc Healpix} elements $N_{_{\rm PIX}}$
and tree maximum opening angle $\theta_\mathrm{lim}$ determining the
maximum angular size of tree nodes. We calculate five models with $N_{_{\rm
PIX}}=12, 48$, and $192$, and two maximum opening angles
$\theta_\mathrm{lim} = 0.5$ and $0.25$ (see Table~\ref{tab:ODtest:cd}). We
define a relative error in the hydrogen column density
\begin{equation}
e_{\mathrm{N_{H}},i_{\rm PIX}} = \frac{|\NHipix 
- \langle N_\mathrm{H}(\theta,\phi)\rangle_{i_{\rm PIX}}|}
{\langle N_\mathrm{H}(\theta,\phi) \rangle_{i_{\rm PIX}} }
\label{eq:eNH}
\end{equation}
where $\langle N_\mathrm{H}(\theta,\phi) \rangle_{i_{\rm PIX}}$ is the mean
value of reference hydrogen column density, $N_\mathrm{H}$, in element
$i_\mathrm{PIX}$. In Table~\ref{tab:ODtest:cd}, we give for each model the
maximum error $e_\mathrm{N_{H},max} \equiv \max(e_{\mathrm{N_{H}},i_{\rm
PIX}})$, where the maximum is taken over the whole sphere.

However, directionally dependent $\NHipix$ does not directly enter calculations
of the gas-radiation interaction. Instead, the {\sc Chemistry} module uses
quantities averaged over all directions, e.g. the mean visual extinction,
$A_\mathrm{V}$, given by Equation~(\ref{eq:av}). Therefore, we further define
a relative error in the mean visual extinction
\begin{equation}
e_{A_\mathrm{V}} = \frac{|A_\mathrm{V} - A_\mathrm{V,ref}|}{A_\mathrm{V,ref}}
\label{eq:eAV}
\end{equation}
where $A_\mathrm{V}$ is the mean visual extinction calculated by the {\sc
OpticalDepth} module and $A_\mathrm{V,ref}$ is the reference value obtained by
averaging the high-resolution reference hydrogen column density
$N_\mathrm{H}(\theta,\phi)$ using Equation~(\ref{eq:av}). Values of
$e_{A_\mathrm{V}}$ for the five calculated runs are also given in
Table~\ref{tab:ODtest:cd} and in top right corners of right panels in
Figure~\ref{fig:odtest}.

The results are summarised in Figure~\ref{fig:odtest} showing, in the Hammer
projection, the reference hydrogen column density $N_\mathrm{H}(\theta,\phi)$
(top panel), values of $\NHipix$ calculated by the {\sc OpticalDepth} module
(left panels), and relative errors, $e_{\mathrm{N_{H}},i_{\rm PIX}}$ (right
panels). Our findings are generally in agreement with those of
\citet{Clark2012a}. Even run (a) with $N_{_{\rm PIX}}=12$ recovers approximately
the overall structure of the cloud and results in $e_\mathrm{N_{H},max} = 0.16$
and $e_{A_\mathrm{V}} = 0.19$. Increasing the {\sc Healpix} angular resolution
to $N_{_{\rm PIX}}=48$ (run (b) with the approximately same size of tree nodes
and {\sc Healpix} elements, see the last column in Table~\ref{tab:ODtest:cd})
leads to a smaller error in $A_\mathrm{V}$ while keeping $e_\mathrm{N_{H},max}$
approximately the same. Since runs (a) and (b) take nearly the same time to
calculate, their comparison shows that it is not worth to degrade the {\sc
Healpix} resolution (by decreasing $N_{_{\rm PIX}}$) below the tree-solver
resolution (given by $\theta_\mathrm{lim}$). Similarly, run (c) with better tree
solver resolution ($\theta_\mathrm{lim} = 0.25$) and the same {\sc Healpix}
resolution results in the approximately same $e_\mathrm{N_{H},max}$ and
$e_{A_\mathrm{V}}$ as run (b), even though the computational costs are much
higher. Run (d) with {\sc Healpix} elements smaller than the tree node size
leads to smaller $e_{A_\mathrm{V}} = 0.01$, however, $e_\mathrm{N_{H},max} =
0.48$ is very high. It is because the approximations adopted when the mass of
relatively large tree nodes is distributed to {\sc Healpix} elements sometimes
result in the assignment of the mass to a different element. This problem is
diminished in run (e), which has again the approximately same angular size of
tree nodes and {\sc Healpix} elements, and in which the visual extinction error
drops to a very small value, $e_{A_\mathrm{V}} = 0.002$ and $e_\mathrm{N_{H},max}$
also decreases in comparison with run (d), even though it is still higher than
in runs (a)-(c).


\subsubsection{Comparison to {\sc RADMC-3D}}
\label{sec:ODtest:radmc}


\begin{figure}
\begin{center}
\includegraphics[width=\columnwidth]{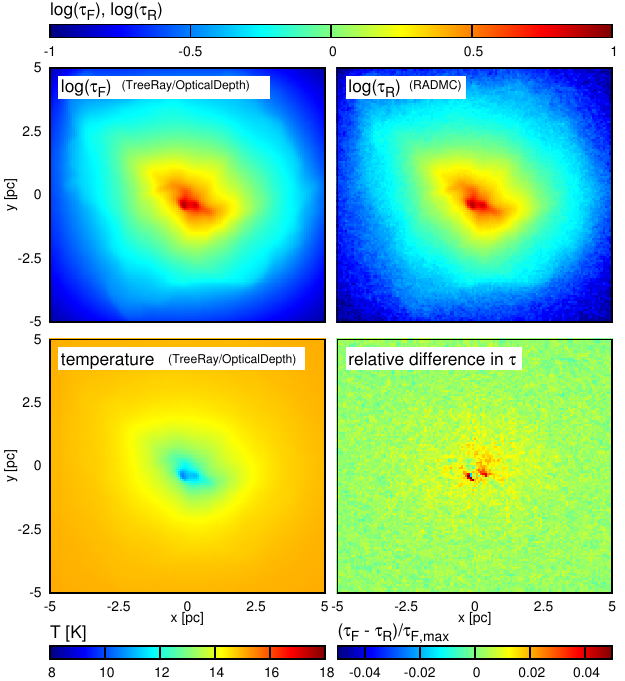}
\end{center}
\caption{Test of the {\sc OpticalDepth} module showing the calculated optical
depths in a slice at $z=0$ through the turbulent sphere at time $t=2$ Myr. {\bf
Top left:} logarithm of the optical depth, $\tau_\mathrm{F}$, at wavelength
$\lambda_0 = 9.36\times 10^{-2}\;\mu{\rm m}$ computed live during the {\sc
FLASH} simulation by the {\sc OpticalDepth} module with $\theta_\mathrm{lim} =
0.5$ and $N_{_{PIX}} = 48$; {\bf top right:} logarithm of the optical depth,
$\tau_\mathrm{R}$, at the same wavelength computed using {\sc RADMC-3D} with 200
million photon packages; {\bf bottom left:} gas temperature in the {\sc Flash}
simulation resulting from the radiative heating taking the optical depth as an
input; and {\bf bottom right:} relative difference between the two optical
depths, $(\tau_\mathrm{F} - \tau_\mathrm{R})/\tau_\mathrm{F}$. The difference is
always $< 10$\% and typically on the level of a few per cent, where most of it
is caused by the noise in the {\sc RADMC-3D} data. }
\label{fig:ODtest}
\end{figure}

Here we compare the spatial distribution of the optical depth,
$\tau_\mathrm{F}$, calculated by the {\sc OpticalDepth} module to the optical
depth, $\tau_\mathrm{R}$, computed using the {\sc RADMC-3D} code\footnote{Note
that the index F in $\tau_{\rm F}$ refers to {\sc FLASH}, i.e. calculation by
the {\sc OpticalDepth} module, and R in $\tau_{\rm R}$ refers to {\sc
RADMC-3D}.}. We use a snapshot at $t = 2$\,Myr from a turbulent sphere
simulation similar to the one discussed in \S\ref{sec:turbsphere} and
\S\ref{sec:ODtest:cd}, but calculated on a uniform grid $128^3$ to make the
{\sc RADMC-3D} calculation feasible. We use $N_{_{\rm PIX}}= 48$ pixels and a
geometric MAC with $\theta_{\rm lim}=0.5$ (see \S\ref{sec:treecol} for details
on the {\sc OpticalDepth} module). Here we assume a constant dust-to-gas ratio
of $0.01$. We select a UV wavelength, $\lambda_0=9.36\times 10^{-2}\;
\mu{\rm m}$, because scattering effects in the UV are minimal, and we can easily
relate the dust column density to the optical depth using the dust opacity at
this wavelength, $\kappa_{{\rm abs}}(\lambda)$. This approach neglects possible
variations of $\kappa_{{\rm abs}}(\lambda)$ along the line of sight, e.g. due to
temperature variations or changes in the dust properties. Using a typical Milky
Way dust opacity provided by \citet{Weingartner2001} (table for
\texttt{MW\_R\_V\_4.0}), we have $\kappa_{{\rm abs}}(\lambda_0) = 6.555 \times
10^{4}\;{\rm cm}^2{\rm g}^{-1}$. We obtain \begin{equation} \tau_{\rm F} =
\kappa_{{\rm abs}}(\lambda_0) \times N_{\rm dust, F}. \end{equation}

Using the dust density field and dust temperature provided by the simulation, we
compute the optical depth at the same wavelength using the {\sc RADMC-3D}
code, $\tau_{\rm R}$. With {\sc RADMC-3D} it is possible to provide an
external radiation field, in which case the photon packages are launched from
the borders of the computational domain and pass through the grid in random
directions. In each cell, they interact with the present dust according to its
opacity. We use the same dust opacity table for {\sc RADMC-3D} as described
above. For the incoming radiation, we use the intensities of a typical ISRF as
provided by \citep{Evans2001}. The incoming intensity at wavelength $\lambda_0$
is $I_0 = 9.547 \times 10^{-21}\;{\rm erg\;s}^{-1}{\rm cm}^{-2}{\rm Hz}^{-1}{\rm
sr}^{-1}$. We run {\sc RADMC-3D} in the mode \texttt{mcmono} to compute the
intensity field at $\lambda_0$ in every cell of the computational domain. Then
we convert this intensity, $I_{\rm cell}$, to $\tau_{\rm R}$ using
\begin{equation}
\tau_{\rm R} = {\rm ln}\left(\frac{I_0}{I_{\rm cell}} \right).
\end{equation}
It is necessary to use a large number of photon packages in order to reduce the
noise in the {\sc RADMC-3D} calculation to an acceptable level. Specifically,
we use 200 million photon packages and therefore it takes $\sim 53$ minutes on
one 10-core Intel-Xeon E5-2650 CPU to simulate one wavelength on the given
uniform grid with $128^3$ resolution, while the calculation with the {\sc
OpticalDepth} module took $24$~seconds on 4-core Intel i7-2600, i.e. it was
$\sim 330\times$ faster when normalising both calculations by number of cores.

In Figure~\ref{fig:ODtest} we show a slice at $z=0$ of the resulting optical
depths (shown in logarithmic scale), $\tau_{\rm F}$ from the {\sc Flash}
calculation (top left panel) and $\tau_{\rm R}$ from the {\sc RADMC-3D}
calculation (top right panel), as well as the difference between the two,
normalized to the maximum $\tau_{\rm F,max}$, of the {\sc Flash} optical depth
in the $xy$-plane (bottom right panel). The resulting gas temperature calculated
by {\sc Flash} in displayed in the bottom left panel. The overall
agreement is very good and on the level of the remaining noise of the {\sc
RADMC-3D} calculation of a few per cent. Although there is a tendency for the
{\sc OpticalDepth} module to slightly overestimate the optical depth in the
densest regions, the difference is always $<10$\% and is $\sim 1$ \% for most
cells in the computational domain. The result improves slightly if we use
$N_{_{\rm PIX}}= 192$ and $\theta_{\rm lim}=0.25$, but the additional expense of
the calculation is generally not worth the effort.


\subsection{Comparison of various MACs}
\label{sec:macs}


\begin{figure}
\includegraphics[width=\columnwidth]{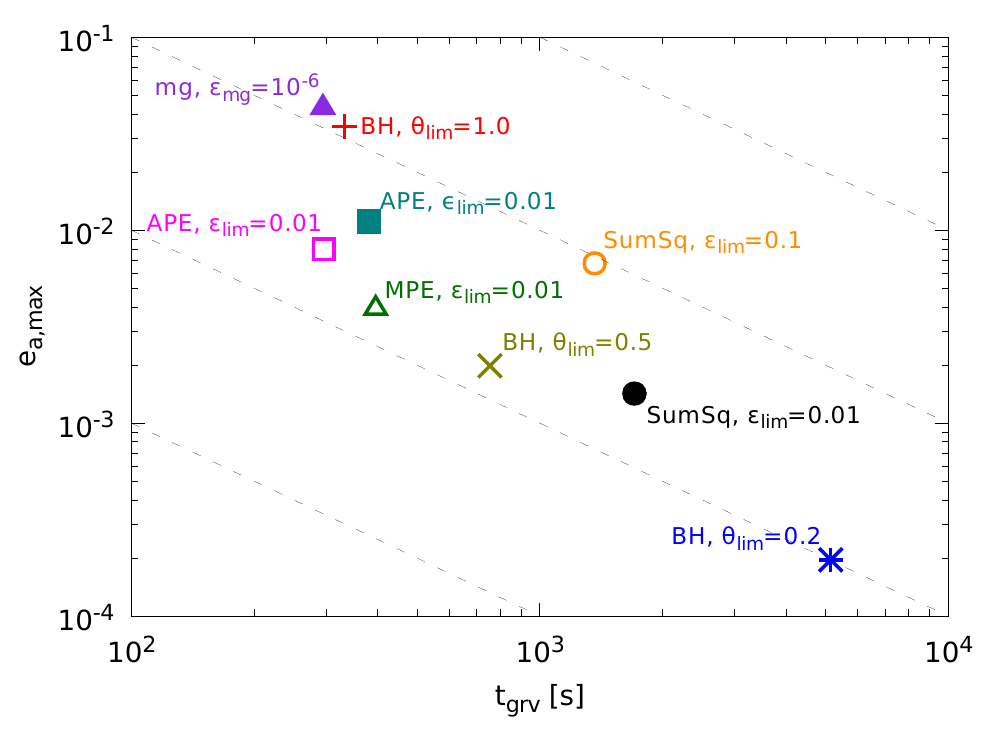}
\caption{Comparison of available MACs for a single time-step calculation of the
Bonnor-Ebert sphere on a uniform grid $128^3$. Each calculation with a different
MAC is plotted in the plane of the gravity calculation duration in seconds
(x-axis) versus the maximum relative error in the gravitational acceleration,
$e_{a,\mathrm{max}}$ (y-axis). The maximum is taken over the whole computational
domain. The tested MACs are: three geometric (BH) MACs with fixed maximum
opening angles $\theta_\mathrm{lim} = 1.0$ (red plus), $\theta_\mathrm{lim} =
0.5$ (olive x), $\theta_\mathrm{lim} = 0.2$ (blue star); two APE MACS with
absolute error limit $a_\mathrm{lim} = \varepsilon_\mathrm{lim}a_\mathrm{max} =
0.01a_\mathrm{max}$ (magenta empty square) and relative error limit
$\epsilon_\mathrm{lim} = 0.01$ (dark cyan filled square); an MPE MAC with
absolute error limit given by $\varepsilon = 0.01$ (dark green empty triangle);
and two SumSquare MACs with absolute error limits $\varepsilon_\mathrm{lim} =
0.01$ (black filled circle) and $\varepsilon_\mathrm{lim} = 0.1$ (orange empty
circle). The violet filled triangle shows the calculation by the multi-grid
solver. The thin dashed lines are iso-lines of constant $t_\mathrm{grv}\times
e_{a\mathrm{max}}$ assessing the code efficiency. 
}
\label{fig:perf:acc}
\end{figure}

We compare all available MACs with their typical parameters for a simple
calculation similar to the static Bonnor-Ebert sphere test described in
\S\ref{sec:bes}, however, carried out on a uniform $128^3$ grid. The aim is to
provide an approximate measure of the code behaviour. A rigorous analysis
of the efficiency of individual MACs, which would need many more tests, 
since it is highly problem dependent, is beyond the scope of this paper. The
time of the gravity calculation, $t_\mathrm{grv}$, was measured on a single
processor core, and since it is a single time-step calculation, the time is
meaningful only for a mutual comparison between individual MACs (as is also the
case for all the static tests in \S\ref{sec:static}). For each
calculation, we determine the relative error in the gravitational acceleration
and find its maximum in the computational domain, $e_{a,\mathrm{max}}$. The
tested MACs are: the geometric (BH) MAC with three maximum opening angles,
$\theta_\mathrm{lim}$, the APE MAC with both absolute and relative error limit,
the MPE MAC, and the experimental SumSquare MAC with two different error limits.
The results are shown in Figure~\ref{fig:perf:acc} which plots the runs in the
$t_\mathrm{grv} - e_{a,\mathrm{max}}$ plane.

In general, the results show an anti-correlation between the computational time
$t_\mathrm{grv}$ and the error $e_{a,\mathrm{max}}$ resulting from the expected
trade-off between computational costs and accuracy. One way how to estimate the
efficiency of the tested MACs is to consider lines of constant
$t_\mathrm{grv}\times e_{a,\mathrm{max}}$. Then, we find that the three most
efficient among the tested MACs are the BH MAC with $\theta_\mathrm{lim} = 0.2$,
BH MAC with $\theta_\mathrm{lim} = 0.5$, and the MPE MAC with
$\varepsilon_\mathrm{lim} = 0.01$, the first one being the slowest and most
accurate, the last one being the fastest of the three. The APE MAC with
$\varepsilon_\mathrm{lim} = 0.01$ is also amongst the most efficient ones while
its relative error is smaller than $e_{a,\mathrm{max}} = 10^{-2}$. Such an
accuracy is generally acceptable and therefore we consider this MAC to be
an optimal choice. Of course, we note that the final decision about the required
accuracy is highly problem dependent and must be made by the user on the basis
of the knowledge of the physical configuration that is being treated.

The comparison between the two APE MACs, one using the
absolute error limit $\varepsilon_\mathrm{lim} = 0.01$, and the second one using
the relative error limit $\epsilon_\mathrm{lim} = 0.01$, shows an interesting, yet not dramatic, difference:
the APE with absolute error limit seems to be more efficient by being
both faster and more accurate. This result seems to support claims by SW94 that setting
the absolute error limit is more appropriate, even though it requires more
effort by the user.

The two SumSquare MACs are not among the most efficient, however, they provide
an additional advantage of guaranteeing that the error will not exceed the
pre-set accuracy limit. It also seems that increasing $\varepsilon_\mathrm{lim}$
to values as high as $0.1$ and above does not result in substantially lower
$t_\mathrm{grv}$.

The multi-grid solver is among the fastest calculations and also among the least
accurate. However, the error is high only in the vicinity of the computational
domain boundaries caused by an inaccurate multipole solver used to calculate
boundary values of the gravitational potential ($m_\mathrm{mp} = 10$). In practice, the high accuracy
is often not needed close to the boundaries and if the region of size $\sim 20$\,\%
around boundaries is excluded from the error calculation, the error
$e_{a,\mathrm{max}}$ drops by approximately one order of magnitude. Then, the
multi-grid solver is comparable to the most efficient and fast APE and MPE MACs.


\subsection{Scaling tests}
\label{sec:scaling}


\begin{figure}
\includegraphics[width=\columnwidth]{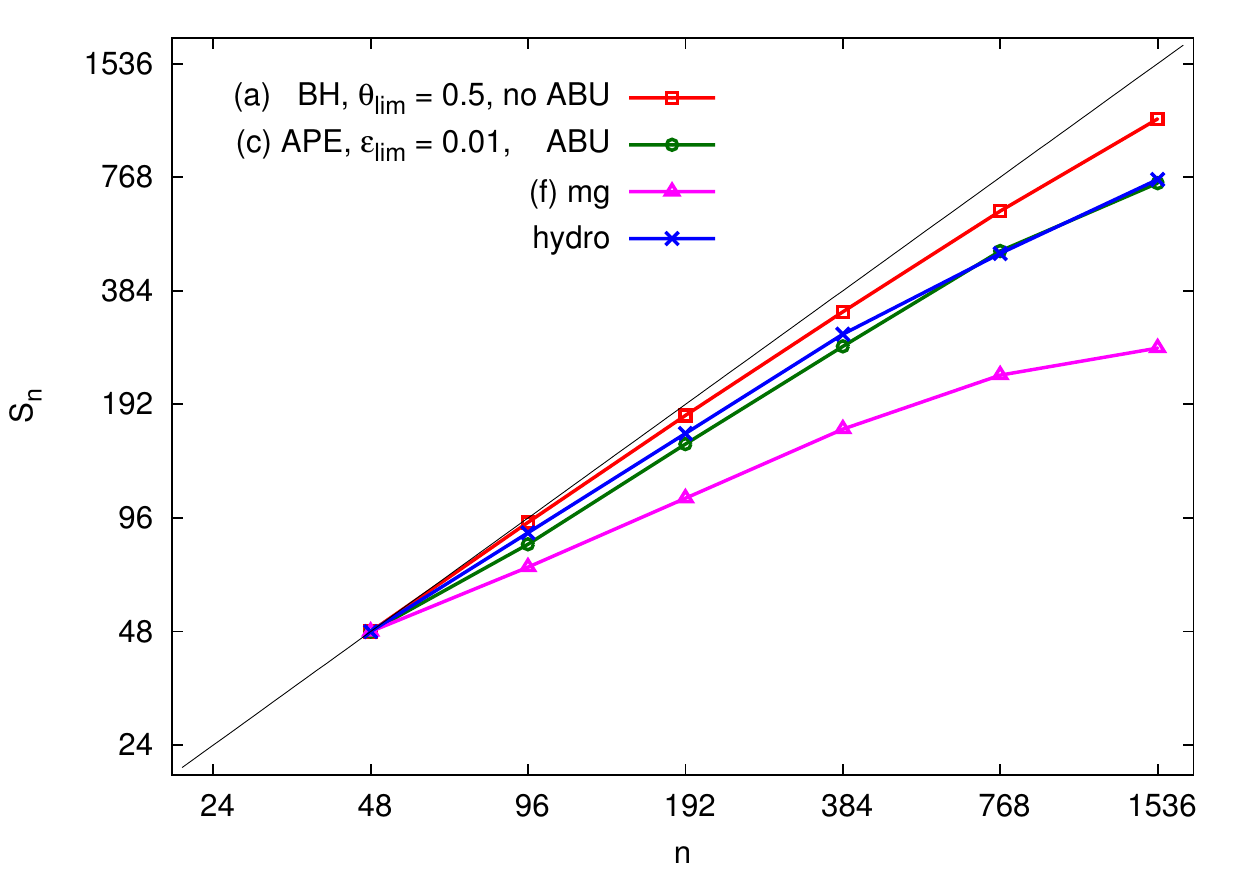}
\caption{Strong scaling test. Speed-up as a function of the number of processor
cores measured for the turbulent sphere test (see \S\ref{sec:turbsphere})
running for 10 time-steps. It compares scaling of the tree-solver with the BH
MAC, $\theta_\mathrm{lim} = 0.5$ with ABU switched off (red squares), the
tree-solver with the APE MAC, $\varepsilon_\mathrm{lim} = 0.01$ and ABU switched
on (green circles), the multi-grid solver with $\varepsilon_\mathrm{mg,lim} =
10^{-6}$ and $m_\mathrm{mp} = 15$ (magenta triangles) and the PPM hydrodynamic
solver measured at the test with the BH MAC tree-solver (blue crosses). The
solid black line shows the (ideal) linear scaling $S_n \sim n$.}
\label{fig:strong}
\end{figure}

\begin{figure}
\includegraphics[width=\columnwidth]{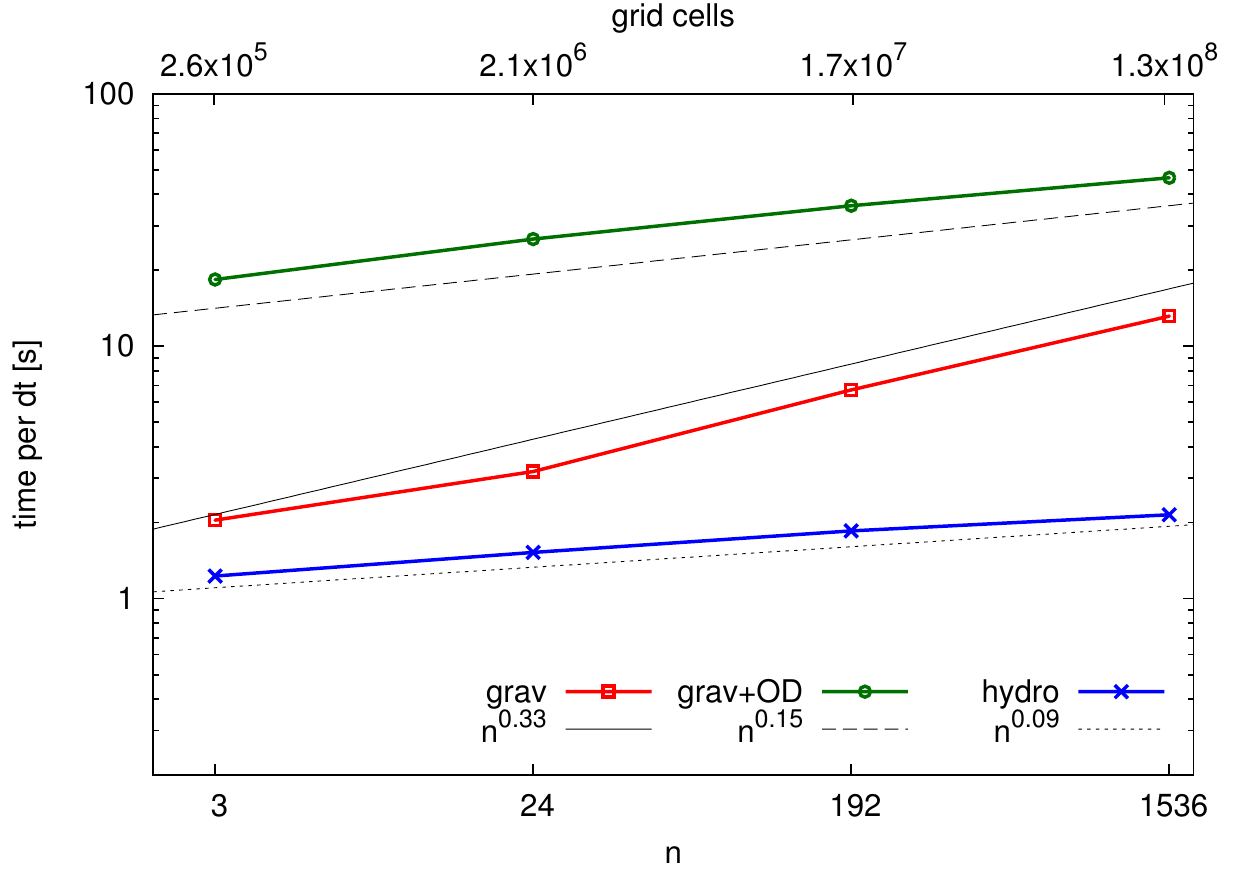}
\caption{Weak scaling test. Time per time-step as a function of number of
cores, $n$, for the setup where the size of the problem (number of blocks or
grid cells -- see top x-axis) is proportional to $n$. Measurements were done for
the turbulent sphere test (see \S\ref{sec:turbsphere} and
\S\ref{sec:ODtest:radmc}) running for 10 time-steps, the grid was uniform using
the following resolutions: $64^3$ running on $3$ cores, $128^3$ on $24$ cores,
$256^3$ on $192$ cores, and $512^3$ on $1536$ cores. The red curve shows times
for the tree solver with the gravity module only using the APE MAC with
$\varepsilon_\mathrm{lim} = 0.01$ and ABU switched on. The green curve shows
scaling for the tree solver using both the gravity and OpticalDepth modules and
BH MAC with $\theta_\mathrm{lim} = 0.5$ with ABU switched off. The blue curve
shows the weak scaling of the Flash internal hydrodynamic module on the given
problem. The black solid, dashed and dotted lines show power-laws with indices
$0.33$, $0.15$ and $0.09$, respectively.
}
\label{fig:weak}
\end{figure}

We perform both strong scaling and weak scaling tests. For that we use the
setup of the turbulent sphere from \S\ref{sec:turbsphere} and
\S\ref{sec:ODtest:radmc}. The strong scaling tests are done for the {\sc
Gravity} module only, the weak scaling is done for both {\sc Gravity} and {\sc
OpticalDepth} modules. 

For the strong scaling tests, we use three code configurations: tree
solver with BH MAC and ABU off (model (a) from Table~\ref{tab:dyn:tsph}), tree
solver with APE MAC and ABU on (model (c)), and multi-grid solver with default
parameters (model (f)). We also show the scaling of the {\sc FLASH} PPM
hydrodynamic solver with default parameters. All tests have been run for 10
time-steps on the IT4I/Salomon supercomputer using $48 - 1536$ cores. The
speed-up on $n$ processor cores, $S_n$ is determined with respect to the run with
$48$ cores
\begin{equation}
S_n = \frac{t_{48}}{t_{n}}
\end{equation}
where $t_{48}$ is the time spent by the evaluated module on $48$ cores and $t_n$
is the time spent by the same module on $n$ cores.

We see that the run with the tree-solver and BH MAC gives the best behaviour
(speed-up closest to linear). On the other hand, this model is also the slowest
one out of the three on $96$ cores (see Table~\ref{tab:dyn:tsph}{ and
Figure~\ref{fig:strong}}). This can be understood by noting that most of
the computational time is spent in the tree-walk, which runs completely in
parallel without any communication. Additionally, without ABU there is no
problem with load balancing, because the computational time is more or less
directly proportional to the number of leaf-blocks, and thus each core receives
the same number of leaf-blocks. The run with the APE MAC and ABU exhibits
slightly worse scaling, however, the test in \S\ref{sec:turbsphere} shows that
on $96$ cores, it is almost $3$ times faster than the BH MAC run. This is
partially because, due to its more efficient MAC the code spends less time
in fully parallel parts and partially because the ABU does not save the time
equally on each processor core. The APE MAC scaling is still very good,
comparable to the scaling of the hydrodynamic solver, which is highly parallel
and needs only to communicate information at the boundaries between domains
belonging to different processor cores. The multi-grid solver is very fast on
$96$ cores, comparable to the tree-solver with APE MAC and ABU, however, its
efficiency decreases on higher number of cores.

The weak scaling test have been done for two configurations: (i) the tree solver
with the gravity only runs using the APE MAC, $\varepsilon_\mathrm{lim} = 0.01$
and ABU switched on, and (ii) runs calculating the gravitational acceleration
and column densities of the three components (total, H$_2$ and CO; see
\S\ref{sec:treecol}) using the BH MAC, $\theta_\mathrm{lim} = 0.5$ with ABU
switched off. Each configuration is run for four different grid resolution
ranging from $64^3$ to $512^3$, with the number of cores, $n$, proportional to
the number of grid cells ($n = 3, 24, 192, 1536$).

Results of the weak scaling tests are shown in Figure~\ref{fig:weak} where a
single time-step runtimes of the two configurations are compared with each other
and with runtimes of the {\sc Flas}h internal hydrodynamic solver. The hydrodynamic
solver times (blue curve) follow approximately the $n^{0.09}$ power-law, which
is slightly worse than the "ideal" constant scaling. The tree solver using both
the {\sc Gravity} and {\sc OpticalDepth} modules (green curve) exhibits a
similar $n^{0.15}$ scaling. On the other hand, runs with only the {\sc Gravity}
module (red curve) show the $n^{0.15}$ scaling only between $3$ and $24$ cores,
and for higher number of cores the scaling gets worse approaching $n^{0.33}$.
This is due to two reasons. Firstly, the gravity only runs are cheaper and the
communication making the scaling worse is relatively more important. Note that
the communication is negligible for runs on up to $24$ cores, since a node on the
Solomon computer comprises $24$ cores with shared memory. Secondly, the load
balancing needed by the ABU becomes worse on a high number of cores. We can
also see that the additional calculations of column densities in the {\sc
OpticalDepth} module make the code approximately $10$ times slower than the
calculation of the gravity on a small number of cores, but it becomes only
a factor of $\sim 3.5$ for $1536$ cores due to better scaling of the more
expensive runs with the {\sc OpticalDepth} module.


\section{Summary}
\label{sec:summary}


We have developed an MPI parallel, general purpose tree solver for the adaptive
mesh refinement hydrodynamic code {\sc FLASH}, that can be used to calculate the
gas gravitational acceleration (or potential), optical depths enabling inclusion
of the ambient diffuse radiation, and in future also general radiation transport
(Paper~II). The code uses an efficient communication strategy predicting which
parts of the tree need to be sent to different processors allowing the whole
tree walk to be executed locally. The advantage of this approach is a relatively
low memory requirement, important in particular for the optical depth
calculation, which needs to process information from different directions. This
also makes the implementation of the general tree-based radiation transport
possible. In addition to commonly implemented, fully isolated and fully periodic
boundary conditions, the code can handle mixed (i.e. isolated in some directions
and periodic in others) boundary conditions using a newly developed
generalisation of the Ewald method. The gravity module implements several
Multipole Acceptance Criteria (MACs) that increase the code efficiency by
selecting which tree-nodes are acceptable for the calculation on the basis of
the mass distribution within them. Using the Adaptive Block Update (ABU)
technique, the code is able to re-use information from the previous time-step
and thus further save computational time.

We have run a series of tests evaluating the code accuracy and
performance, and compared them to the in-built multi-grid gravity solver of {\sc
FLASH}. The simpler "static" tests of the gravity module show that the code
provides good accuracy for all combinations of boundary conditions. Comparison
with the {\sc FLASH} default multi-grid solver suggests that the tree-solver
provides better accuracy for the same computational costs in the case of fully
isolated boundary conditions, while with fully periodic boundary conditions the
multi-grid solver seems to be more efficient.

Further, we run two more complex dynamical tests. The Evrard test (gravitational
collapse and re-bounce of the adiabatic, initially cold gaseous sphere) shows
that it is critical to resolve well the dense centre, in order to ensure energy
conservation. We found that in order to limit the error in the total energy to
less than a few percent, it is necessary to resolve the Jeans length with at
least $4$ grid cells, a result similar to that of \citet{1997ApJ...489L.179T}
where the same resolution is needed to prevent artificial fragmentation. In
general, the Evrard test turns out to be harder for grid codes in comparison
with Lagrangian (e.g. SPH) hydrodynamic codes which reach almost perfect energy
conservation with very small numbers of particles.

In the second dynamical test we simulate a turbulent sphere which collapses,
fragments and forms sink particles (representing newly formed stars). We find
the the tree-solver performs well and runs with accuracy of order several
percent if it calculates the accelerations directly, and if it is used with the
BH or APE MAC with typical parameters. Calculating the gravitational potential
instead of the acceleration results in high (up to $80$\,\%) errors due to
numerical discretisation, and may result in numerical artifacts. The Adaptive
Block Update accelerates the calculation by a factor of several ($\sim 2$) 
for a given test. The multi-grid solver exhibits relatively high error ($\sim
20$\,\%) close to computational domain boundaries, resulting from an inaccurate
multipole solver. If the boundary regions are excluded, the accuracy of the
tree-solver and multi-grid solver are comparable, while the tree-solver is
approximately $2$ times faster for the given test.

We run two tests of the {\sc OpticalDepth} module. In the first one we measure
the direction-dependent optical depths as a function of the angular resolution,
and we find (in agreement with \citep*{Clark2012a}) that the code runs most
efficiently if the angular resolution given by the number of {\sc Healpix} rays
is similar to the opening angle used by the tree-solver. In the second test we
compare the optical depth calculated with the tree-solver with those calculated
with the accurate radiation transport code RADMC-3D and we find an excellent
agreement even for relatively low angular resolution -- 48 {\sc Healpix} rays.

Further, using a simplified turbulent sphere test with uniform resolution, we
compare the efficiency of all available MACs with their typical parameters.
Generally, the BH MAC provides better accuracy for higher computational costs,
while APE and MPE MACs result in lower (but often still acceptable) accuracy and
are substantially faster. For applications, where an accuracy of order
$10^{-2}$ is sufficient, the fastest choice seems to be the APE MAC with the
absolute limit on the error.

Finally, we run strong scaling tests and show that the code scales up very well
up to at least 1536 processor cores. We conclude that the presented tree-solver
is a viable method for calculating self-gravity and other processes in
astrophysics and that it is competitive with more commonly used iterative
multi-grid methods.


\section*{Acknowledgments}%

This study has been supported by project 15-06012S of the Czech Science
Foundation and by the institutional project RVO:67985815. SW acknowledges the
Deutsche Forschungsgemeinschaft (DFG) for funding through the SPP 1573 ``The
Physics of the Interstellar Medium'', the Bonn-Cologne-Graduate School, the SFB
956 "The conditions and impact of star formation", and the funding from the
European Research Council under the European Community's Framework Programme FP8
via the ERC Starting Grant RADFEEDBACK (project number 679852). APW gratefully
acknowledges the support of a consolidated grant (ST/K00926/1) from the UK
Science and Technology Facilities Council. This work was supported by The
Ministry of Education, Youth and Sports from the Large Infrastructures for
Research, Experimental Development and Innovations project "IT4Innovations
National Supercomputing Center – LM2015070". The software used in this work was
in part developed by the DOE NNSA-ASC OASCR Flash Center at the University of
Chicago.

\bibliographystyle{mn2e}
\bibliography{references}%

\begin{thebibliography}{}

\bibitem[\protect\citeauthoryear{{Ahn} \& {Lee}}{{Ahn} \&
  {Lee}}{2008}]{Ahn2008}
{Ahn} C.-O.,  {Lee} S.~H.,  2008, Computer Physics Communications, 178, 121

\bibitem[\protect\citeauthoryear{{Anderson}}{{Anderson}}{1999}]{Anderson1999}
{Anderson} R.~J.,  1999, SIAM J. Comput., 28, 1923

\bibitem[\protect\citeauthoryear{{Bagla}}{{Bagla}}{2002}]{Bagla2002}
{Bagla} J.~S.,  2002, Journal of Astrophysics and Astronomy, 23, 185

\bibitem[\protect\citeauthoryear{{Bagla} \& {Khandai}}{{Bagla} \&
  {Khandai}}{2009}]{Bagla2009}
{Bagla} J.~S.,  {Khandai} N.,  2009, \mnras, 396, 2211

\bibitem[\protect\citeauthoryear{{Barnes} \& {Hut}}{{Barnes} \&
  {Hut}}{1986}]{Barnes1986}
{Barnes} J.,  {Hut} P.,  1986, \nat, 324, 446

\bibitem[\protect\citeauthoryear{{Barnes}}{{Barnes}}{1990}]{Barnes1990}
{Barnes} J.~E.,  1990, Journal of Computational Physics, 87, 161

\bibitem[\protect\citeauthoryear{{Becciani}, {Antonuccio-Delogu} \&
  {Comparato}}{{Becciani} et~al.}{2007}]{2007CoPhC.176..211B}
{Becciani} U.,  {Antonuccio-Delogu} V.,    {Comparato} M.,  2007, Computer
  Physics Communications, 176, 211

\bibitem[\protect\citeauthoryear{{Bentley}}{{Bentley}}{1979}]{Bentley1979}
{Bentley} J.~L.,  1979, IEEE Trans. Softw. Eng., 4, 333

\bibitem[\protect\citeauthoryear{{Benz}, {Cameron}, {Press} \& {Bowers}}{{Benz}
  et~al.}{1990}]{Benz1990}
{Benz} W.,  {Cameron} A.~G.~W.,  {Press} W.~H.,    {Bowers} R.~L.,  1990, \apj,
  348, 647

\bibitem[\protect\citeauthoryear{{Bode} \& {Ostriker}}{{Bode} \&
  {Ostriker}}{2003}]{BodeOstriker2003}
{Bode} P.,  {Ostriker} J.~P.,  2003, \apjs, 145, 1

\bibitem[\protect\citeauthoryear{{Bode}, {Ostriker} \& {Xu}}{{Bode}
  et~al.}{2000}]{Bode2000}
{Bode} P.,  {Ostriker} J.~P.,    {Xu} G.,  2000, \apjs, 128, 561

\bibitem[\protect\citeauthoryear{{Bonnor}}{{Bonnor}}{1956}]{Bonnor1956}
{Bonnor} W.~B.,  1956, \mnras, 116, 351

\bibitem[\protect\citeauthoryear{{Brown}, {Byrne} \& {Hindmarsh}}{{Brown}
  et~al.}{1989}]{Brown1989}
{Brown} P.~N.,  {Byrne} G.~D.,    {Hindmarsh} A.~C.,  1989, SIAM J. Sci. Stat.
  Comput., 10, 1038

\bibitem[\protect\citeauthoryear{{Clark} \& {Glover}}{{Clark} \&
  {Glover}}{2014}]{Clark2014}
{Clark} P.~C.,  {Glover} S.~C.~O.,  2014, \mnras, 444, 2396

\bibitem[\protect\citeauthoryear{{Clark}, {Glover} \& {Klessen}}{{Clark}
  et~al.}{2012}]{Clark2012a}
{Clark} P.~C.,  {Glover} S.~C.~O.,    {Klessen} R.~S.,  2012, \mnras, 420, 745

\bibitem[\protect\citeauthoryear{{Clark}, {Glover}, {Klessen} \&
  {Bonnell}}{{Clark} et~al.}{2012}]{Clark2012b}
{Clark} P.~C.,  {Glover} S.~C.~O.,  {Klessen} R.~S.,    {Bonnell} I.~A.,  2012,
  \mnras, 424, 2599

\bibitem[\protect\citeauthoryear{{Dehnen}}{{Dehnen}}{2000}]{Dehnen2000}
{Dehnen} W.,  2000, \apjl, 536, L39

\bibitem[\protect\citeauthoryear{{Dehnen}}{{Dehnen}}{2002}]{Dehnen2002}
{Dehnen} W.,  2002, Journal of Computational Physics, 179, 27

\bibitem[\protect\citeauthoryear{{Draine} \& {Bertoldi}}{{Draine} \&
  {Bertoldi}}{1996}]{Draine1996}
{Draine} B.~T.,  {Bertoldi} F.,  1996, \apj, 468, 269

\bibitem[\protect\citeauthoryear{{Dubinski}}{{Dubinski}}{1996}]{Dubinski1996}
{Dubinski} J.,  1996, \na, 1, 133

\bibitem[\protect\citeauthoryear{{Ebert}}{{Ebert}}{1955}]{Ebert1955}
{Ebert} R.,  1955, Zeitschrift fur Astrophysik, 37, 217

\bibitem[\protect\citeauthoryear{{Evans} II, {Rawlings}, {Shirley} \&
  {Mundy}}{{Evans} et~al.}{2001}]{Evans2001}
{Evans} II N.~J.,  {Rawlings} J.~M.~C.,  {Shirley} Y.~L.,    {Mundy} L.~G.,
  2001, \apj, 557, 193

\bibitem[\protect\citeauthoryear{{Evrard}}{{Evrard}}{1988}]{Evrard1988}
{Evrard} A.~E.,  1988, \mnras, 235, 911

\bibitem[\protect\citeauthoryear{{Ewald}}{{Ewald}}{1921}]{Ewald1921}
{Ewald} P.~P.,  1921, Annalen der Physik, 369, 253

\bibitem[\protect\citeauthoryear{{Federrath}, {Banerjee}, {Clark} \&
  {Klessen}}{{Federrath} et~al.}{2010}]{Federrath2010}
{Federrath} C.,  {Banerjee} R.,  {Clark} P.~C.,    {Klessen} R.~S.,  2010,
  \apj, 713, 269

\bibitem[\protect\citeauthoryear{{Fryxell}, {Olson}, {Ricker}, {Timmes},
  {Zingale}, {Lamb}, {MacNeice}, {Rosner}, {Truran} \& {Tufo}}{{Fryxell}
  et~al.}{2000}]{Fryxell2000}
{Fryxell} B.,  {Olson} K.,  {Ricker} P.,  {Timmes} F.~X.,  {Zingale} M.,
  {Lamb} D.~Q.,  {MacNeice} P.,  {Rosner} R.,  {Truran} J.~W.,    {Tufo} H.,
  2000, \apjs, 131, 273

\bibitem[\protect\citeauthoryear{{Gatto}, {Walch}, {Naab}, {Girichidis},
  {W{\"u}nsch}, {Glover}, {Klessen}, {Clark}, {Peters}, {Derigs}, {Baczynski}
  \& {Puls}}{{Gatto} et~al.}{2017}]{Gatto2017}
{Gatto} A.,  {Walch} S.,  {Naab} T.,  {Girichidis} P.,  {W{\"u}nsch} R.,
  {Glover} S.~C.~O.,  {Klessen} R.~S.,  {Clark} P.~C.,  {Peters} T.,  {Derigs}
  D.,  {Baczynski} C.,    {Puls} J.,  2017, \mnras, 466, 1903

\bibitem[\protect\citeauthoryear{{Girichidis}, {Walch}, {Naab}, {Gatto},
  {W{\"u}nsch}, {Glover}, {Klessen}, {Clark}, {Peters}, {Derigs} \&
  {Baczynski}}{{Girichidis} et~al.}{2016}]{Girichidis2016}
{Girichidis} P.,  {Walch} S.,  {Naab} T.,  {Gatto} A.,  {W{\"u}nsch} R.,
  {Glover} S.~C.~O.,  {Klessen} R.~S.,  {Clark} P.~C.,  {Peters} T.,  {Derigs}
  D.,    {Baczynski} C.,  2016, \mnras, 456, 3432

\bibitem[\protect\citeauthoryear{{Glover} \& {Clark}}{{Glover} \&
  {Clark}}{2012}]{GloverClark2012a}
{Glover} S.~C.~O.,  {Clark} P.~C.,  2012, \mnras, 421, 116

\bibitem[\protect\citeauthoryear{{Glover}, {Federrath}, {Mac Low} \&
  {Klessen}}{{Glover} et~al.}{2010}]{Glover2010}
{Glover} S.~C.~O.,  {Federrath} C.,  {Mac Low} M.-M.,    {Klessen} R.~S.,
  2010, \mnras, 404, 2

\bibitem[\protect\citeauthoryear{{Glover} \& {Mac Low}}{{Glover} \& {Mac
  Low}}{2007}]{Glover2007}
{Glover} S.~C.~O.,  {Mac Low} M.-M.,  2007, \apjs, 169, 239

\bibitem[\protect\citeauthoryear{{G{\'o}rski}, {Hivon}, {Banday}, {Wandelt},
  {Hansen}, {Reinecke} \& {Bartelmann}}{{G{\'o}rski} et~al.}{2005}]{Gorski2005}
{G{\'o}rski} K.~M.,  {Hivon} E.,  {Banday} A.~J.,  {Wandelt} B.~D.,  {Hansen}
  F.~K.,  {Reinecke} M.,    {Bartelmann} M.,  2005, \apj, 622, 759

\bibitem[\protect\citeauthoryear{{Hernquist}, {Bouchet} \& {Suto}}{{Hernquist}
  et~al.}{1991}]{Hernquist1991}
{Hernquist} L.,  {Bouchet} F.~R.,    {Suto} Y.,  1991, \apjs, 75, 231

\bibitem[\protect\citeauthoryear{{Hockney} \& {Eastwood}}{{Hockney} \&
  {Eastwood}}{1981}]{HockneyEastwood1981}
{Hockney} R.~W.,  {Eastwood} J.~W.,  1981, {Computer Simulation Using
  Particles}

\bibitem[\protect\citeauthoryear{{Hubber}, {Batty}, {McLeod} \&
  {Whitworth}}{{Hubber} et~al.}{2011}]{Hubber2011}
{Hubber} D.~A.,  {Batty} C.~P.,  {McLeod} A.,    {Whitworth} A.~P.,  2011,
  \aap, 529, A27

\bibitem[\protect\citeauthoryear{{Hubber}, {Rosotti} \& {Booth}}{{Hubber}
  et~al.}{2018}]{Hubber2018}
{Hubber} D.~A.,  {Rosotti} G.~P.,    {Booth} R.~A.,  2018, \mnras, 473, 1603

\bibitem[\protect\citeauthoryear{{Jeans}}{{Jeans}}{1902}]{Jeans1902}
{Jeans} J.~H.,  1902, Royal Society of London Philosophical Transactions Series
  A, 199, 1

\bibitem[\protect\citeauthoryear{{Jernigan} \& {Porter}}{{Jernigan} \&
  {Porter}}{1989}]{JerniganPorter1989}
{Jernigan} J.~G.,  {Porter} D.~H.,  1989, \apjs, 71, 871

\bibitem[\protect\citeauthoryear{{Khandai} \& {Bagla}}{{Khandai} \&
  {Bagla}}{2009}]{Khandai2009}
{Khandai} N.,  {Bagla} J.~S.,  2009, Research in Astronomy and Astrophysics, 9,
  861

\bibitem[\protect\citeauthoryear{{Kiessling}}{{Kiessling}}{1999}]{Kiessling1999}
{Kiessling} M.~K.~.,  1999, ArXiv Astrophysics e-prints

\bibitem[\protect\citeauthoryear{{Klessen}}{{Klessen}}{1997}]{Klessen1997}
{Klessen} R.,  1997, \mnras, 292, 11

\bibitem[\protect\citeauthoryear{{Lukat} \& {Banerjee}}{{Lukat} \&
  {Banerjee}}{2016}]{2016NewA...45...14L}
{Lukat} G.,  {Banerjee} R.,  2016, \na, 45, 14

\bibitem[\protect\citeauthoryear{{MacNeice}, {Olson}, {Mobarry}, {de
  Fainchtein} \& {Packer}}{{MacNeice} et~al.}{2000}]{MacNeice2000}
{MacNeice} P.,  {Olson} K.~M.,  {Mobarry} C.,  {de Fainchtein} R.,    {Packer}
  C.,  2000, Computer Physics Communications, 126, 330

\bibitem[\protect\citeauthoryear{{Makino}}{{Makino}}{1990}]{Makino1990}
{Makino} J.,  1990, Journal of Computational Physics, 88, 393

\bibitem[\protect\citeauthoryear{{Merlin}, {Buonomo}, {Grassi}, {Piovan} \&
  {Chiosi}}{{Merlin} et~al.}{2010}]{Merlin2010}
{Merlin} E.,  {Buonomo} U.,  {Grassi} T.,  {Piovan} L.,    {Chiosi} C.,  2010,
  \aap, 513, A36

\bibitem[\protect\citeauthoryear{{Nelson}, {Wetzstein} \& {Naab}}{{Nelson}
  et~al.}{2009}]{Nelson2009}
{Nelson} A.~F.,  {Wetzstein} M.,    {Naab} T.,  2009, \apjs, 184, 326

\bibitem[\protect\citeauthoryear{{Ostriker}}{{Ostriker}}{1964}]{Ostriker1964}
{Ostriker} J.,  1964, \apj, 140, 1056

\bibitem[\protect\citeauthoryear{{Peters}, {Naab}, {Walch}, {Glover},
  {Girichidis}, {Pellegrini}, {Klessen}, {W{\"u}nsch}, {Gatto} \&
  {Baczynski}}{{Peters} et~al.}{2017}]{Peters2017}
{Peters} T.,  {Naab} T.,  {Walch} S.,  {Glover} S.~C.~O.,  {Girichidis} P.,
  {Pellegrini} E.,  {Klessen} R.~S.,  {W{\"u}nsch} R.,  {Gatto} A.,
  {Baczynski} C.,  2017, \mnras, 466, 3293

\bibitem[\protect\citeauthoryear{{Potter}, {Stadel} \& {Teyssier}}{{Potter}
  et~al.}{2017}]{2017ComAC...4....2P}
{Potter} D.,  {Stadel} J.,    {Teyssier} R.,  2017, Computational Astrophysics
  and Cosmology, 4, \#2

\bibitem[\protect\citeauthoryear{{Press}}{{Press}}{1986}]{Press1986}
{Press} W.~H.,  1986, in {Hut} P.,  {McMillan} S.~L.~W.,  eds, The Use of
  Supercomputers in Stellar Dynamics Vol.~267 of Lecture Notes in Physics,
  Berlin Springer Verlag, {Techniques and Tricks for N-Body Computation}.
p.~184

\bibitem[\protect\citeauthoryear{{Ricker}}{{Ricker}}{2008}]{Ricker2008}
{Ricker} P.~M.,  2008, \apjs, 176, 293

\bibitem[\protect\citeauthoryear{{Salmon} \& {Warren}}{{Salmon} \&
  {Warren}}{1994}]{SalmonWarren1994}
{Salmon} J.~K.,  {Warren} M.~S.,  1994, Journal of Computational Physics, 111,
  136

\bibitem[\protect\citeauthoryear{{Smith}, {Glover} \& {Klessen}}{{Smith}
  et~al.}{2014}]{Smith2014}
{Smith} R.~J.,  {Glover} S.~C.~O.,    {Klessen} R.~S.,  2014, \mnras, 445, 2900

\bibitem[\protect\citeauthoryear{{Spitzer} Jr.}{{Spitzer}}{1942}]{Spitzer1942}
{Spitzer} Jr. L.,  1942, \apj, 95, 329

\bibitem[\protect\citeauthoryear{{Springel}}{{Springel}}{2005}]{Springel2005}
{Springel} V.,  2005, \mnras, 364, 1105

\bibitem[\protect\citeauthoryear{{Springel}}{{Springel}}{2010}]{Springel2010}
{Springel} V.,  2010, \mnras, 401, 791

\bibitem[\protect\citeauthoryear{{Springel}, {Yoshida} \& {White}}{{Springel}
  et~al.}{2001}]{Springel2001}
{Springel} V.,  {Yoshida} N.,    {White} S.~D.~M.,  2001, \na, 6, 79

\bibitem[\protect\citeauthoryear{{Stadel}}{{Stadel}}{2001}]{Stadel2001}
{Stadel} J.~G.,  2001, PhD thesis, UNIVERSITY OF WASHINGTON

\bibitem[\protect\citeauthoryear{{Truelove}, {Klein}, {McKee}, {Holliman} II,
  {Howell} \& {Greenough}}{{Truelove} et~al.}{1997}]{1997ApJ...489L.179T}
{Truelove} J.~K.,  {Klein} R.~I.,  {McKee} C.~F.,  {Holliman} II J.~H.,
  {Howell} L.~H.,    {Greenough} J.~A.,  1997, \apjl, 489, L179

\bibitem[\protect\citeauthoryear{{Wadsley}, {Stadel} \& {Quinn}}{{Wadsley}
  et~al.}{2004}]{Wadsley2004}
{Wadsley} J.~W.,  {Stadel} J.,    {Quinn} T.,  2004, \na, 9, 137

\bibitem[\protect\citeauthoryear{{Walch}, {Girichidis}, {Naab}, {Gatto},
  {Glover}, {W{\"u}nsch}, {Klessen}, {Clark}, {Peters}, {Derigs} \&
  {Baczynski}}{{Walch} et~al.}{2015}]{Walch2015}
{Walch} S.,  {Girichidis} P.,  {Naab} T.,  {Gatto} A.,  {Glover} S.~C.~O.,
  {W{\"u}nsch} R.,  {Klessen} R.~S.,  {Clark} P.~C.,  {Peters} T.,  {Derigs}
  D.,    {Baczynski} C.,  2015, \mnras, 454, 238

\bibitem[\protect\citeauthoryear{{Waltz}, {Page}, {Milder}, {Wallin} \&
  {Antunes}}{{Waltz} et~al.}{2002}]{Waltz2002}
{Waltz} J.,  {Page} G.~L.,  {Milder} S.~D.,  {Wallin} J.,    {Antunes} A.,
  2002, Journal of Computational Physics, 178, 1

\bibitem[\protect\citeauthoryear{{Weingartner} \& {Draine}}{{Weingartner} \&
  {Draine}}{2001}]{Weingartner2001}
{Weingartner} J.~C.,  {Draine} B.~T.,  2001, \apj, 548, 296

\bibitem[\protect\citeauthoryear{{Wetzstein}, {Nelson}, {Naab} \&
  {Burkert}}{{Wetzstein} et~al.}{2009}]{Wetzstein2009}
{Wetzstein} M.,  {Nelson} A.~F.,  {Naab} T.,    {Burkert} A.,  2009, \apjs,
  184, 298

\bibitem[\protect\citeauthoryear{{Xu}}{{Xu}}{1995}]{Xu1995}
{Xu} G.,  1995, \apjs, 98, 355

\end{thebibliography}
\clearpage

\appendix

\section{Equations for acceleration in computational domains with periodic and mixed boundary conditions}
\label{ap:accEwald}

In this appendix, we provide formulae for acceleration in computational domains with periodic and mixed 
boundary conditions. 
These formulae might be interesting particularly for the reader who intends to implement the Ewald method 
or its modification to a computational domain with mixed BCs
\footnote{The formulae are organised so as to avoid problems with floating point representation.}.
The orientation of symmetric axes is the same as in Section \ref{sec:gravity}.

In analog to the equation for potential (\ref{eq:evfld}), we write acceleration $\mathbf{a} (\mathbf{r})$ 
at target point $\mathbf{r}$ as
\begin{equation}
\mathbf{a} (\mathbf{r}) = - G \sum_{a=1}^N m_a \Aew (\mathbf{r} - \mathbf{r_a}),
\label{accdefine}
\end{equation}
where 
\begin{equation}
\Aew = - \nabla \phi .
\label{aewdefine}
\end{equation}

\subsection{Periodic boundary conditions}

Defining
\begin{eqnarray}
\eit & = & \frac{\exp{(-\zeta (l_1^2 + (l_2/b)^2 + (l_3/c)^2))}}{l_1^2 + (l_2/b)^2  + (l_3/c)^2}, \label{a2subst3d0} \\
\uit & = & (x-x_a-i_1 L_x)^2 + (y-y_a-i_2 b L_x)^2 \nonumber \\ 
&& + (z-z_a - i_3 c L_x)^2, \label{a2subst3d1} \\
\vlt & = & \frac{2 \pi l_1 (x - x_a)}{L_x} + \frac{2 \pi l_2 (y - y_a)}{b L_x} + \frac{2 \pi l_3 (z - z_a)}{c L_x}, \label{a2subst3d2}
\label{a2subst3d}
\end{eqnarray}
one obtains by differencing Equation~(\ref{ewpotential}) the components of function $\mathbf{A}$ in the form of
\begin{eqnarray}
\Aewa{x} = \sum_{\substack{i_1,i_2,i_3 \\ i_1^2+(b i_2)^2 + (c i_3)^2 \leq 10}} 
\Bigg\{ \Bigg( \frac{2 \alpha}{\sqrt{\pi}} \frac{\exp(-\alpha^2 \uit)}{\uit} + 
\frac{ \erfc(\alpha \sqrt{\uit})}{\uit^{3/2}} \Bigg) \times \nonumber \\ 
(x-x_a-i_1 L_x) \Bigg\} 
 +\frac{2}{bc L_x^2} \sum_{\substack{l_1,l_2,l_3 \\ l_1^2+(l_2/b)^2 + (l_3/c)^2 \leq 10}} 
l_1 \, \eit \sin{(\vlt)},
\label{a2accx3p}
\end{eqnarray}
\begin{eqnarray}
\Aewa{y} = \sum_{\substack{i_1,i_2,i_3 \\ i_1^2+(b i_2)^2 + (c i_3)^2 \leq 10}} 
\Bigg\{ \Bigg( \frac{2 \alpha}{\sqrt{\pi}} \frac{\exp(-\alpha^2 \uit)}{\uit} + 
\frac{ \erfc(\alpha \sqrt{\uit})}{\uit^{3/2}} \Bigg) \times \nonumber \\ 
(y-y_a-i_2 b L_x) \Bigg\} 
 +\frac{2}{b^2 c L_x^2} \sum_{\substack{l_1,l_2,l_3 \\ l_1^2+(l_2/b)^2 + (l_3/c)^2 \leq 10}} 
l_2 \, \eit \sin{(\vlt)},
\label{a2accy3p}
\end{eqnarray}
\begin{eqnarray}
\Aewa{x} = \sum_{\substack{i_1,i_2,i_3 \\ i_1^2+(b i_2)^2 + (c i_3)^2 \leq 10}} 
\Bigg\{ \Bigg( \frac{2 \alpha}{\sqrt{\pi}} \frac{\exp(-\alpha^2 \uit)}{\uit} + 
\frac{ \erfc(\alpha \sqrt{\uit})}{\uit^{3/2}} \Bigg) \times \nonumber \\ 
(z-z_a-i_3 c L_x) \Bigg\} 
 +\frac{2}{b c^2 L_x^2} \sum_{\substack{l_1,l_2,l_3 \\ l_1^2+(l_2/b)^2 + (l_3/c)^2 \leq 10}} 
l_3 \,\eit \sin{(\vlt)}.
\label{a2accz3p}
\end{eqnarray}

\subsection{Mixed boundary conditions of type 2P1I}

To simplify the formulae below, we define
\begin{eqnarray}
\uid & = & (x-x_a-i_1 L_x)^2 + (y-y_a-i_2 b L_x)^2 + (z - z_a)^2, \label{a2subst2p1i1} \\
\vld & = & \frac{2 \pi l_1 (x - x_a)}{L_x} + \frac{2 \pi l_2 (y - y_a)}{b L_x}, \label{a2subst2p1id2}
\end{eqnarray}
and
\begin{eqnarray}
\widetilde{I}(l_1,l_2,z-z_a) & \equiv & I(l_1,l_2,z-z_a) \exp{(-\zeta (l_1^2 + (l_2/b)^2))}  \nonumber \\
& =&   \frac{\pi}{2\sqrtij} \Bigg\{\exp(-\frac{\gamma^2}{4 \zeta}) \exp(-\zeta (l_1^2 +(l_2/b)^2)) \nonumber \\
&& \times \erfcx\left( \frac{\zeta \sqrtij + \gamma/2}{\sqrt{\zeta}}  \right) \nonumber \\
&& + \exp(-\gamma \sqrtij) \nonumber\\
&& \times \erfc \left(  \frac{\zeta \sqrtij - \gamma/2}{\sqrt{\zeta}}    \right)\Bigg\}, \\
I'(l_1,l_2,z-z_a)  & \equiv &  \mathrm{d} \widetilde{I}(l_1,l_2, \gamma) / \mathrm{d} \gamma \nonumber \\
& = &  \frac{\pi}{2} \Bigg\{\exp(-\frac{\gamma^2}{4 \zeta}) \exp(-\zeta (l_1^2 +(l_2/b)^2)) \nonumber \\
&& \times \erfcx\left( \frac{\zeta \sqrtij + \gamma/2}{\sqrt{\zeta}}  \right) \nonumber \\
&& - \exp(-\gamma \sqrtij) \nonumber\\
&& \times \erfc \left(  \frac{\zeta \sqrtij - \gamma/2}{\sqrt{\zeta}} \right)\Bigg\},
\label{intijscaled}
\end{eqnarray}
where $I(l_1,l_2,z-z_a)$ is defined by Equation~(\ref{intij}).

Function $\mathbf{A}$ then takes the form
\begin{eqnarray}
\Aewa{x} & = & \sum_{\substack{i_1,i_2 \\ i_1^2+(b i_2)^2  \leq 10}}
\left\{ \frac{2 \alpha}{\sqrt{\pi}} \frac{\exp(-\alpha^2 \uid)}{\uid} + 
\frac{ \erfc(\alpha \sqrt{\uid})}{\uid^{3/2}} \right\} (x-x_a-i_1 L_x) \nonumber \\
&& + \frac{2}{b L_x^2} \sum_{\substack{l_1,l_2 \\ l_1^2+(l_2/b)^2  \leq 10}} l_1 \sin (\vld) \widetilde{I} (l_1,l_2,z-z_a),
\label{a2accx2p1i}
\end{eqnarray}
\begin{eqnarray}
\Aewa{y} & = & \sum_{\substack{i_1,i_2 \\ i_1^2+(b i_2)^2  \leq 10}}
\left\{ \frac{2 \alpha}{\sqrt{\pi}} \frac{\exp(-\alpha^2 \uid)}{\uid} + 
\frac{ \erfc(\alpha \sqrt{\uid})}{\uid^{3/2}} \right\} (y-y_a-i_2 b L_x) \nonumber \\
&& + \frac{2}{b^2 L_x^2} \sum_{\substack{l_1,l_2 \\ l_1^2+(l_2/b)^2  \leq 10}} l_2 \sin (\vld) \widetilde{I} (l_1,l_2,z-z_a),
\label{a2accy2p1i}
\end{eqnarray}
\begin{eqnarray}
\Aewa{z} &  = &  \sum_{\substack{i_1,i_2 \\ i_1^2+(b i_2)^2  \leq 10}}
\left\{ \frac{2 \alpha}{\sqrt{\pi}} \frac{\exp(-\alpha^2 \uid)}{\uid} + 
\frac{ \erfc(\alpha \sqrt{\uid})}{\uid^{3/2}} \right\} (z-z_a) \nonumber \\
&& - \frac{2}{b L_x^2} \sum_{\substack{l_1,l_2 \\ l_1^2+(l_2/b)^2  \leq 10}} \cos (\vld) I'(l_1,l_2,z-z_a).
\label{a2accz2p1i}
\end{eqnarray}

\subsection{Mixed boundary conditions of type 1P2I}

Here, we introduce
\begin{eqnarray}
\uij &  = & (x-x_a-i_1 L_x)^2 + (y-y_a)^2 + (z - z_a)^2, \label{a2subst1p2i1} \\
\vlj & = & \frac{2 \pi l_1 (x - x_a)}{L_x}, \label{a2subst1p2id2}
\label{a2subst1p2i}
\end{eqnarray}
which simplifies the formula for function $\mathbf{A}$ to
\begin{eqnarray}
\Aewa{x} & = &  \sum_{i_1, i_1^2  \leq 10}
\left\{ \frac{2 \alpha}{\sqrt{\pi}} \frac{\exp(-\alpha^2 \uij)}{\uij} + 
\frac{ \erfc(\alpha \sqrt{\uij})}{\uij^{3/2}} \right\} (x-x_a-i_1 L_x) \nonumber \\
&& + \frac{4 \pi}{L_x^2} \sum_{l_1, l_1^2 \leq 10} l_1 \exp{(-\zeta l_1^2)} \sin (\vlj) K(l_1,y-y_a,z-z_a),
\label{a2accx1p2i}
\end{eqnarray}
\begin{eqnarray}
\Aewa{y} &  = &  \sum_{i_1, i_1^2  \leq 10}
\left\{ \frac{2 \alpha}{\sqrt{\pi}} \frac{\exp(-\alpha^2 \uij)}{\uij} + 
\frac{ \erfc(\alpha \sqrt{\uij})}{\uij^{3/2}} \right\} (y-y_a) \nonumber \\
&& + \frac{4 \pi}{L_x^2} \frac{y - y_a}{\sqrt{(y - y_a)^2 + (z - z_a)^2}} \times \nonumber \\
&& \sum_{l_1, l_1^2 \leq 10} \exp{(-\zeta l_1^2)} \cos (\vlj) M(l_1,y-y_a,z-z_a),
\label{a2accy1p2i}
\end{eqnarray}
\begin{eqnarray}
\Aewa{z} &  = & \sum_{i_1, i_1^2  \leq 10}
\left\{ \frac{2 \alpha}{\sqrt{\pi}} \frac{\exp(-\alpha^2 \uij)}{\uij} + 
\frac{ \erfc(\alpha \sqrt{\uij})}{\uij^{3/2}} \right\} (z-z_a) \nonumber \\
&& + \frac{4 \pi}{L_x^2} \frac{z - z_a}{\sqrt{(y - y_a)^2 + (z - z_a)^2}} \times \nonumber \\
&& \sum_{l_1, l_1^2 \leq 10} \exp{(-\zeta l_1^2)} \cos (\vlj) M(l_1,y-y_a,z-z_a),
\label{a2aewy1p2i}
\end{eqnarray}
where $K(l_1,y-y_a,z-z_a)$ is given by Equation~(\ref{int00p1}), 
and function $M(l_1,y-y_a,z-z_a) \equiv -\dd K(l_1, \eta (y-y_a,z-z_a)) /\dd \eta$ is
\begin{equation}
M(l_1,y-y_a,z-z_a) = \int_0^{\infty} \frac{J_1 (\eta q) \exp(-\zeta q^2) q^2}{l_1^2+q^2} \dq, 
\label{e4fceM}
\end{equation}
where $J_1$ is the Bessel function of the first kind and first order.
Note that variables $\zeta$, $\gamma$ and $\eta$ are defined in section \ref{sec_BND_mixed}.

\section{Code runtime parameters}
\label{ap:runpar}

Here we list runtime parameters of the tree solver and {\sc Gravity} and {\sc
OpticalDepth} modules that can be set in the {\tt flash.par} configuration file.
Apart from parameters discussed in the main body of this work (e.g. MAC
selection, accuracy limit), the code should work well with the default
parameters. Additional information is provided in the Flash Users Guide and
directly in the source code as comments.


\subsection{Tree solver parameters}


\begin{description}
\item[{\tt gr\_bhPhysMACTW}] -- indicates whether MACs of physical modules (e.g.
Gravity) are used during tree-walks; if false, the geometric BH MAC is used
instead (type: logical, default: false)
\item[{\tt gr\_bhPhysMACComm}] -- indicates whether MACs of physical modules (e.g.
Gravity) are used for communication of block-trees; if false, the geometric BH
MAC is used instead (type: logical, default: false)
\item[{\tt gr\_bhTreeLimAngle}] -- maximum opening angle, $\theta_\mathrm{lim}$, of
the geometric BH MAC (type: real, default: $0.5$)
\item[{\tt gr\_bhTreeSafeBox}] -- relative (w.r.t. to the block size) size of a
cube around each block, $\eta_\mathrm{SB}$, in which the target point cannot be
located (type: real, default: $1.2$)
\item[{\tt gr\_bhUseUnifiedTW}] -- obsolete, will be deleted in future versions
\item[{\tt gr\_bhTWMaxQueueSize}]: maximum number of elements in the priority
queue (type: integer, default: 10000)
\item[{\tt gr\_bhAcceptAccurateOld}] -- indicates whether Adaptive Block Update
(see \S\ref{sec:abu}) is active; will be renamed to {gr\_bhABU} in future
versions (type: logical, default: false)
\item[{\tt gr\_bhLoadBalancing}] -- indicates whether Load Balancing (see
\S\ref{sec:abu}) is active (type: logical, default: false)
\item[{\tt gr\_bhMaxBlkWeight}] -- maximum workload weight, $\omega_\mathrm{wl}$
(type: real, default: $10$)
\end{description}

\subsection{Gravity module parameters}

\begin{description}
\item[{\tt grv\_bhNewton}] -- Newton's constant of gravity; if
negative, the value is obtained from the Flash internal database of physical
constants (type: real, default: -1)
\item[{\tt grv\_bhMAC}] -- type of Multipole Acceptance Criterion
(MAC) calculated by the Gravity module if {\tt gr\_bhPhysMACTW} or {\tt
gr\_bhPhysMACComm} is set true; currently accepted values are:
"ApproxPartialErr", "MaxPartialErr" and "SumSquare" (experimental), (type:
string, default: "ApproxPartialErr")
\item[{\tt grv\_bhMPDegree}] -- degree of multipole expansion used to estimate
the error of a single node contribution with APE and MPE MACs; {\tt
grv\_bhMPDegree} corresponds to $p+1$ used in Equations~(\ref{eq:mpe} and
(\ref{eq:ape}); (type: integer, default: $2$)
\item[{\tt grv\_bhUseRelAccErr}] -- indicates whether the {\tt grv\_bhAccErr}
parameter (below) should be interpreted as a relative error limit,
$\epsilon_\mathrm{lim}$ (true), or an absolute error limit, $a_\mathrm{lim}$
(false); see Equations~(\ref{eq:abs_err}) and (\ref{eq:rel_err}); (type:
logical, default: false)
\item[{\tt grv\_bhAccErr}] -- maximum allowed error set either relatively with
respect to the acceleration from the previous time-step,
$\epsilon_\mathrm{lim}$, or absolutely, $a_\mathrm{lim}$; (type: real, default:
$0.1$)

\item[{\tt grav\_boundary\_type}] -- type of boundary conditions for gravity for
all directions; the accepted values are: "isolated","periodic" and "mixed"; if
set to "mixed", BCs in individual directions are set by the parameters below
(type: string, default: "mixed")
\item[{\tt grav\_boundary\_type\_x}] -- type of gravity BCs in the
$x$-direction; the accepted values are: "isolated" and "periodic" (type: string,
default: "isolated")
\item[{\tt grav\_boundary\_type\_y}] -- same as {\tt grav\_boundary\_type\_x}
but in the $y$-direction
\item[{\tt grav\_boundary\_type\_z}] -- same as {\tt grav\_boundary\_type\_x}
but in the $z$-direction
\item[{\tt grv\_bhEwaldSeriesN}] -- number of terms used in the expansion given by
Equation~(\ref{ewpotential}) to calculate the Ewald field (type: integer,
default: $10$)
\item[{\tt grv\_bhEwaldAlwaysGenerate}] -- indicates whether the Ewald field
should be regenerated at the simulation start; if false, it is read from file
with name given by parameters {\tt grv\_bhEwaldFName} or {\tt
grv\_bhEwaldFNameAccV42} and {\tt grv\_bhEwaldFNamePosV42} (type: logical,
default: true)

\item[{\tt grv\_bhEwaldFieldNxV42}] -- number of points of the Ewald field
lookup table in the $x$-direction when the first approach described in
\S\ref{sec:ewald:lookup} is used (default in {\sc Flash} versions up to 4.2);
(type: integer, default: $32$)
\item[{\tt grv\_bhEwaldFieldNyV42}] -- same as the preceeding parameter but for the $y$-direction
\item[{\tt grv\_bhEwaldFieldNzV42}] -- same as the preceeding parameter but for the $z$-direction
\item[{\tt grv\_bhEwaldNRefV42}] -- number of nested grid levels of the Ewald
field when the first approach described in \S\ref{sec:ewald:lookup} is used;
if negative, the number of nested grid levels is calculated automatically from
the minimum cell size (type: integer, default: $-1$)
\item[{\tt grv\_bhLinearInterpolOnlyV42}] -- indicates whether the linear
interpolation in the Ewald field is used (with the first approach described in
\S\ref{sec:ewald:lookup}); if false, then the more expensive and accurate quadratic
interpolation is used for some calculations (type: logical, default: true)
\item[{\tt grv\_bhEwaldFNameAccV42}] -- name of file to store the Ewald field
accelerations when the first approach described in \S\ref{sec:ewald:lookup}
is used (type: string, default: "ewald\_field\_acc")
\item[{\tt grv\_bhEwaldFNamePotV42}] -- name of file to store the Ewald field
potential when the first approach described in \S\ref{sec:ewald:lookup}
is used (type: string, default: "ewald\_field\_pot")
\item[{\tt grv\_bhEwaldNPer}] -- number of points in each direction of the Ewald
field coefficients when the second approach described in \S\ref{sec:ewald:lookup}
is used (type: integer, default: $32$)
\item[{\tt grv\_bhEwaldFName}] -- name of file to store the Ewald field
coefficients in the case the second approach described in \S\ref{sec:ewald:lookup}
is used (type: string, default: "ewald\_coeffs")

\item[{\tt grv\_useExternalPotential}] -- indicates whether the external
time-independent gravitational potential read from file is used (type: logical,
default: false)
\item[{\tt grv\_usePoissonPotential}] -- indicates whether the potential (or
accelerations) computed by the (tree) Poisson solver is used (type: logical,
default: true)
\item[{\tt grv\_bhExtrnPotFile}] -- name of file with the external gravitational
potential (type: string, default: "external\_potential.dat"
\item[{\tt grv\_bhExtrnPotType}] -- symmetry of the external gravitational
potential; currently, two options are available: "spherical" and "planez"
(plane-parallel, varying along the $z$-direction); (type: string, default: "planez")
\item[{\tt grv\_bhExtrnPotCenterX}] -- center of the external potential
$x$-coordinate given in the {\sc Flash} internal coordinates (type: real, default: $0$)
\item[{\tt grv\_bhExtrnPotCenterY}] -- same as the preceeding parameter but for the $y$-direction
\item[{\tt grv\_bhExtrnPotCenterZ}] -- same as the preceeding parameter but for the $z$-direction

\end{description}

\subsection{OpticalDepth module parameters}

\begin{description}

\item[{\tt tr\_nSide}] -- level of the {\sc HealPix} grid; number of pixels is
$N_{_{\rm PIX}} = 12\times 4^{({\mathtt tr\_nSide}-1)}$ (type: integer, default:
$1$)
\item[{\tt tr\_ilNR}] -- number of points in the radial direction for the
calculation of the fraction of node that intersects with a given ray (type:
integer, default: $50$)
\item[{\tt tr\_ilNTheta}] -- number of points in the $\theta$-direction of the
table recording a fraction of the node that intersects with a ray at a given
$\theta$ (type: integer, default: $25$)
\item[{\tt tr\_ilNPhi}] -- number of points in the $\phi$-direction of the
node-ray intersection table (type: integer, default: $50$)
\item[{\tt tr\_ilNNS}] -- number of points describing the angular node size in
the node-ray intersection table (type: integer, default: $25$)
\item[{\tt tr\_ilFinePix}] -- number of additional pixels in each angular
directions used to calculate the node-ray intersection table (type: integer,
default: $4$)
\item[{\tt tr\_bhMaxDist}] -- maximum distance from a target point up to which
the optical depth is calculated (type: real, default: $10^{99}$)
\item[{\tt tr\_odCDTOIndex}] -- exponent relating the gas density to the
absorption coefficient used during the calculation of the optical depth in a
given direction (type: real, default: $1$)

\end{description}


\label{lastpage}

\end{document}